\documentclass[10pt]{article}

\usepackage[UKenglish]{babel}
\usepackage{graphicx}
\usepackage{subcaption}
\usepackage{amsmath}
\usepackage{booktabs}
\usepackage{geometry}
\usepackage{cleveref}
\usepackage{setspace}
\usepackage{authblk}
\usepackage{parskip}
\usepackage[sorting=none,autocite=superscript]{biblatex}
\usepackage{titlesec}
\usepackage{pdfpages}

\crefname{figure}{Fig.}{Figs.}
\Crefname{figure}{Fig.}{Figs.}
\crefname{table}{Table}{Tables}
\Crefname{table}{Table}{Tables}

\titleformat*{\section}{\large\bfseries}
\titleformat*{\subsection}{\normalsize\bfseries}
\titleformat*{\subsubsection}{\normalsize\itshape}

\DeclareMathOperator\arctanh{arctanh}

\begin{document}

\pagenumbering{gobble}  %

\title{Generalisation of automatic tumour segmentation in histopathological whole-slide images across multiple cancer types}

\author[1]{Ole-Johan Skrede\textsuperscript{*}}
\author[1]{Manohar Pradhan}
\author[1]{Maria Xepapadakis Isaksen}
\author[1]{Tarjei Sveinsgjerd Hveem}
\author[1]{Ljiljana Vlatkovic}
\author[3,7]{Arild Nesbakken}
\author[2,7]{Kristina Lindemann}
\author[1,2]{Gunnar B Kristensen}
\author[15]{Jenneke Kasius}
\author[16]{Alain G Zeimet}
\author[7,8]{Odd Terje Brustugun}
\author[10,11]{Lill-Tove Rasmussen Busund}
\author[10,11]{Elin H Richardsen}
\author[1,12]{Erik Skaaheim Haug}
\author[4]{Bjørn Brennhovd}
\author[13,14]{Emma Rewcastle}
\author[13,14]{Melinda Lillesand}
\author[5]{Vebjørn Kvikstad}
\author[13,14]{Emiel Janssen}
\author[17]{David J Kerr}
\author[1,6]{Knut Liestøl}
\author[1,6]{Fritz Albregtsen}
\author[1,6,9]{Andreas Kleppe}

\affil[1]{Institute for Cancer Genetics and Informatics, Oslo University Hospital, Oslo, Norway}
\affil[2]{Department of Gynaecological Oncology, Oslo University Hospital, Oslo, Norway}
\affil[3]{Department of Gastrointestinal, Oslo University Hospital, Oslo, Norway}
\affil[4]{Department of Urology, Oslo University Hospital, Oslo, Norway}
\affil[5]{Department of Forensic Medicine, Oslo University Hospital, Oslo, Norway}
\affil[6]{Department of Informatics, University of Oslo, Oslo, Norway}
\affil[7]{Institute of Clinical Medicine, University of Oslo, Oslo, Norway}
\affil[8]{Section of Oncology, Vestre Viken Hospital Trust, Drammen, Norway}
\affil[9]{Centre for Research-based Innovation Visual Intelligence, UiT The Arctic University of Norway, Tromsø, Norway}
\affil[10]{Department of Medical Biology, UiT The Arctic University of Norway, Tromsø, Norway}
\affil[11]{Department of Clinical Pathology, University Hospital of North Norway, Tromsø, Norway}
\affil[12]{Department of Urology, Vestfold Hospital Trust, Tønsberg, Norway}
\affil[13]{Department of Pathology, Stavanger University Hospital, Stavanger, Norway}
\affil[14]{Department of Chemistry, Bioscience and Environmental Engineering, University of Stavanger, Stavanger, Norway}
\affil[15]{Department of Gynecological Oncology, Amsterdam University Medical Centres, Centre for Gynecological Oncology Amsterdam, Amsterdam, The Netherlands}
\affil[16]{Department of Obstetrics and Gynaecology, Comprehensive Cancer Center Innsbruck, Innsbruck Medical University, Innsbruck, Austria}
\affil[17]{Nuffield Division of Clinical Laboratory Sciences, University of Oxford, Oxford, UK}

\date{}

\makeatletter\def\@maketitle{%
  \newpage
  \null%
  \vskip 2em%
    \let \footnote \thanks
    {\large \bf \@title \par}
    \vskip 1.5em
    {
      \normalsize
      \lineskip .5em
        \@author
      \par
    }
    \vskip 1em
    {\large \@date }
  \par
  \vskip 1.5em
}

\maketitle

\textsuperscript{*}Corresponding author:\\
Ole-Johan Skrede\\
Institute for Cancer Genetics and Informatics, Oslo University Hospital\\
NO-0424 Oslo, Norway\\
olejohas@ifi.uio.no

\clearpage
\begin{abstract}
\noindent Deep learning is expected to aid pathologists by automating tasks such as tumour
segmentation.
We aimed to develop one universal tumour segmentation model for histopathological images and
examine its performance in different cancer types.
The model was developed using over 20\,000 whole-slide images from over 4\,000 patients with
colorectal, endometrial, lung, or prostate carcinoma.
Performance was validated in pre-planned analyses on external cohorts with over 3\,000
patients across six cancer types.
Exploratory analyses included over 1\,500 additional patients from The Cancer Genome Atlas.
Average Dice coefficient was over 80\% in all validation cohorts with \emph{en bloc}
resection specimens and in The Cancer Genome Atlas cohorts.
No loss of performance was observed when comparing the universal model with models
specialised on single cancer types.
In conclusion, extensive and rigorous evaluations demonstrate that generic tumour
segmentation by a single model is possible across cancer types, patient populations,
sample preparations, and slide scanners.

\end{abstract}

\clearpage
\pagenumbering{arabic}  %
\section{Introduction}

The rapidly increasing adoption of digital pathology enables workflows that contribute
towards the realization of precision medicine.\autocite{hanna2022integrating}
In particular, the introduction of methods based on modern artificial intelligence (AI)
promises for improved selection among therapeutic options.\autocite{rajpurkar2022ai}
Provided sufficient representative data, these AI-based methods outperform earlier
automatic procedures, and may help ease routines, improving the precision in diagnostic
tasks,
and allowing the pathologists to focus on especially challenging problems.
Advances in technology also enable efficient collection of
multiple samples from each patient.
However, the subsequent analyses of these samples will further increase the pressure on
pathological services already affected by increasing cancer incidences and a general
shortage of pathologists.\autocite{soerjomataram2021planning,world2023world}
Thus, automatic procedures also in the analytic steps of the diagnostic processes may be
vital to bring diagnostic advances to practical use in the clinic.
The development of bright-field slide scanners enabled the production of
high-quality whole-slide images (WSI) of tumour slides.
A first step in many analyses of such WSIs, especially automatic analyses, is the
segmentation of the tumour areas from the background.
The modern deep learning AI-techniques have demonstrated high efficiency
in recognizing patterns in images, thus making automatic tumour segmentation an attractive and
realistic alternative to manual tumour segmentation.\autocite{niazi2019digital,skrede2020deep}
Such automatic procedure may also produce heat maps highlighting regions of particular
interest, assisting pathologists in their assessment.

In this study, we aimed to develop a universal deep learning model for automatic tumour
segmentation in whole-slide images of haematoxylin and eosin (H\&E) stained tissue
sections from formalin-fixed, paraffin-embedded (FFPE) tissue blocks.
Most previously published models are developed and tested using data from a single
cancer type (e.g.\ lung, prostate, or breast cancer).
Some relevant studies presents results from multiple cancer types, but none of them
present performance estimates in cancer types different from the one used to train the
model.\autocite{van2021hooknet,ciga2021learning,schmitz2021multi,khened2021generalized,ho2023deep,frank2023accurate}
Recently, we have seen \emph{foundation models}\autocite{bommasani2021opportunities}
published for computational pathology which are trained by self-supervised learning on
histological images from multiple cancer types.\autocite{%
chen2024towards,  %
xu2024whole,  %
campanella2024computational, %
nechaev2024hibou, %
filiot2024phikon, %
zimmermann2024virchow2, %
alber2025atlas, %
}
These models are pan-cancer by design, and can be utilised for tumour segmentation if
additional segmentation-specific components that also needs to be trained are attached.

Although focusing on a single cancer type can ease utilization
features characteristic to the specific cancer type, this also limits the
applicability of the resulting model.
A pan-cancer model may tend to focus on more general characteristics
in cancer tissues, can be trained and validated on large data volumes, and can be
applied on multiple cancer types, including rare cancers with insufficient amounts of
available data to train a specialised model.
A universal tumour segmentation model trained on much and varied data may also be expected to
be more robust and generalise better than specialised models.
There is, however, also a need to study the limits of these models in terms of elements
such as technical quality of the input,
different tissue types, and sociodemographic variations.

\begin{figure}[h]
  \centering
  \includegraphics[width=\textwidth]{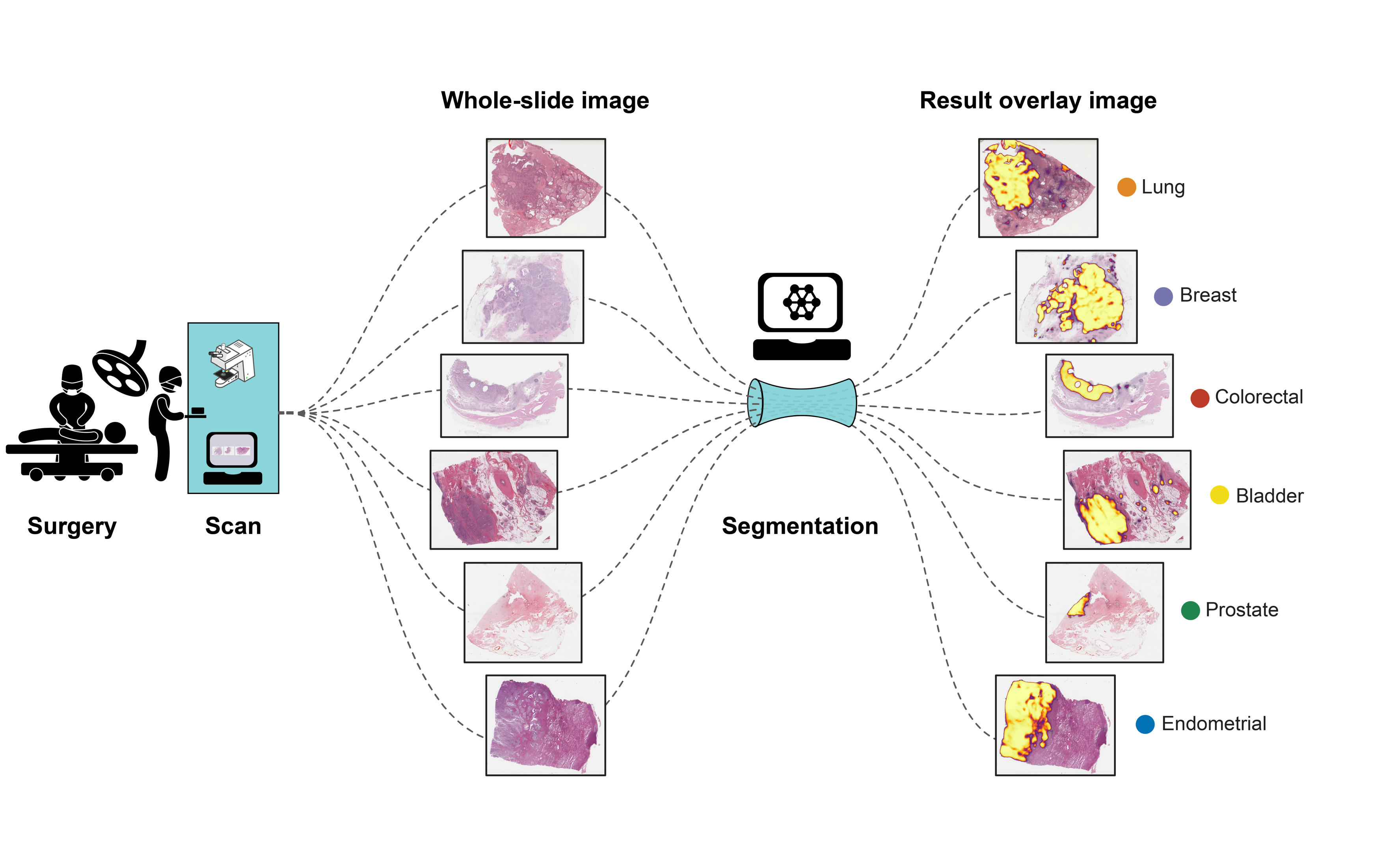}
  \caption[Overview]{%
    \textbf{Overview of input and corresponding result}\\
  }\label{fig:segmentation-overview}
\end{figure}

To examine pan-cancer segmentation in WSIs of H\&E stained tissue sections, we here
present the performance of a single segmentation model developed using cohorts from
colorectal, endometrial, lung, and prostate carcinoma.
The performance of the model is validated in a pre-specified primary analysis using
independent cohorts from the same four cancer types as well as from breast and bladder
carcinoma (\cref{fig:segmentation-overview}).
Combined, the included cancer types represent about 40\textendash45\% of both new cancer cases and cancer
deaths worldwide in 2020.\autocite{sung2021global}
Pre-specified secondary analyses compare the model with single-cancer models and
evaluate its robustness on images from different slide scanners.
To illustrate the level of uncertainty in the segmentation task, we also report the
intra- and inter-observer variability of two pathologists on a breast cancer cohort.
Further exploratory analyses include performance evaluation in four different cohorts
from The Cancer Genome Atlas (TCGA) and examination of factors that may lead to
suboptimal segmentation results.

\section{Results}

\subsection{Materials}

\begin{figure}[h]
  \centering
  \includegraphics[width=\textwidth]{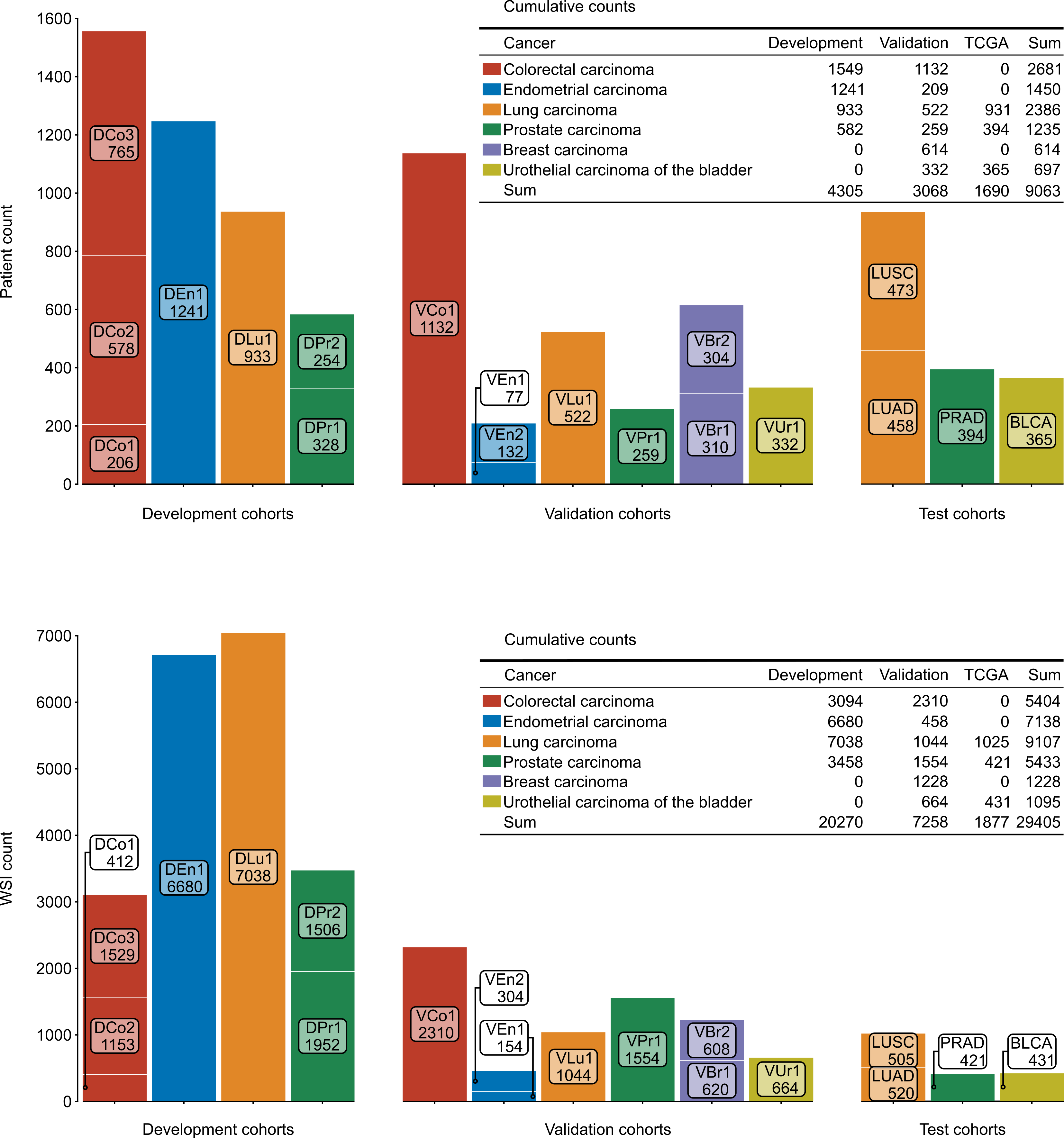}
  \caption[Material count]{%
    \textbf{Included patient and WSI count}\\
  }\label{fig:material-count}
\end{figure}

To develop the primary segmentation model, we used 20\,270 WSIs from 4\,305 patients
encompassing four types of cancer obtained using two different microscope scanners.
The pre-planned primary analysis assessed the performance by comparing the automatic
segmentation with a manual reference segmentation in 3\,629 WSIs from 3\,068 patients
and six cancer types.
Additional exploratory analyses included evaluating 1\,877 WSIs from 1\,690 patients and three
cancer types from TCGA.\
See \cref{fig:material-count} for patient and WSI counts stratified by cohort and grouped
by cancer type and use, and Methods for further details about the included patient
cohorts.

\subsection{Primary analysis of model performance}

\begin{figure}[h]
  \centering
  \includegraphics[width=\textwidth]{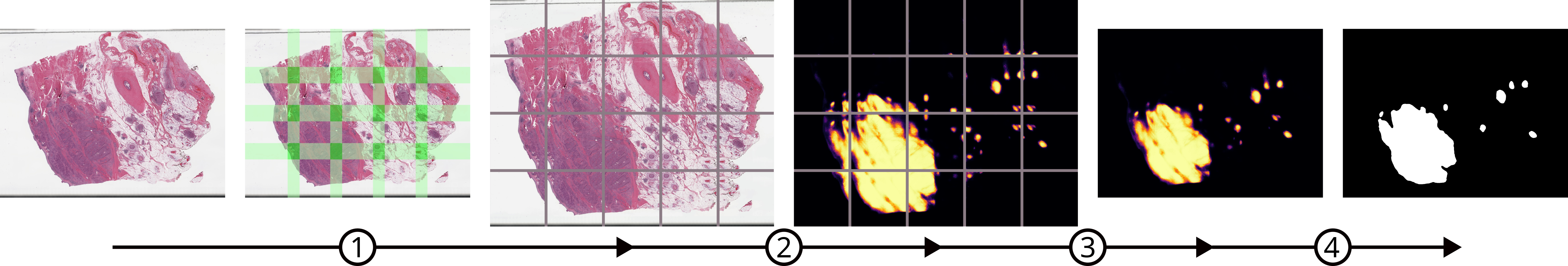}
  \caption[Pipeline]{%
    \textbf{Segmentation method pipeline}\\
  }\label{fig:inference-pipeline}
\end{figure}

Figure~\ref{fig:inference-pipeline} illustrates how the method segments a WSI
by creating an image indicating the probability of a pixel being part of a region
displaying tumour, before the final dichotomous tumour segmentation is
obtained through a thresholding procedure.
The performance of the automatic segmentation is evaluated by comparing it with the
manual segmentation using the Dice similarity coefficient (DSC), and the regions
involved in this computation are illustrated by example in \cref{fig:segmentation-example}c. 

The primary segmentation model achieved a mean DSC of 82\% to 94\% in the validation
cohorts with \emph{en bloc} specimens of solid tumours imaged with the Aperio AT2 scanner
(\cref{fig:full-run-0-aperio-dice}).
This includes two cohorts from breast carcinoma, a cancer type not present in the
development materials.
Especially good performance was observed for endometrial carcinoma (cohorts VEn1 and
VEn2) with DSC well over 90\%.
The exception from these satisfactory results was for the VUr1 cohort with transurethral
resection (TUR) specimens from early-stage urothelial carcinoma of the bladder.

Additional performance evaluation metrics were also computed (see Supplementary
Table S1 and Supplementary Fig. S1).
Except for the TUR-sample bladder cohort, the proportion of tissue marked as tumour
was similar when segmented automatically and manually, with only a slight tendency for
the automatic procedure to mark more as tumour in colorectal (VCo1) and lung cancer
(VLu1).
For VLu1, this tendency is also reflected in a high true positive rate (sensitivity) of
91\% compared with a slightly lower true negative rate (specificity) of 88\%, indicating some
over-segmentation in this cohort.
A difference between sensitivity and specificity is also seen in the cohorts from endometrial cancer where
(sensitivity, specificity) are (96\%, 96\%) and (92\%, 93\%) for VEn1 and VEn2, respectively.
Conversely, the automatic segmentation displays higher specificity than sensitivity in the
prostate and breast cancer cohorts, with differences ranging from 0.08 to 0.15.

The area of every segmented region in the validation cohorts was measured and the
distributions are presented in Supplementary Fig. S18.
The manually annotated tumour region size distribution is similar in all validation
cohorts, except for the lung cohort (VLu1) where there are more small regions
(area less than 1 mm\textsuperscript{2}) per WSI, and in the bladder cohort (VUr1) where there are
even more small regions per WSI\@.
The size distributions for the regions automatically segmented by the primary model are
similar in all validation cohorts, with a tendency towards more small regions in the
prostate, breast and bladder cohorts.
This results in many small false negative regions in VLu1 and VUr1, and some more small
false positive regions in the prostate and breast cohorts compared with the other
validation cohorts.
The average DSC is high in all validation cohorts when only considering true positive
regions, but the fraction of images containing true positive regions is considerably
lower in the bladder cancer cohort than in the other validation cohorts (Supplementary
Table S9).

\subsection{Factors affecting segmentation performance}

The DSC of the primary model correlated positively with both the manually segmented tumour area
and prevalence of tumour in all validation cohorts (Supplementary
Fig. S9) with Spearman's rank correlation coefficient $\rho >
0.36$ and p value $< 0.0001$ except for in endometrial carcinoma, where only VEn2 was
significantly correlated with area ($\rho = 0.29, \text{p} = 0.0004$) and prevalence
($\rho = 0.38, \text{p} < 0.0001$).
DSC also showed some correlation with known risk predictors, e.g.\ with pathological T
stage (pT) in VLu1 ($\rho = 0.10$, $\text{p} = 0.017$), VPr1 ($\rho = 0.11$, $\text{p} = 0.021$)
and VBr1 ($\rho = 0.28$, $\text{p} < 0.0001$), with Nottingham prognostic
index in the breast cancer cohorts VBr1 ($\rho = 0.18$, $\text{p} = 0.0017$) and VBr2
($\rho = 0.25$, $\text{p} < 0.0001$), and with Gleason score in VPr1 ($\rho = 0.12$,
$\text{p} = 0.0010$).
See Supplementary Section 2.1 for additional correlations
between the DSC of the primary model and selected variables in the validation cohorts.

\subsection{Comparison with models specialised on single cancer types}

Four cancer type specialised models were developed on subsets of the full training data;
the specialised colorectal model was trained using only the cohorts from colorectal
carcinoma (DCo1, DCo2, DCo3), and vice versa for the specialised endometrial, lung, and
prostate models.
The four cancer type specialised models achieved similar results as the
general primary model when tested on validation cohorts from the same cancer types they were
trained on (see performance overview in
\cref{fig:all-models-validation-test-aperio-dice} with details in Supplementary
Section 1.3), with a mean difference in DSC model below 0.007
between the general and specialised model for all specialised models
(Supplementary Fig. S17).
Thus, the larger and more varied training data of the pan-cancer model seems to have
compensated for the specific features that the specialised models may utilize.
In general, the cancer type specialised models failed to generalise beyond their respective
cancer types, with exceptions including the lung model that performed well in cohorts
from endometrial carcinoma.

\subsection{Robustness to variations in sample origin, preparation and imaging}

The performance of the primary model in images from the Aperio AT2 scanner was
preserved the images from the NanoZoomer XR scanner (\cref{fig:primary-model-validation-aperio-vs-xr-dice}).
Scan-by-scan comparisons reveal no particular shift between scans from Aperio AT2 and
NanoZoomer XR, with a mean difference DSC below 0.006 in cohorts from colorectal,
endometrial, and lung cancer, and a mean difference DSC below 0.014 in cohorts from
breast cancer (Supplementary Fig. S17).
Moreover, from \cref{fig:primary-model-vco1-five-scanners-dice}, we see that the
evaluation of the primary model on colorectal cancer (VCo1) scanned with five different
scanners show no particular performance shift between the scanner models, with the
largest reduction from the DSC of 84.6\% in AperioAT2 seen on Aperio GT450dx with DSC of
82.9\% (more details in Supplementary Section 2.5).
Robustness to both external laboratory sample preparation and imaging was
demonstrated by achieving a mean DSC over 83\% in all included TCGA cohorts BLCA, LUAD,
LUSC and PRAD (\cref{fig:full-run-0-aperio-dice}).

When comparing the performance in the validation cohorts with the cohorts used to
develop the model, we see from \cref{fig:full-run-0-aperio-dice} that the performance is
better in the development cohorts from colorectal (91.18\%, 90.96\% and 89.88\% vs
84.54\%) and lung cancer (87.69\% vs 82.22\%), while it is more similar in the cohorts
from endometrial (94.17\% vs 95.28\% and 93.40\%) and prostate cancer (83.75\% and
84.30\% vs 84.36\%).

Finally, the performance of the primary model in all validation and test cohorts was
similar in two models trained with an identical setup as the primary model, only differing by
using different random seeds which affects model weight initialisation and input order
(see result overview in \cref{fig:all-models-validation-test-aperio-dice} with details
in Supplementary Section 1.4).
Per-scan comparisons show an absolute mean difference in DSC between the primary and the
replicated results below 0.005 in all non-TUR validation cohorts except for VLu1 where the
difference is 0.009 and 0.013 for the first and second replication, respectively
(Supplementary Fig. S17).

\subsection{Intra- and inter pathologist variability}

All 304 Aperio AT2 scans in the breast carcinoma validation cohort VBr2 were annotated
for tumour a second time by pathologists MP and LV, about two years after this cohort
had been annotated the first time by MP.\
The DSC between the first and second segmentations of MP was 91\%, while the DSC between
LV and the second segmentation of MP was 77\%.
For comparison, the DSC between the primary automatic segmentation and both the first
and the second segmentations by MP was 88\% (Supplementary Fig. S19).
MP and LV had segmented overlapping regions in all 304 scans in VBr2, while the
automatic primary model did not segment any regions in four scans (1.32\%).

\subsection{Failure to segment fragmented samples from early-stage tumours in bladder}

The primary model did not segment any regions in 108 (33\%) of the 332 scans in the
bladder validation cohort (VUr1) even though all of them show tumour tissue.
A likely reason is that the tumour samples obtained through TUR are generally small
and from early-stage tumours.
In the 342 (79\%) scans without fragmented tissue in the TCGA bladder cohort BLCA, the
model failed to detect any cancerous regions in only 8 scans (2\%) with a mean DSC of
91\% in the 334 (98\%) scans with predictions.
The performance degraded when considering the 87 (20\%) scans with fragmented tissue: 12 (14\%)
scans had no predicted cancerous regions and the mean DSC was 77\% in the 75 (86\%) scans
with predictions.

In VUr1, 255 (77\%) scans are from pTa, 1 (less than 1\%) scan is from pTis, and 76 (23\%) are from pT1
(study protocol Table 8 in Supplementary Section 6).
In the pTa or pTis group, no regions were segmented in 38\% of the scans, but this proportion
decreased to 13\% in pT1.
For the more advanced stage cases of the BLCA cohort (pT2: n = 112 (26\%),
pT3: n = 203 (47\%), pT4: n = 58 (13\%)), results were markedly better with a mean DSC of
84\% in the whole cohort (Supplementary Table S10).

\subsection{Performance comparison with MedSAM}

To provide context for our results, we evaluated the performance of MedSAM in all
validation datasets.\autocite{ma2024segment}
MedSAM is presented as a foundation model for medical image segmentation, developed by
finetuning the segment anything model (SAM) on a large dataset of medical
images.\autocite{kirillov2023segment}
MedSAM requires prompting, that is, some marker in the image to indicate where the
regions to segment are located.
For this reason, we evaluated two versions of MedSAM, one where the prompt was the
bounding box of the tissue foreground region, and another where we used the bounding box
of the manually segmented tumour areas as prompts.

MedSAM prompted by a bounding box of the manually segmented tumour achieved a mean DSC
of 79\%, 89\%, 87\%, 72\%, 66\%, 81\%, 83\% and 75\% when applied on the Aperio AT2
scans from validation cohorts VCo1, VEn1, VEn2, VLu1, VPr1, VBr1, VBr2 and VUr1,
respectively (Supplementary Fig. S22).
When prompted by a bounding box of the whole tissue foreground, MedSAM achieved a mean
DSC of 48\%, 63\%, 53\%, 47\%, 28\%, 34\%, 42\% and 64\% in the same datasets.
The bounding boxes of the manually segmented tumour without refined segmentation by
MedSAM achieved a mean DSC of 74\%, 82\%, 78\%, 67\%, 60\%, 70\%, 73\% and 74\%.

\section{Discussion}

A deep learning model developed to automatically delineate cancerous regions in WSIs of
conventionally H\&E-stained tissue sections demonstrated good overall performance in
external validation cohorts from different cancer types, including breast cancer not represented in the
development set.
Comparing the pan-cancer model to specialised models developed and tested on cohorts of
one cancer type indicated no loss of performance, neither overall nor scan-by-scan.
This might be because the specialised models have been trained on a subset of the
general model's training data, and that the network has sufficient capacity to make
efficient use of the more comprehensive data.
However, that this good performance also extends to cancer types not present in the
training set (an ability we did not generally observe in the specialised models),
indicates that the pan-cancer model also can utilise more general features to
distinguish between cancerous and non-cancerous tissue.

Our primary performance evaluation metric, DSC, is a purely overlap-based metric which
is independent of true negative counts, and also invariant to a reference-prediction
swap.
To further examine nuances in the behaviour of the segmentation model, we included
additional analyses and performance metrics.
All additional statistics included in the first secondary analysis are derived from the
pixel overlap contingency table.
This provides insight into the kind of overlap (e.g.\ over-segmentation or
under-segmentation), but does not distinguish between disconnected regions in an image,
nor does it consider the shape of the regions.
Shape similarity between reference and predicted regions is not explicitly evaluated in
this study, but we designed the model to produce visually similar results as the
reference segmentation.

Another perspective of the segmentation performance is provided by the analysis of
individual segmented regions (connected pixels annotated as tumour).
This reveals that the size distributions of manual segmentations are different between
the cancer types, which, to a lesser extent, also is observed in the automatic
segmentations.
In particular, the behaviour of the automatic segmentation in cohorts from prostate and
breast carcinoma is similar, which might be explained by the similarities of the two
diseases.\autocite{risbridger2010breast}

From \cref{fig:full-run-0-aperio-dice}, we see that the mean DSC is determined by a
majority of scans with high DSC and a small minority with very low DSC, and that the
median DSC is substantially higher than the mean.
Related to this is the high DSC when only considering true positive regions (Supplementary
Table S9).
This can suggest that the dominant failure type is few completely failed segmentations,
rather than many partly failed segmentations.

The model's performance is highly correlated with the annotated tumour area size in
most cohorts.
Moreover, results with DSC less than 50\% (and in particular 0\%) mostly appear in
scans with an aggregated tumour area less than (10 mm)\textsuperscript{2}.
Inspection of the prediction probability images shows that these regions often have a
positive signal which is discarded in the final dichotomisation into tumour and
background (see \cref{fig:segmentation-example}d).
In general, if a more sensitive model is desired, one can lower the post processing
thresholds without requiring retraining of the underlying neural network.

Image preparation is a source of variation that might cause worse performance in
settings external to the development settings.
The segmentation model is developed and externally validated on scans from samples originating
from many institutions in many different countries and has been shown to perform
consistently across the differences in sample preparation and patient population.
Since both development and validation cohorts in lung and prostate are from
Norwegian hospitals, we evaluated the model on lung and prostate cohorts from TCGA
(LUAD, LUSC, and PRAD).
The performance was maintained, increasing our confidence that the model generalises
well.
We also included scans acquired using two different scanners to create a model that
produced similar results across scanners.
The model behaved similarly in the two included scanners,
both when considering individual slides and performance averaged over cohorts.
This result is corroborated with the similar performance of the
model when evaluated on VCo1 scanned with five different scanners.

We evaluated the model on all scans from the development cohorts as a check to see if
any substantial over-fitting had occurred.
Even though the performance is good in these cohorts, they are not out of line
compared with the results in the validation cohorts, suggesting that over-fitting is
limited.

For simplicity, we did not employ any hyperparameter tuning or model selection, nor did
we combine models to form an ensemble model.
This can come with a cost of repeatability, but the results from the replicated models,
both overall performance and scan-by-scan comparisons,
show that the performance of models resulting from our method is stable.

In the experiment where VBr2 was manually annotated a second time, the intra-observer
similarity was greater than the inter-observer variability, which we find reasonable.
What is perhaps more surprising is that the average DSC was higher between the automatic
segmentation and pathologist MP, than between pathologists LV and MP.\

It was challenging to find published automatic tumour segmentation methods that we could
apply without additional training and that would meaningful to compare against, and
MedSAM is not ideal since it requires prompting and is not developed primarily for
tumour segmentation in histological images.
The tissue prompted version is an example of a truly automatic method, while
the tumour prompted version could represent a scenario where a human expert use MedSAM
to segment areas of interest.
A more relevant approach to compare against would be a computational pathology
foundation model adapted for segmentation using methods such as ViT-Adapter and
Mask2Former, but this would have required additional
training.\autocite{chen2022vision,cheng2022masked}

MedSAM prompted by tissue bounding boxes performs substantially worse than the tumour
prompted version, which again performs substantially worse than our primary method on
VLu1 and VPr1.
In the other validation cohorts, its performance is lower but comparable to our primary
method, except for in VUr1, where its performance is substantially better.
However, the performance of the tumour prompted MedSAM in VUr1 can largely be explained
by the prompting bounding boxes, which without MedSAM, achieves almost the same
performance.

The poor performance in the validation set from urothelial carcinoma (VUr1) was probably
related to its origin from the TUR procedure, often resulting in scans of small, fragmented samples.
Regions where manual and automatic segmentations overlap are often correctly segmented, and
the poor general performance is dominated by regions completely missed by the automatic
segmentation.
Challenges with fragmented tumours and small fragments were also seen for
the TCGA urothelial carcinoma cohort BLCA.\
The observed correlation between the size of the tumour region and DSC confirms the problems
with detecting small areas of the tumour, which partly explains the poor performance in
VUr1, since it contains many small annotated tumour regions compared with the other
cohorts.
Additionally, VUr1 contained many scans from stage 0 cancers where the model failed to
detect any tumour.
This improved in the stage 1 cancers, in line with the observation in other cancer types
indicating more challenges with segmenting early-stage cancers.
It is, therefore, reasonable to conclude that the model is not inferior in samples from
bladder carcinoma as such, but that the poor performance in VUr1 is rather
explained by its fragmented tissue samples and the high proportion of early-stage cancer.

This study relates explicitly to segmentation of images into regions with and without predicted
tumours.
This dichotomisation is useful for evaluating the method's performance, and the
resulting masks can readily be used in subsequent analyses.
However, as a visual aid for pathologists in the clinic, the non-dichotomised probability
image displayed as a heat map might be more useful (see \cref{fig:segmentation-example}b).

A possible limitation of this study is that all scans included were manually annotated by the same
pathologist (MP), potentially biasing the reported performance compared to
the performance in cohorts annotated by other pathologists.
All scans in the development and validation materials were scanned at the same
laboratory at the Institute for Cancer Genetics and Informatics in Oslo, Norway, which might also impose a systematic bias that could
cause results to be overoptimistic.
However, the results in TCGA cohorts scanned elsewhere suggests that this is not a
substantial issue.

It should be noted that the method has been developed and mainly validated in materials
from resections, and that we can not anticipate how it will behave in biopsy samples.
The performance in the validation cohort with TUR samples, and the suggested causes of
this problem, might indicate that the method with the presented settings is not suited
for biopsies.
Also, all samples are from carcinomas, and we have not evaluated the performance in
other histological super-categories such as sarcoma.
Exploring this is required before this model is applied to cancers other than
carcinomas.

With the advent of publicly available foundation models for computational pathology
which are pan-cancer in nature, it would be natural to adapt them to tumour segmentation
and evaluate their performance on the
materials included in this study.
This is not currently done, but is subject to future studies.

We emphasise our use of pre-planned validation in external cohorts and our extensive
performance evaluation.
All planned analyses together with information required to define these analyses were
specified in a study protocol that adheres to the PIECES (\emph{Protocol Items for
External Cohort Evaluation of a deep learning System}) recommendations, and this
protocol was fixed prior to validation (Supplementary Section 6).\autocite{kleppe2021designing}
That no adjustments was done to the primary model after validation, and that the
validation was pre-specified and performed only once, means that we can trust that the
primary analysis gives an unbiased and realistic assessment of the model's performance
that is not overly optimistic and actually reflect how the model will perform in real
usage on new data.\autocite{kleppe2021designing,dhiman2023tripod}

We conclude that it was possible to develop an automatic segmentation model that
generalises well to multiple cancer types, without sacrificing performance compared with
specialised models only trained on a single cancer type.
Small, fragmented tumours are a challenge, but otherwise the model was observed to
perform well on tumour types not present in the development cohorts, on
different scanners, on slides prepared at different laboratories and in patients from
different countries.
Thus, we conclude that such pan-cancer segmentation models can serve as a first step for
subsequent automatic analyses of tumour areas and be implemented in digital pathology
platforms for a more streamlined and effective diagnostic pipeline.

\clearpage
\section{Methods}

\subsection{Materials}

In the following, a brief description of all included cohorts is presented.
A detailed description of the development and validation cohorts, including
acquisition flow diagrams and baseline characteristics is available in the study
protocol section 1 (Supplementary Section 6).

A simple naming scheme is used for the development and validation cohorts.
The first letter is either \emph{D} or \emph{V}, signifying whether the cohort was used
for development or validation, respectively.
Then, two letters identify the type of cancer: \emph{Co} for colorectal
carcinoma, \emph{En} for endometrial carcinoma, \emph{Lu} for lung carcinoma, \emph{Pr}
for prostate carcinoma, \emph{Br} for breast carcinoma, and \emph{Ur} for urothelial
carcinoma of the bladder.
A final integer distinguishes cohorts of the same kind.
The TCGA cohorts retain their original names.

\subsubsection{Development cohorts}
We used seven cohorts from four cancer types for method development.
DCo1 is based on a consecutive series of patients with colonic adenocarcinoma treated
between 1988 and 2000 at Akershus University Hospital,
Norway.\autocite{bondi2005expression}
DCo2 is based on a consecutive series of patients with stage I to III colorectal
carcinoma treated between 1993 and 2003 at Aker University Hospital (now part of Oslo
University Hospital (OUH)), Norway.\autocite{merok2013microsatellite}
DCo3 originates from the VICTOR trial (ISRCTN registry, ISRCTN98278138)
which recruited patients with stage II and III colorectal cancer from 151 hospitals in
the UK between 2002 and 2004.\autocite{kerr2007rofecoxib}
DEn1 comprises patients referred to the Department of Gynecological Oncology at OUH,
Norway, and diagnosed or operated for endometrial carcinoma between 2006 and
2017.
DLu1 consists of patients resected for primary lung cancer as part of primary treatment
between 2006 and 2018 at OUH, Norway.\autocite{dyrbekk2023evaluation}
DPr1 comprises patients who underwent radical prostatectomy (RP) between 1999 and 2010
at Vestfold Hospital Trust, Norway.
DPr2 consists of patients who underwent RP between 1987 and 2005 at the Norwegian Radium
Hospital (now part of OUH), Norway.\autocite{waehre2014fifteen}

\subsubsection{Validation cohorts}
We used eight cohorts from six cancer types for the pre-planned method validation.
VCo1 comprises patients with stage II and III colorectal carcinoma enrolled between 2005
and 2010 from 170 hospitals in seven countries for the QUASAR 2 trial (ISRCTN registry,
ISRCTN45133151).\autocite{kerr2016adjuvant}
VEn1 consists of patients with endometrial carcinoma collected between 2001 and 2016 at
Amsterdam Medical Center, The Netherlands.
VEn2 comprises patients with endometrial carcinoma collected between 1999 and 2018 at
the Department of Obstetrics and Gynaecology, Innsbruck Medical University, Austria.
VLu1 includes a consecutive series of patients with stage I to III non-small cell lung
carcinoma operated between 1990 and 2010 at the University Hospital of North Norway
and Nordland Hospital Trust, Norway.\autocite{hald2018lag}
VPr1 consists of patients who underwent RP between 2001 and 2006 at the Norwegian
Radium Hospital, Norway.\autocite{cyll2021pten}
Note that although DPr2 and VPr1 both originates from the Norwegian Radium Hospital and
have some overlap in time, they comprise a disjoint set of patients with different
responsible surgeons.
VBr1 are patients registered with lymph node negative breast cancer between 1990 and
1998 at Stavanger University Hospital, Norway, while
VBr2 are patients from the same hospital registered with breast cancer between 2000 and
2004.\autocite{skaland2009validating,egeland2019validation}
VUr1 comprises patients diagnosed with early-stage non-muscle invasive urothelial
carcinoma of the bladder and without upper urinary tract urothelial carcinoma
between 2002 and 2010 at Stavanger University Hospital,
Norway.\autocite{lillesand2020mitotic}
All samples in VUr1 are from TURs which result in
glass slides typically containing fragmented tissue sections rather than a larger single
tissue section typical of the other development and validation cohorts.

\subsubsection{Test cohorts}
We used four cohorts from three cancer types from TCGA for additional exploratory
analyses: from lung (LUAD and LUSC), prostate (PRAD) and bladder carcinoma
(BLCA).\autocite{tcga-blca,tcga-luad,tcga-lusc,tcga-prad}
See Supplementary Section 4 for acquisition flow
diagrams and baseline characteristics.

\subsection{Sample acquisition and preparation}

A 3 \textmu{}m section is cut from a FFPE tumour tissue block, mounted on a glass slide
and stained with H\&E before imaging with a microscope scanner to form a WSI.\
For some cohorts (DCo1, DCo2, DEn1, DLu1, DPr1, DPr2, VCo1, VEn1, VEn2, VPr1), we
received FFPE blocks and prepared tissue slides locally.
For the rest of the cohorts (DCo1, VLu1, VBr1, VBr2, VUr1), we received H\&E-stained
tissue slides.
All cohorts, except those from TCGA, were scanned locally using the highest available
resolution in two scanners, an Aperio AT2 and a NanoZoomer
XR, resulting in WSIs with a size on the order of
$100\,000\times 100\,000$ pixels with about 0.24 \textmu{}m per pixel.
WSIs from TCGA were downloaded from the TCGA Research Network
(\url{https://www.cancer.gov/tcga}).
For TCGA, we don't know how samples were prepared, nor which scanner models were used
for imaging.
Clinical data are from the TCGA Pan-Cancer Clinical Data Resource which publication
should be consulted when interpreting the included variables and their
values.\autocite{liu2018integrated}
Manual tumour annotations were created by a pathologist (MP) for all included WSIs.

\subsection{Automatic tumour segmentation}

A WSI is read and partitioned into image tiles processed by a segmentation network to
form probability images of the same size as the input tiles and with the network's
prediction of tumour presence.
The segmentation network has an \emph{encoder-decoder} structure where the encoder is a
\emph{Normalizing-free Network}, and the decoder is a \emph{DeepLabV3+}
network.\autocite{brock2021high,chen2018encoder}
All trainable network parameters are randomly initialised and only adjusted using images
from the development set.
Loss curves from the network optimisation are displayed in Supplementary
Section 3.
The tile results of a WSI are merged to form a prediction for the entire WSI,
which is then dichotomised by hysteresis thresholding, finalising the segmentation (see
\cref{fig:segmentation-example} b and c).
A visual summary is provided in \cref{fig:inference-pipeline}, and a detailed description
can be found in the study protocol section 2 (Supplementary Section 6).

Each WSI is read at a magnification of 1 \textmu{}m per pixel and sampled in a grid of
overlapping tiles.
Single nuclei are easily distinguished at this magnification, and even the nucleolus can
be visible (see \cref{fig:segmentation-example} d and e).
For inference, tiles have a size of $7\,680\times 7\,680$ pixels which corresponds to a
physical area of $7.68\times 7.68$ mm$^2$, where $7.68$ mm is about one-third of the
width of a typical glass slide.
The tile size was determined by hardware constraints and the sampling magnification was
chosen to balance high resolution and large physical area.
We used multiple tiles per batch during training, limiting the tile size to
$2\,024\times 2\,024$ pixels.
Although the image size difference between training and inference is quite large in our
study, we and others have found that such differences can be
beneficial.\autocite{khened2021generalized,touvron2019fixing}
In training, there is an overlap between adjacent tiles of minimum 1024 pixels in both
horizontal and vertical directions, and for inference the minimum overlap is 0 pixels.

\subsection{Planned analyses}

A study protocol (Supplementary Section 6) was written following our
previously published PIECES recommendations and fixed prior to all
investigations that could reveal associations between the predicted and target
segmentation masks in the validation cohorts.\autocite{kleppe2021designing}
It includes a description of the materials (study protocol section 1 in Supplementary Section 6),
a technical account of the method (study protocol section 2 in Supplementary Section 6), and the
set of analyses we commit to report on (study protocol section 3 in Supplementary Section 6).

\subsubsection{Primary analysis}
The primary analysis evaluates the segmentation method trained on all scans in the
development cohorts.
The resulting \emph{primary model} is evaluated in all Aperio AT2 (Leica Biosystems,
Germany) scans in each validation cohort.
The DSC was selected as the primary performance metric since it commonly used and
suitable for measuring overall segmentation quality.\autocite{dice1945measures,
maier2018rankings, maier2024metrics}
The DSC equals two times the number of foreground pixels common in the predicted mask
and the corresponding reference mask, divided by the sum of foreground pixels in the
predicted mask and the foreground pixels in the reference mask.
It ranges from 0 (no common foreground pixels) to 1 (all pixels are classified equally
in the prediction and reference).
Performance is reported per cohort as the cohort-average DSC with an accompanying 95\%
confidence interval (CI) computed using a Student's $t$-statistic.

\subsubsection{Secondary analyses}
Four secondary analyses were planned.
The first analysis further illuminates the performance of the primary model
in the validation cohorts by computing 11 additional contingency table summary
statistics.
The second analysis investigates how the primary model performs in scans from
NanoZoomer XR (Hamamatsu Photonics, Japan).\
The third analysis compares the primary model with models specialised on a single cancer
type.
The specialised colorectal model was trained using only the cohorts from colorectal
carcinoma (DCo1, DCo2, DCo3), and vice versa for the specialised endometrial, lung, and
prostate models.
Finally, the primary model is compared to replication models that are developed
identically as the primary model, except with different random seeds.

\subsection{Exploratory analyses}
A set of exploratory analyses were performed \emph{post-hoc} after the study protocol
was fixed and validation results were ready.

\subsubsection{Correlation between segmentation performance and cohort characteristics}
Associations between the resulting Dice similarity coefficient and other data
characteristics are measured using Spearman's rank correlation coefficient, $\rho$.
See \emph{statistical analysis} section for elaboration.

\subsubsection{Per-scan performance comparison}
Per-scan performance comparisons between the primary model results and the other models
were conducted to supplement the average results obtained from the pre-planned secondary
analyses.
Results are presented in Supplementary Fig. S17.

\subsubsection{Region area analyses}
In this section we include analyses on a region level.
A region is a set of 4-connected foreground pixels in the predicted or reference
segmentation mask.
For each detected reference region, we locate predicted regions that are overlapping.
We say that a reference region and a predicted region \emph{correspond} if they have an
intersection over union (Jaccard index) greater than 50\%.
This ensures that if a reference region correspond with a predicted region, it cannot
correspond with any other predicted regions.
Also, this guarantees that the predicted region also only corresponds with the same
reference region.
Note that the above definition of corresponding regions only considers single regions,
which labels predicted regions that would correspond to a union of smaller reference
regions as false positive, and reference regions that would correspond to a union of
smaller predicted regions as false negative.

The true positive reference regions are then the set of reference regions that have a
corresponding predicted region, and vice versa.
A false negative reference region is a reference region not included in the set of true
positive reference regions.
A false positive predicted region is a predicted region not included in the set of true
positive predicted regions.

Supplementary Table S9 shows pixel overlap measured with Dice
similarity coefficient between the prediction and reference when only considering true
positive regions.
We get a Dice similarity coefficient for each image by adding the contingency tables for
all pairs of corresponding regions in the image (we therefore only get a result for an
image if this image contains at least one pair of corresponding regions).
The Dice similarity coefficient is then averaged over all images within a cohort with at
least one pair of corresponding regions.

Supplementary Fig. S18 shows the distribution of regions and their
size in an image.
Reference regions smaller than 1600 pixels are discarded since they are artefacts of the
background segmentation (study protocol section 2.2.4 in Supplementary Section 6).

\subsubsection{Primary model performance on TCGA and development cohorts}
We also evaluated the primary model on all included TCGA scans and Aperio AT2 scans from
the development cohorts.

\subsubsection{Bladder subgroup analyses}
In BLCA, a pathologist (MP) noted for each scan whether it was likely to originate from
TUR or not by considering the presence of fragmented tissue sections in the imaged glass
slide.
Performance was measured in pT stage groups and fragmented tissue groups with DSC
averaged both over all scans and only in scans with a prediction.

\subsubsection{Intra- and inter-observer variability}
We tasked pathologists Manohar Pradhan (MP) and Ljiljana Vlatkovic (LV) to annotate
tumour regions in the VBr2 validation cohort.
MP had already annotated the scans in this cohort, about two years prior to this second
annotation round.
In this section, we let \emph{MP-1} refer to MP's first set of annotations, and
\emph{MP-2} to his second set of annotations.

LV is a retired uropathologist currently serving as a consultant at the Institute for
Cancer Genetics and Informatics, Oslo University Hospital, Norway.
She holds a master's degree in cytology from the University Hospital Centre Zagreb,
Croatia, in addition to her specialisation in pathology.
She has over 40 years of experience, and has contributed to over 50 research papers
throughout her career.

MP is a pathologist employed at the Institute for Cancer Genetics and Informatics, Oslo
University Hospital.
He holds a PhD in image cytometry from the University of Oslo, Norway, in addition to
his specialisation in pathology.
He has over 20 years of experience as a pathologist, and has contributed to over 30
research papers throughout his career.
MP was involved in all manual annotations of the development and validation materials
used in this study (study protocol section 1.1 and 1.2 in Supplementary Section 6).

Both were given instructions to provide a rough delineation of all tumour areas,
including infiltrating tumour areas and intraductal carcinoma.
In situ carcinoma, atypical ductal hyperplasia, and lobular hyperplasias were also
included.
These are the same instructions given to MP in his initial annotation round.
In this experiment, MP and LV did not look at the existing annotations, and they did not
consult each other on how to annotate if they encountered uncertainties.

Measured differences will capture where the pathologists disagree, where they chose
differently in decisions on doubtful regions, and their general difference in annotation
``style''.
The result of this experiment is simply a quantification of the similarities between the
annotations in this cohort, and does not give any indication of which annotation is the
most ``correct''.
Although this result will give a measure of intra- and inter-observer variability, it
was performed primarily to conceptualise the values of the Dice similarity coefficient.

Referring to Supplementary Fig. S19, \emph{MP-2 vs MP-1} will give an indication
of the intra-observer variability with a separation of two years, while \emph{LV vs
MP-1} and \emph{LV vs MP-2} will indicate inter-observer variability.
Measured similarity with the primary automatic model presented in this study (labelled
\emph{Auto}) is also included for reference.

\subsubsection{Performance evaluation in five different scanners}
Slides from VCo1 were scanned on three different scanners in addition to Aperio AT2 and
NanoZoomer XR\: Aperio GT 450 DX (Leica Biosystems, Germany), KF-PRO-400 (KFBIO, China)
and Pannoramic 1000 (3DHISTECH, Hungary)

All slides were scanned on the KF-PRO-400 scanner, two were not scanned on
Aperio GT 450 DX, and an additional slide was not scanned on Pannoramic 1000.
In all three cases, the reason for not scanning was that parts of the glass slide were
broken.

Before scanning on the three additional scanners, 39 included tissue sections were
restained because of weak staining in the original sections.
The 39 glass slides with restained tissue sections were also scanned on the Aperio AT2
and NanoZoomer XR scanners.
The Dice similarity coefficient was similar between the original and the restained
version in all 39 sections on both Aperio AT2 and NanoZoomer XR, except for one section
that originally incorrectly produced no predicted tumour regions (see
Supplementary Fig. S20).

All experiments presented in this section evaluate the 1\,152 slides that were scanned
on all scanners, with 39 tissue sections that were restained and therefore differ from
the corresponding 39 original tissue sections from VCo1 evaluated elsewhere in this study.
Note that, although the same glass slides were scanned on the different scanners, they
were not scanned at the same time.
In general, the slides were scanned on Aperio AT2 and NanoZoomer XR in 2018, on Aperio
GT 450 DX and KF-PRO-400 in 2023, and on Pannoramic 1000 in 2024.

Per-scan differences in Dice similarity coefficient are presented in Supplementary
Fig. S21 and statistics on the performance per
scanner is summarised in Supplementary Table S12.

\subsubsection{MedSAM segmentation performance in validation cohorts}

Whole-slide images are downscaled to 5 \textmu{}m per pixel before input to MedSAM, and the
resulting probability image is dichotomised with the same hysteresis thresholding used
by the primary method in this study.

\subsection{Statistical analysis}

The Spearman's rank correlation coefficient, $\rho$ is computed using the Pearson's
sample correlation coefficient $r$ applied on the rank of the variables.
P values are computed using
\begin{equation*}
  t = r \sqrt{\frac{n - 2}{1 - r^2}}
\end{equation*}
which is approximately Student's $t$-distributed with $n - 2$ degrees of freedom under the
null hypothesis $\rho = 0$ where $n$ is the number of samples.
A two-sided p value below 0.05 was considered statistically significant.
Correlation and p values are computed using the \texttt{scipy.stats.spearmanr} function
with \texttt{scipy} version 1.10.1 and Python version 3.11.3.\autocite{virtanen2020scipy}

The confidence interval is computed using that
\begin{equation*}
  z = \arctanh(r)
\end{equation*}
is approximately normally distributed and has a standard error of approximately $1 /
\sqrt{n - 3}$.\autocite{laake2012medical}
A $100(1 - \alpha)\%$ confidence interval is then
\begin{equation*}
  \tanh\left(\arctanh(r) \pm z_{1 - \alpha/2} \frac{1}{\sqrt{n - 3}} \right).
\end{equation*}

\clearpage
\subsection{Contributors}

O-JS and FA initiated the project.
AN, KL, GBK, JK, AGZ, OTB, L-TRB, EHR, ESH, BB, ER, ML, VK, EJ, and DJK provided access to samples, and clinical and pathological data.
MP annotated all the WSIs used in the study.
LV annotated the validation cohort VBr2 from breast carcinoma.
O-JS, MP, MXI, TSH, and AK decided on inclusions and exclusions of samples.
O-JS developed and implemented the segmentation method, conducted the statistical
analyses, and wrote the first draft of the manuscript.
O-JS, TSH, KL, FA, and AK revised the manuscript draft.
All authors reviewed, contributed to, and approved the manuscript.
All authors had full access to all the data in the study. 
O-JS had the final responsibility for the decision to submit for publication.

\subsection{Declaration of interests}

O-JS, MP, MXI, TSH, DJK, KL, FA, and AK report having shares in DoMore Diagnostics.
KL reports being a board member in DoMore Diagnostics.
O-JS, TSH, and KL report filing a patent application titled ``Histological image
analysis'' with International Patent Number PCT/EP2018/080828.
O-JS, TSH, KL, and AK report filing a patent application titled ``Histological image
analysis'' with International Patent Application Number PCT/EP2020/076090.

\subsection{Code availability}

The source code is made available to reviewers as a submitted zip archive file, and can
be made otherwise available upon publication.

\subsection{Data availability}

Materials from TCGA can be downloaded from the TCGA Research Network
(\url{https://www.cancer.gov/tcga}).
Individual patient-level data from the other materials can be made available to other researchers
upon reasonable request by contacting the corresponding author, subject
to approval by the relevant people or review board at the institutions that
provided the original data.

\subsection{Funding}

This study was funded by The Research Council of Norway through its IKTPLUSS Lighthouse
program (grant number 259204).
The Research Council of Norway had no role in study design, data collection, data analysis, data
interpretation, writing the report, or the decision to submit the paper for publication.

\subsection{Acknowledgements}

The authors thank our dear friend and colleague Håvard Emil Danielsen that passed away
in the autumn of 2023; he was among the initiators of this study, facilitated access to
materials, and contributed to early versions of the manuscript draft.
We thank the laboratory and technical personnel at the Institute for Cancer Genetics and
Informatics for essential sample preparation and assistance;
Marian Seiergren and Paul Callaghan for assisting with figures.
Akershus University hospital, Oslo University Hospital (Aker Hospital,
Rikshospitalet, and the Norwegian Radium Hospital), Amsterdam Medical Center, Innsbruck Medical
University, University Hospital of North Norway, Nordland Hospital Trust, Vestfold
Hospital Trust, and Stavanger University Hospital for access to materials, the personnel
at said institutions for sample preparation, and all contributing patients;
the participating centres in the QUASAR 2 trial and the VICTOR trial, and all
participating patients.

\clearpage

\printbibliography%

\clearpage
\section{Figure captions}

\begin{figure}[h]
  \centering
  \includegraphics[width=\textwidth]{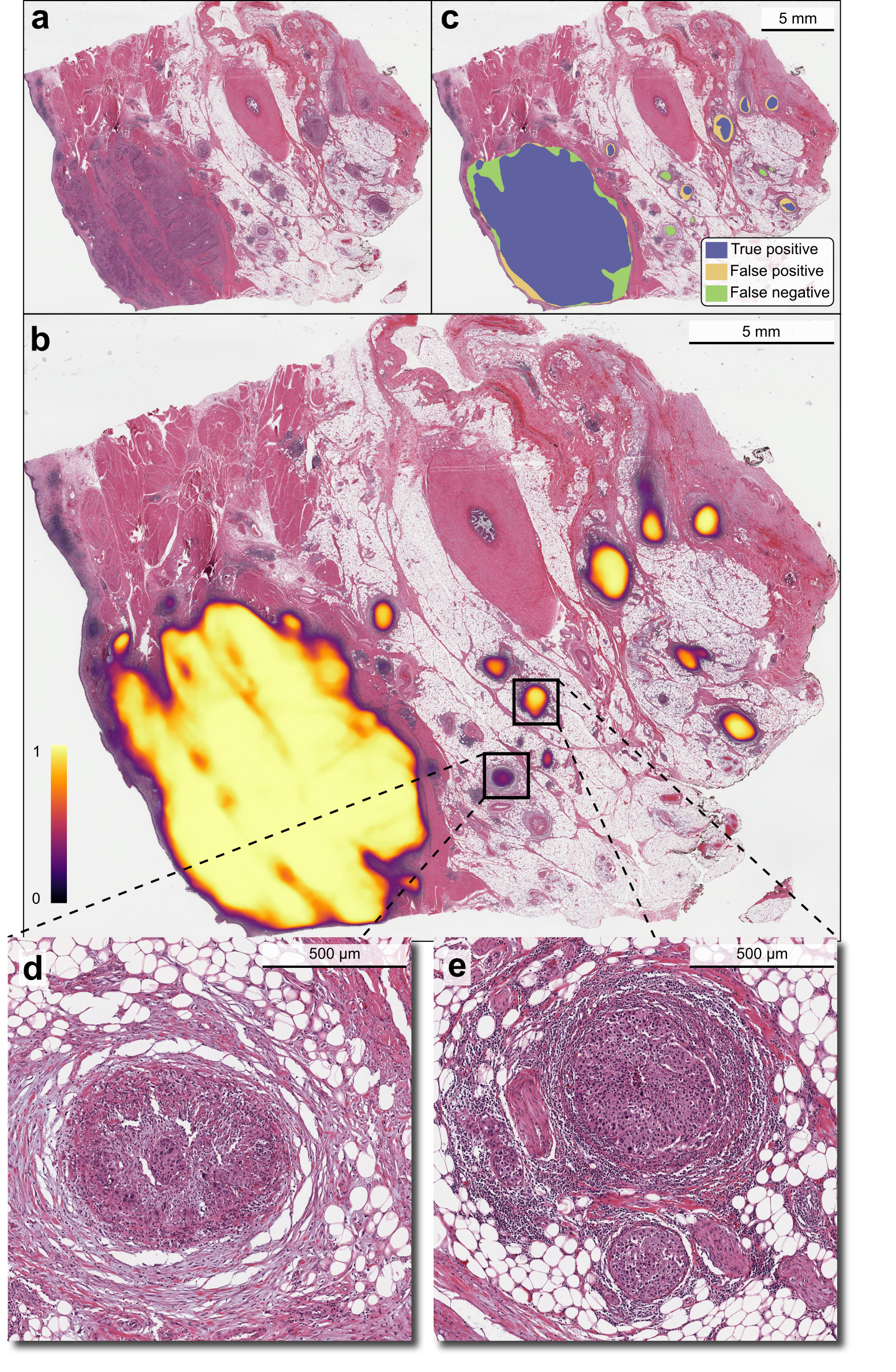}
  \caption[Segmentation example]{%
    \textbf{Example result in TCGA-FD-A6TE-01Z-00-DX1 from BLCA}\\
  }\label{fig:segmentation-example}
\end{figure}

\textbf{\cref{fig:segmentation-overview}:}
  \textbf{Overview of input and corresponding result}\\
  WSIs of H\&E stained tissue from different cancer types are all
  segmented by the same deep learning-based segmentation method.
  This figure show input images on the left and result heatmaps on the right overlain
  the corresponding input images.
  Heatmaps show the output of the segmentation network as a score image coloured
  as in \cref{fig:segmentation-example}b from transparent (value 0) to yellow (value
  100\%).

\textbf{\cref{fig:material-count}:}
  \textbf{Included patient and WSI count}\\
  The charts show counts stratified by patient cohort and grouped by cancer type and
  cohort use (for method development, validation or test), for patients (top panel)
  and WSIs (bottom panel).
  The tables in the respective panels display cumulative counts aggregated by cancer
  type and cohort use.

\textbf{\cref{fig:inference-pipeline}:}
  \textbf{Segmentation method pipeline}\\
  Illustrated with example scan TCGA-FD-A6TE-01Z-00-DX1 from BLCA, the same that is
  used in \cref{fig:segmentation-example}.
  1: Downscale the input scan to resolution 1 \textmu{}m per pixel and partition it
  into tiles of size 7680$\times$7680 pixels with minimum 1024 pixels overlap in each
  direction.
  In the second image from the left, green opacity signify overlap.
  2: Process each tile with the segmentation network to produce score tiles.
  3: Merge score tiles to score image with linear weight based on distance in
  overlapping regions.
  4: Segment the score image into foreground and background regions.

\textbf{\cref{fig:segmentation-example}:}
  \textbf{Example result in TCGA-FD-A6TE-01Z-00-DX1 from BLCA}\\
  Input WSI (\textbf{a}), annotated with the probability image (\textbf{b}) and with the
  segmentation result (\textbf{c}).
  The resulting DSC is 92.16\% which is similar to the BLCA median DSC of 92.31\%.
  With reference to panel \textbf{c}: the DSC is computed as two times the true
  positive area (blue) divided by the sum of the automatically segmented areas (blue and
  yellow) and the manually segmented areas (blue and green).
  The detailed crops show one false negative region (\textbf{d}) and one true positive
  region (\textbf{e}).
  We see that the false negative region (\textbf{d}) has a signal in the probability
  image (\textbf{b}), but that it is too weak to be included in the final segmentation
  (\textbf{c}).
  Comparing \textbf{d} with \textbf{e}, both show clusters of tumour cells and
  aggregates of lymphocytes surrounded by adipose tissue, but the area of the largest
  tumour cell cluster in \textbf{e} is about ten times larger than the area of the
  largest cluster in \textbf{d}, which might explain the weaker response.

\begin{figure}[h]
  \centering
  \includegraphics[width=\textwidth]{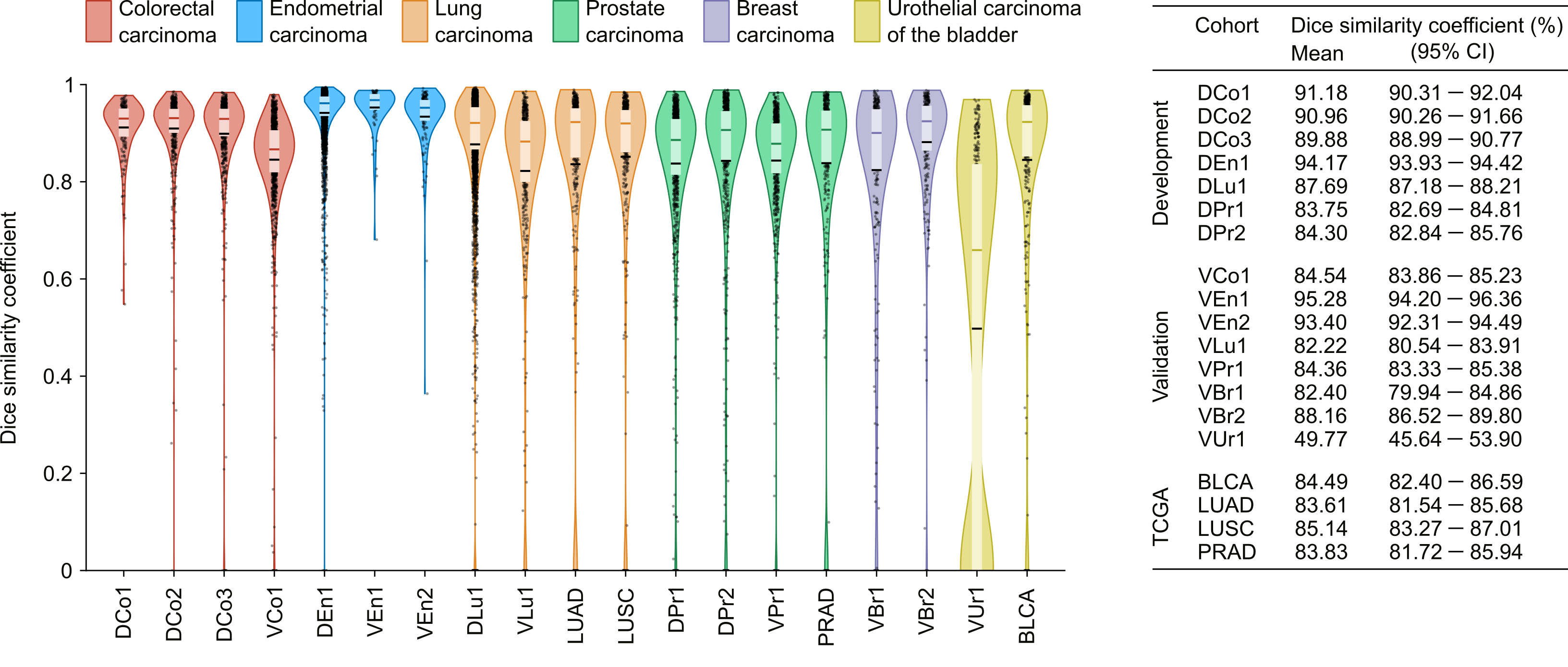}
  \caption[Primary model result]{%
    \textbf{Primary model results in cohorts from development, validation, and TCGA}\\
  }\label{fig:full-run-0-aperio-dice}
\end{figure}

\textbf{\cref{fig:full-run-0-aperio-dice}:}
  \textbf{Primary model results in cohorts from development, validation, and TCGA}\\
  For each cohort, the chart in the left panel show the DSC for individual scans (black dots) and the
  approximate DSC distribution as a violin plot.
  It also summarises the DSC with interquartile range (light box), mean value (black
  horizontal line), median value (coloured horizontal line).
  The table in the right panel shows the mean DSC per cohort and corresponding 95\% confidence interval (CI).

\textbf{\cref{fig:all-models-validation-test-aperio-dice}:}
  \textbf{Performance comparison of all presented models}\\
  Results from test and validation cohorts with the Aperio AT2 scanner.
  See \cref{fig:full-run-0-aperio-dice} for display legend.

\textbf{\cref{fig:primary-model-validation-aperio-vs-xr-dice}:}
  \textbf{Performance comparison between Aperio AT2 and NanoZoomer XR}\\
  The results show the DSC of the primary model evaluated on all validation cohorts.
  In the top panel, results are summarised in violin plots (see
  \cref{fig:full-run-0-aperio-dice} for display legend), while the bottom panel show
  scatter plots where the diagonal line trace equal score in scans from Aperio AT2 and
  NanoZoomer XR.\
  Markers in the scatter plots are coloured by estimated density using the same
  colourmap as in \cref{fig:segmentation-example}b, using Gaussian kernel density
  estimation from \texttt{skipy.stats.gaussian\_kde} in Python.

\textbf{\cref{fig:primary-model-vco1-five-scanners-dice}:}
  \textbf{Performance comparison on slides scanned with five different scanners}\\
  DSC of the primary model evaluated on the VCo1
  validation cohort with five different scanners.
  See \cref{fig:full-run-0-aperio-dice} for violin plot legend and
  \cref{fig:primary-model-validation-aperio-vs-xr-dice} for scatter plot legend.

\clearpage
\section{Figures}

\clearpage

\clearpage

\clearpage
\newgeometry{footskip=4.8cm}
\restoregeometry%

\clearpage

\clearpage
\begin{figure}[h]
  \centering
  \includegraphics[width=\textwidth]{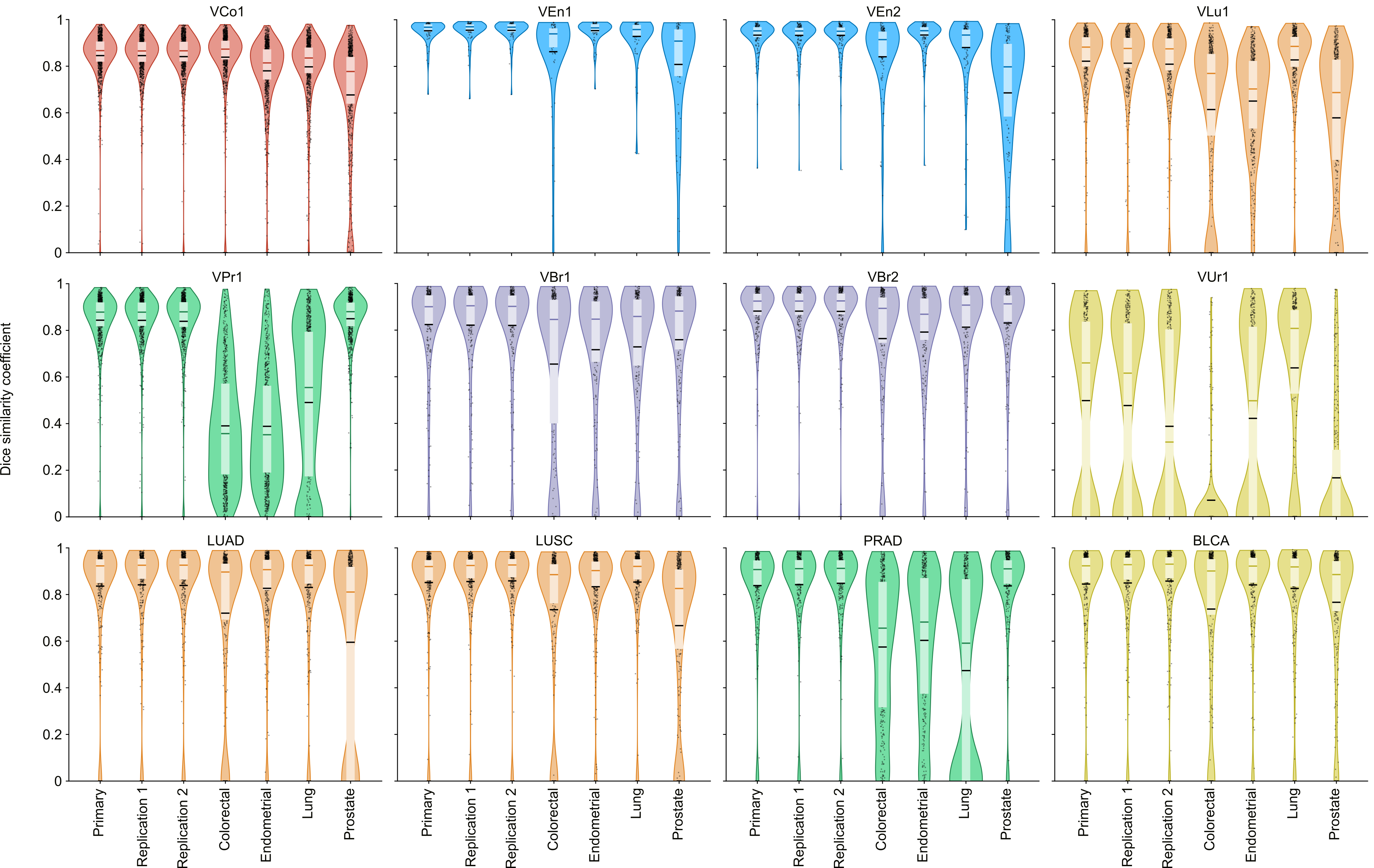}
  \caption[Model comparisons]{%
    \textbf{Performance comparison of all presented models}\\
  }\label{fig:all-models-validation-test-aperio-dice}
\end{figure}

\clearpage
\begin{figure}[h]
  \centering
  \includegraphics[width=\textwidth]{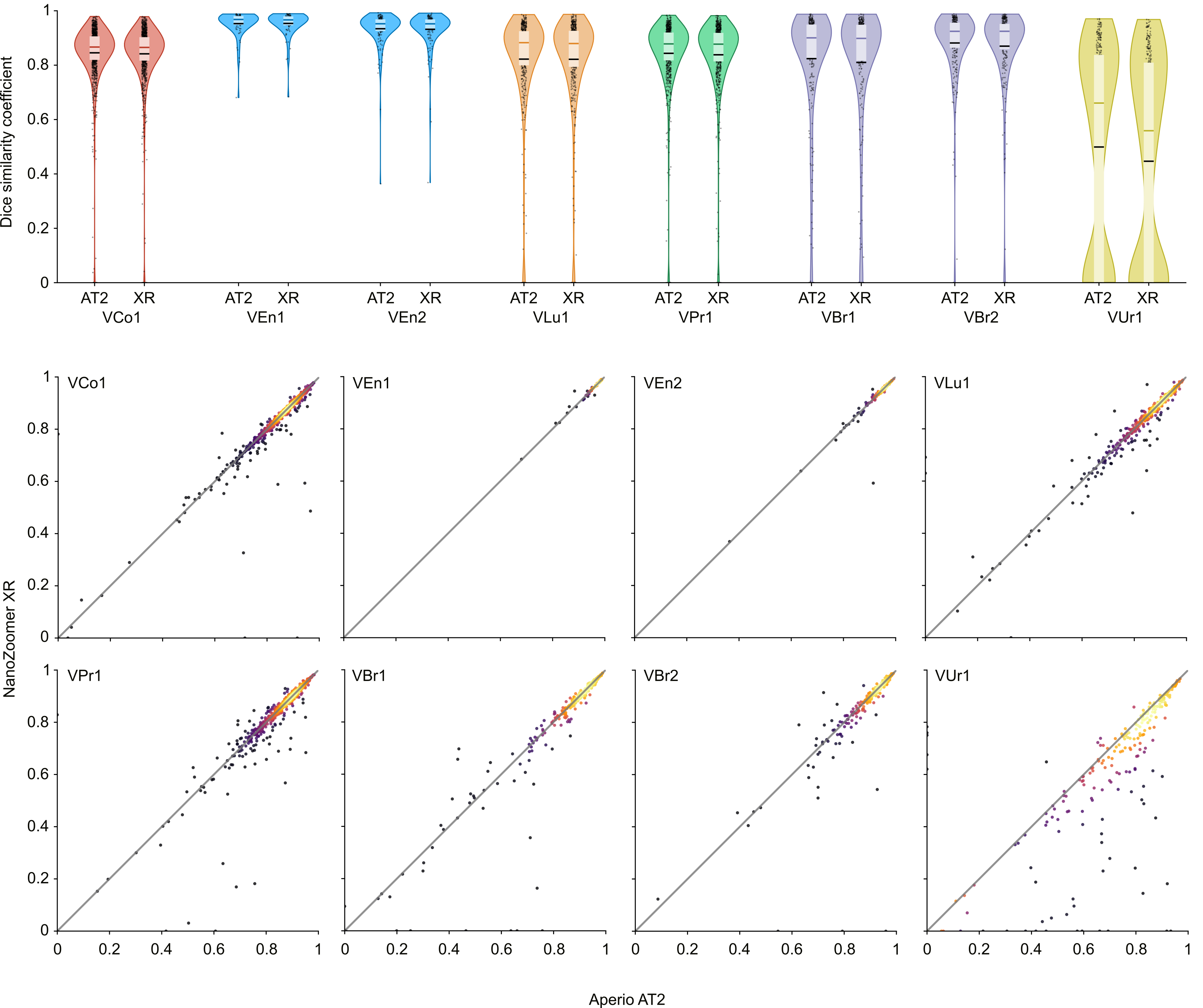}
  \caption[Aperio AT2 vs NanoZoomer XR]{%
    \textbf{Performance comparison between Aperio AT2 and NanoZoomer XR}\\
  }\label{fig:primary-model-validation-aperio-vs-xr-dice}
\end{figure}

\clearpage
\begin{figure}[h]
  \centering
  \includegraphics[width=\textwidth]{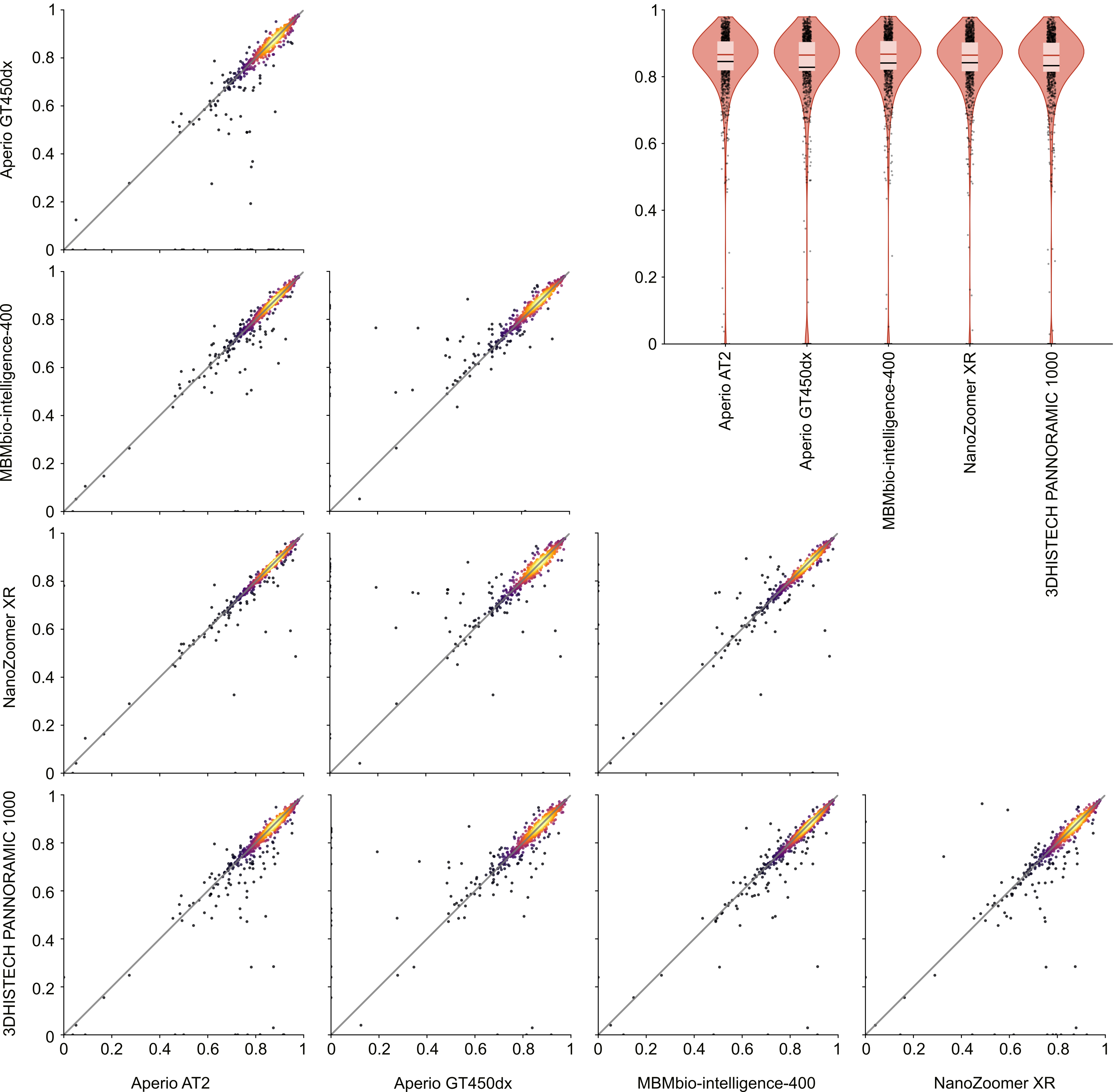}
  \caption[VCo1 scanned on five different scanners]{%
    \textbf{Performance comparison on slides scanned with five different scanners}\\
  }\label{fig:primary-model-vco1-five-scanners-dice}
\end{figure}

\clearpage
\appendix


\section*{Supplementary material}

**6  Study protocol**

# Protocol for external validation of a pan cancer deep learning method for automatic tumour segmentation in digital histology images

This protocol was last modified March 24, 2023, prior to all investigations that could reveal associations between the predicted and target segmentation masks in the method validation cohorts. At that time the method validation cohorts had been scanned, annotated and tiled; baseline characteristics, colour statistics and annotation statistics had been computed; the neural network part of all methods listed in the primary and secondary analyses had been applied on all tiles, but no further processing had been performed. All of the above interactions with the validation cohorts were performed blindly and did not inform choices made in the method development.

## Contents

80

i      47

81

# List of Figures

82

# List of Tables

83

# 1 Materials

Images analysed in this study are images of thin slices of resected solid tumours. Details about how these images are acquired follows below, before a per-cohort characterisation is presented.

The mass of interest is surgically removed from the patient and placed in containers with formalin for fixation. Pathologists examine the formalin-fixed specimen and regions of interest are cut into blocks and embedded in paraffin. The formalin-fixed paraffin-embedded (FFPE) blocks are sliced into thin tissue sections using a microtome and mounted on slides. Unless otherwise specified, the sections used in this study have thickness 3 µm. The tissue section is then stained with conventional haematoxylin and eosin (H&E), prepared as a tissue slide and imaged with a microscope scanner to form a so-called whole slide image (WSI).

This scan is then examined by a pathologist, and the tumour area, if any, is delineated. When the tissue slide is scanned by multiple different scanners, the digital tumour annotation is usually created on a scan from one scanner and digitally transformed to match the corresponding scan from different scanners. In these cases the transferred tumour annotation is verified by a pathologist, who corrects the annotation when necessary. If the transfer fails, the destination scan is manually annotated. To transform the annotation from the source scan to the destination scan, we first downsample the scans by a factor of 32. Then an image registration from the source image to the destination image is computed using a scale-invariant feature transform (SIFT).[1] This transformation is then used to transfer the polygons of the source annotation to the destination image.

In some of the cohorts, FFPE blocks are received at the Institute for Cancer Genetics and Informatics (ICGI), Oslo University Hospital, Norway, and sectioned, stained, scanned, and annotated by laboratory technicians. For other cohorts, we receive H&E stained tissue slides which we scan. In the remaining cohorts we receive digital scan files. What kind of material we received for each particular cohort studied will be specified in the cohort description (section 1.1 and section 1.2).

In this study, we use two scanners; the Aperio AT2 (Leica Biosystems, Germany) and the NanoZoomer XR (Hamamatsu Photonics, Japan). Digital scan files are read using the *Python* interface of the *OpenSlide C* library version 3.4.1.[2]

All cohorts are presented in section 1.1 and section 1.2 for method development and validation cohorts, respectively. Further descriptive analyses of the included materials are presented in section 1.3.

It should be noted that the materials were chiefly acquired for other projects, and not this seg-

mentation study. The exclusion reasons are therefore, in general, not consistent between materials. Also, some exclusion reasons would perhaps not have been applied if the material were acquired specifically for tumour segmentation.

Unless otherwise specified, all included scans in this study were manually annotated by pathologist Manohar Pradhan (MP) with over fifteen years of experience at the time the first cohorts in this study were annotated.

## 1.1 Materials for method development

20 270 scans from two scanners and 4 305 patients from 7 different cohorts were used in developing the tumour segmentation method; 3 from colorectal carcinoma, 1 from endometrial carcinoma, 1 from lung carcinoma and 2 from prostate carcinoma (protocol table 1).

**Protocol Table 1:** Number of patients, annotated scans and tiles in the development cohorts. Scans are from two different scanners. Tiles are produced at resolution 1 μm per pixel and have a size of 2 048 × 2 048 pixels.

| Cancer type | Cohort | Patients | Scans | | Tiles | |
|---|---|---|---|---|---|---|
| | | | Aperio AT2 | NanoZoomer XR | Aperio AT2 | NanoZoomer XR |
| Colorectal carcinoma | DCo1 | 206 | 206 | 206 | 28 911 | 30 552 |
| | DCo2 | 578 | 578 | 575 | 81 651 | 82 294 |
| | DCo3 | 765 | 765 | 764 | 108 429 | 109 720 |
| | Sum | 1 549 | 1 549 | 1 545 | 218 991 | 222 566 |
| Endometrial carcinoma | DEn1 | 1 241 | 3 340 | 3 340 | 514 341 | 566 105 |
| Lung carcinoma | DLu1 | 933 | 3 519 | 3 519 | 492 150 | 551 767 |
| Prostate carcinoma | DPr1 | 328 | 976 | 976 | 158 001 | 163 527 |
| | DPr2 | 254 | 753 | 753 | 95 015 | 97 867 |
| | Sum | 582 | 1 729 | 1 729 | 253 016 | 261 394 |
| Sum | | 4 305 | 10 137 | 10 133 | 1 478 498 | 1 601 832 |

### 1.1.1 DCo1 — Colorectal carcinoma

224 patients with colonic adenocarcinoma and adjacent normal mucosa treated between 1988 and 2000 at Akershus University Hospital, Norway.[3, 4] Tissue blocks were received at ICGI, prepared as tissue slides and scanned. After exclusions, (see protocol figure 1) 206 annotated scans remained from both the Aperio AT2 the NanoZoomer XR scanner.

Scans from Aperio AT2 were manually annotated for tumour by a pathologist (MP), and these tumour annotations were transferred to the corresponding NanoZoomer XR scans.

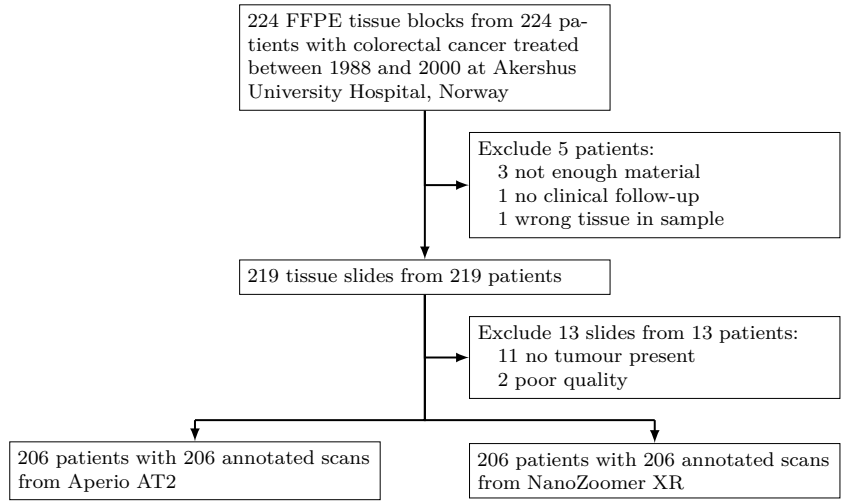

**Protocol Figure 1:** Flow from received blocks to annotated scans for the DCo1 cohort

### 1.1.2 DCo2 — Colorectal carcinoma

The material origin are patients with colorectal cancer treated between 1993 and 2003 at Aker Hospital (now part of Oslo University Hospital), Norway. 578 patients with resected tissue section and stages I, II, and III colorectal cancer that were analysed in previous studies were included.[4, 5, 6] One tissue slide per patient was prepared at ICGI, and scanned with the Aperio AT2 and NanoZoomer XR scanners. Three slides that had already been scanned with the Aperio AT2 scanner could not be scanned with the NanoZoomer XR scanner because of damaged cover glass. This resulted in 578 annotated scans from the Aperio AT2 scanner and 575 scans from the NanoZoomer XR scanner (summarised in protocol figure 2).

Scans from Aperio AT2 were manually annotated for tumour by a pathologist (MP), and these tumour annotations were transferred to the corresponding NanoZoomer XR scans.

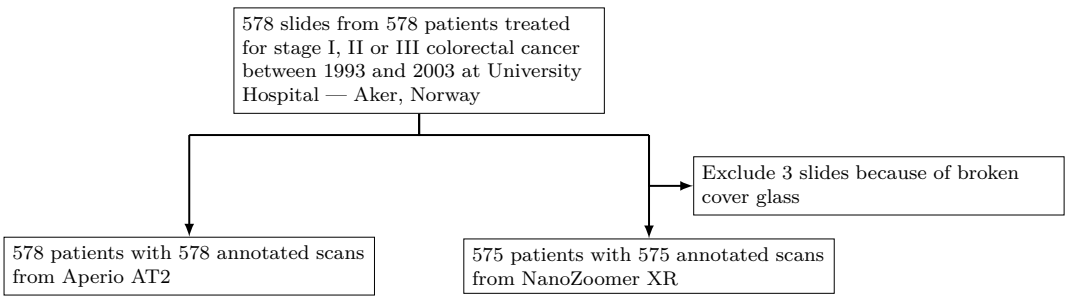

**Protocol Figure 2:** Flow from inclusions in previous studies (reference [5, 6]) to annotated scans for the DCo2 cohort

### 1.1.3 DCo3 — Colorectal carcinoma

2327 patients with histologically proven stage II or III colorectal cancer and resected primary tumour recruited from 151 hospitals in the United Kingdom between 2002 and 2004 for the *VICTOR* trial and randomly assigned to receive either rofecoxib (1167 patients) or placebo (1160 patients).[7] From 795 patients that we have included in a previous study, 795 H&E stained tissue sections were obtained at ICGI.[4] Some tissue sections where sectioned from FFPE blocks at ICGI, and some elsewhere. After exclusions as in reference [4] we had 768 scans from the Aperio AT2 scanner and 768 scans from the NanoZoomer XR scanner (see protocol figure 3). Two additional patients were excluded since their slides were of poor quality, and the scans of the new sections did not have manual annotations. One additional slide was also excluded because of no presence of tumour in the tissue section. A final scan from the NanoZoomer XR scanner was excluded since the annotation did not match the scan.

In this cohort, scans from NanoZoomer XR were manually annotated for tumour by a pathologist (MP) and these annotations were transferred to the corresponding Aperio AT2 scans.

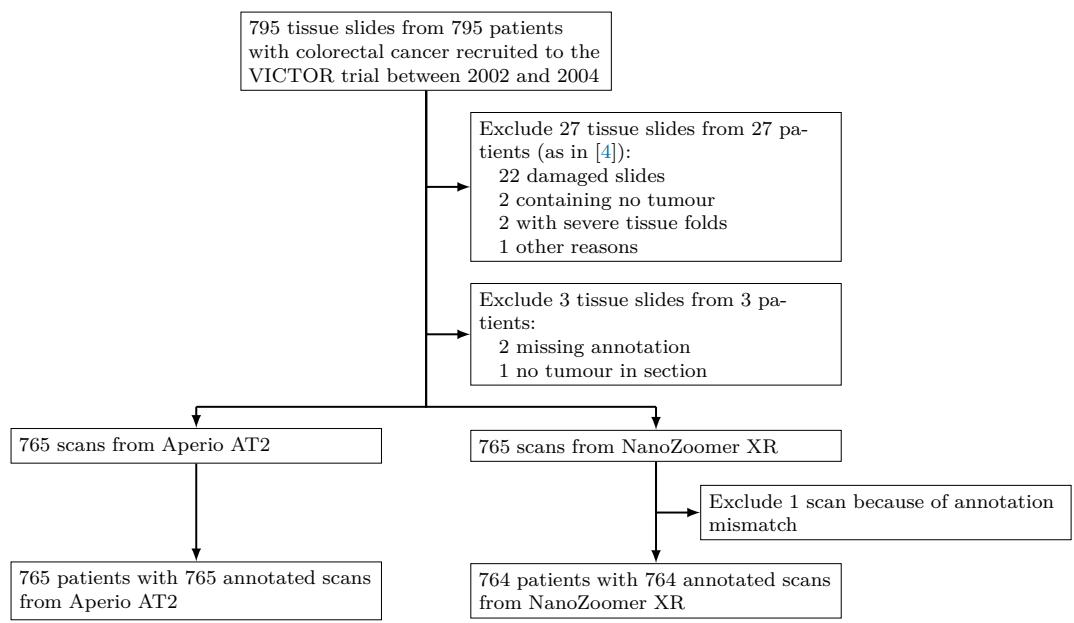

**Protocol Figure 3:** Flow from received slides to annotated scans for the DCo3 cohort

### 1.1.4 DEn1 — Endometrial carcinoma

We considered 1795 patients who underwent surgery for endometrial carcinoma between 2006 and 2018 at Oslo University Hospital, Norway.

We considered 1795 patients referred to the Department of Gynecological Oncology at Oslo University Hospital (OUH), Norway, and diagnosed *or* operated for endometrial carcinoma at OUH between 2006 and 2017. This material originates from the MoMaTEC (Molecular Markers in Treatment of Endometrial Cancer) trial (NCT number NCT00598845).[8, 9]

Note that among the 1795 patients, 11 patients (with 28 FFPE tumour blocks) had neuroendocrine tumours, which was removed from the World Health Organization classification of endometrial carcinoma between the 2014 edition and the 2020 edition.[10, 11]

FFPE blocks were collected by ICGI and prepared as tissue slides. Large sections were split and placed on two slides. After exclusions (see protocol figure 4), there remained 1229 patients with 4760 FFPE tumour blocks. At most three FFPE blocks were randomly selected from each patient, except for patients with tumours with mixed histology where all blocks were selected. This reduced the number of included blocks to 3331. 11 sections from 11 blocks were too large for a single slide and were placed on two slides. With this we have 1241 patients and 3331 blocks with 3340 annotated scans from Aperio AT2 and NanoZoomer XR.

After an update of the source of this material (after training but before validation), the following inconsistencies were noticed: 2 patients with 3 blocks each should have been excluded due to previous irradiation to a pelvic field including the uterus, 2 patients with 3 blocks each should have been excluded due to neoadjuvant treatment, 1 patient with 3 blocks was wrongly excluded due to complications after surgery. Protocol Figure 4 describe the material that was used in this study.

Scans from the NanoZoomer XR scanner were manually annotated for tumour by a pathologist (MP) and automatically transferred to the corresponding scans from the Aperio AT2 scanner.

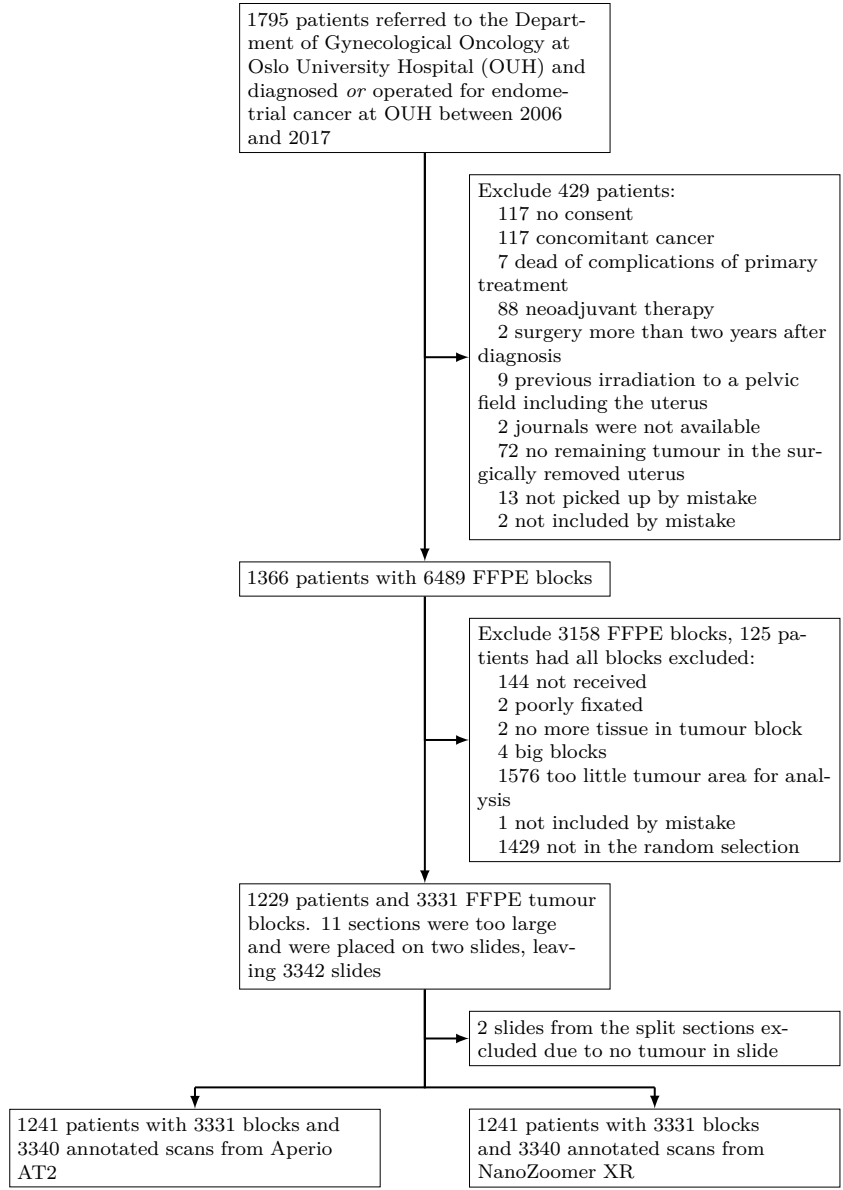

**Protocol Figure 4:** Flow from operated patients to annotated scans for the DEn1 cohort

### 1.1.5  DLu1 — Lung carcinoma

967 patients resected for primary lung cancer as part of primary treatment between March 2006 and December 2018 at Oslo University Hospital, Norway. FFPE tissue blocks were requested from all but 5 excluded patients (see protocol figure 5) from pathology departments at Oslo University Hospital. 3519 H&E slides were successfully prepared at ICGI from 3519 FFPE tissue blocks from 933 patients.

Scans from the NanoZoomer XR scanner were manually annotated by a pathologist (MP) and trained laboratory technician Jonathan Gullesen (JG). Annotations were verified and corrected if needed by a pathologist (MP). Large areas of fibrous tissue and necrosis were not included in the annotation. The manual annotations were automatically transferred to the corresponding Aperio AT2 scans.

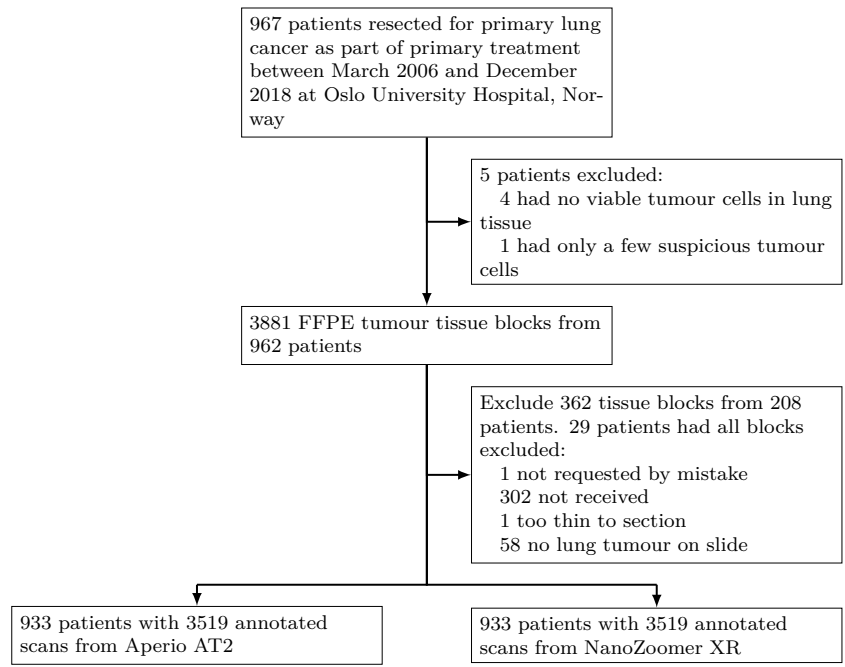

**Protocol Figure 5:** Flow from eligible patients to annotated scans for the DLu1 cohort

### 1.1.6  DPr1 — Prostate carcinoma

The cohort comprised 389 patients who underwent radical prostatectomy (RP) between 1999 and 2010 at Vestfold Hospital Trust, Norway. 61 patients were excluded: 26 for missing FFPE blocks, 6 for no tumour material and 29 patients for failing one or more FFPE block selection criteria. The

criteria were: one block with the highest Gleason score, a second block with the largest tumour area, a third block selected randomly from the remaining blocks with a tumour area > 16 mm² on a diagnostic H&E section.[12] At scan preparation, 7 blocks were excluded, and a further slide was excluded since this slide was missing when Aperio scanning was performed, leaving us with 976 annotated scans from both Aperio AT2 and NanoZoomer XR originating from 328 patients (summarised in protocol figure 6).

Tumour areas in scans from Aperio AT2 were manually annotated by a pathologist (MP) and large benign areas were avoided. These manual annotations were automatically transferred to NanoZoomer XR scans.

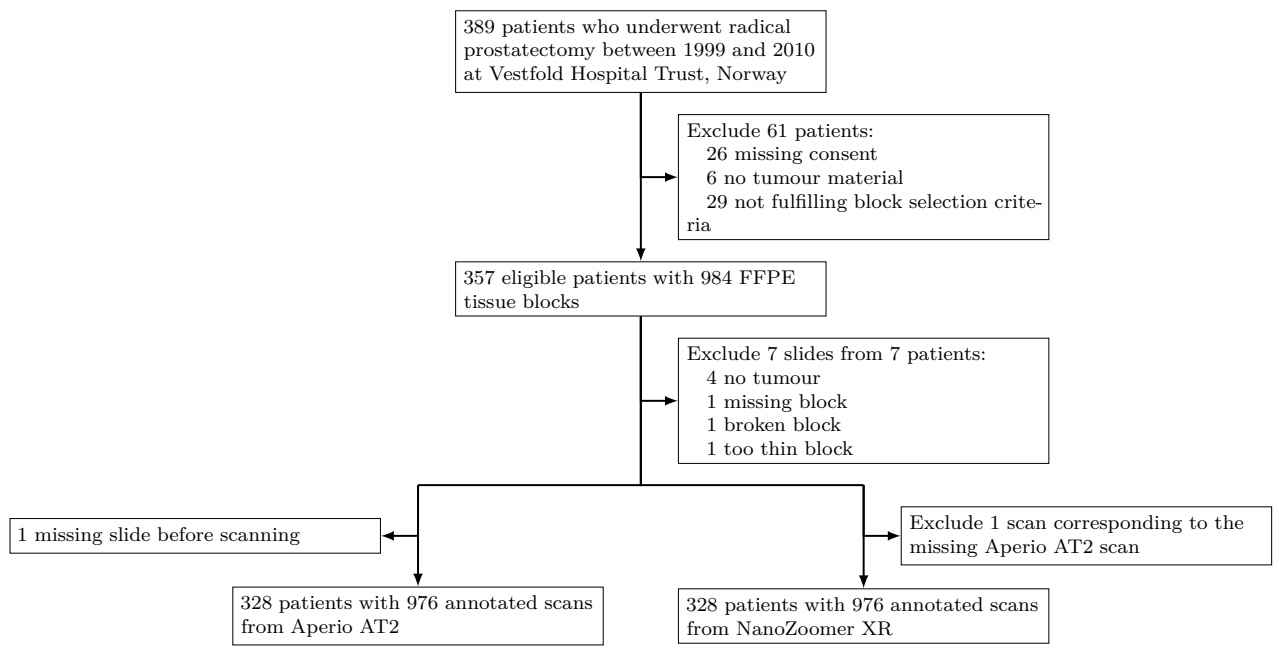

**Protocol Figure 6:** Flow from operated patients to annotated scans for the DPr1 cohort

### 1.1.7 DPr2 — Prostate carcinoma

The cohort originally comprised 317 patients who underwent RP between 1987 and 2005 at the Norwegian Radium Hospital (now part of Oslo University Hospital), Norway. Inclusion criteria for RP were preoperative absence of known metastases, age less than 75 years and life expectancy of at least 10 years. Adjuvant therapy was started in cases with elevated PSA after surgery and/or metastatic disease. All patients were operated by one surgeon (Håkon Wæhre).[13]

After exclusions (see protocol figure 7) there remained 307 eligible patients, of which 255 patients had three available tumour-containing blocks and were therefore included for further analyses. The assessment was based on the highest Gleason sum and/or previously assessed non-diploid DNA ploidy status, that had tumour areas measuring $> 4\,\mathrm{mm}^2$ on a diagnostic H&E section.[14] After further exclusions, we had 753 annotated scans from 254 patients (245 patients with 3 scans each and 9 patients with 2 scans each) from both the Aperio AT2 scanner and the NanoZoomer XR scanner.

Tumour areas were manually annotated avoiding large benign areas. 50% of the Aperio AT2 scans were manually annotated by a pathologist (MP) and automatically transferred to the corresponding NanoZoomer XR scans. The other 50% of the NanoZoomer XR scans were manually annotated by a laboratory technician (JG). Annotations were verified and corrected if needed by a pathologist (MP). These annotations were automatically transferred to the corresponding Aperio AT2 scans.

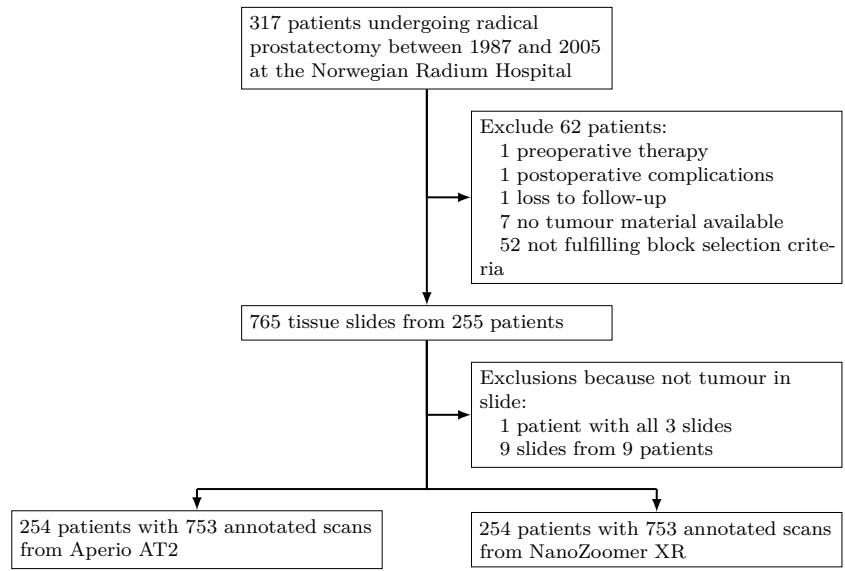

**Protocol Figure 7:** Flow from operated patients to annotated scans for the DPr2 cohort

## 1.2 Materials for method validation

7 258 scans from two scanners and 3 068 patients from 8 different cohorts were used in validating

the tumour segmentation method; 1 from colorectal carcinoma, 2 from endometrial carcinoma, 1

from lung carcinoma, 1 from prostate carcinoma, 1 from bladder carcinoma and 2 from breast

carcinoma (see protocol table 2).

All included scans from all validation cohorts were manually segmented prior to any investiga-

tion that could reveal correlations between predicted segmentation masks and manually segmented

target masks. All validation cohorts were primarily manually segmented for purposes other than

serving as validation cohorts for this project, and had already been manually segmented when

this project was initiated. These manual segmentations did not inform the development of the

presented segmentation method, nor did the development of the presented segmentation method

inform the manual segmentation.

**Protocol Table 2:** Number of patients, annotated scans and tiles in the validation cohorts. Scans are from two different scanners. Tiles are produced at resolution 1 μm per pixel and have a size of $7\,680 \times 7\,680$ pixels

| Cancer type | Cohort | Patients | Scans | | Tiles | |
| | | | Aperio AT2 | NanoZoomer XR | Aperio AT2 | NanoZoomer XR |
|---|---|---|---|---|---|---|
| Colorectal carcinoma | VCo1 | 1 132 | 1 155 | 1 155 | 17 686 | 18 635 |
| Endometrial carcinoma | VEn1 | 77 | 77 | 77 | 1 279 | 1 408 |
| | VEn2 | 132 | 152 | 152 | 2 179 | 2 383 |
| | Sum | 209 | 229 | 229 | 3 458 | 3 791 |
| Lung carcinoma | VLu1 | 522 | 522 | 522 | 7 100 | 7 649 |
| Prostate carcinoma | VPr1 | 259 | 777 | 777 | 12 568 | 1 4072 |
| Breast carcinoma | VBr1 | 310 | 310 | 310 | 4 165 | 4 689 |
| | VBr2 | 304 | 304 | 304 | 4 098 | 4 576 |
| | Sum | 614 | 614 | 614 | 8 263 | 9 265 |
| Urothelial carcinoma | VUr1 | 332 | 332 | 332 | 3 446 | 3 814 |
| Sum | | 3 068 | 3 629 | 3 629 | 52 521 | 57 226 |

### 1.2.1 VCo1 — Colorectal carcinoma

This cohort comprises participants in the QUASAR 2 (QUick And Simple And Reliable) trial

(ISRCTN registry number ISRCTN45133151). Between 2005 and 2010, 1952 eligible patients were

enrolled from 170 hospitals in seven countries (Australia, Austria, Czeck Republic, New Zealand,

Serbia, Slovenia, and the UK). The trial investigated the effect of bevacizumab on disease-free survival after potentially curative surgery of primary tumour.[15] FFPE tissue blocks were collected from 1 251 patients with either stage II or III colorectal cancer, of whom usable tissue blocks from 1 140 patients were received at ICGI.[4] After exclusions (see protocol figure 8) we had 1132 eligible patients with 1155 annotated scans from the Aperio AT2 scanner and the NanoZoomer XR scanner. Note that there were originally one section per patient, but some large sections had to be split before scanning in order to fit the cover slip.

Scans from both Aperio AT2 and NanoZoomer XR were manually annotated for tumour by a pathologist (MP).

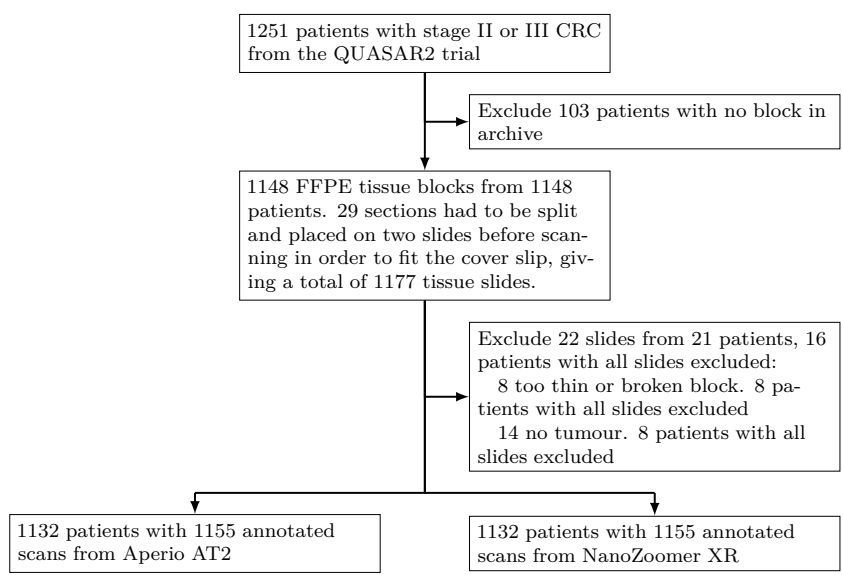

**Protocol Figure 8:** Flow from eligible patients to annotated scans for the VCo1 cohort

### 1.2.2 VEn1 — Endometrial carcinoma

95 blocks from 95 patients collected between 2001 and 2016 at Amsterdam Medical Center, The Netherlands. Six patients were excluded for clinical reasons, after which a further twelve blocks were excluded, leaving 77 patients with 77 annotated scans from both the Aperio AT2 and NanoZoomer XR scanner (see protocol figure 9).

Scans from the NanoZoomer XR scanner were manually annotated by a pathologist (MP) and transferred to the corresponding Aperio AT2 scans.

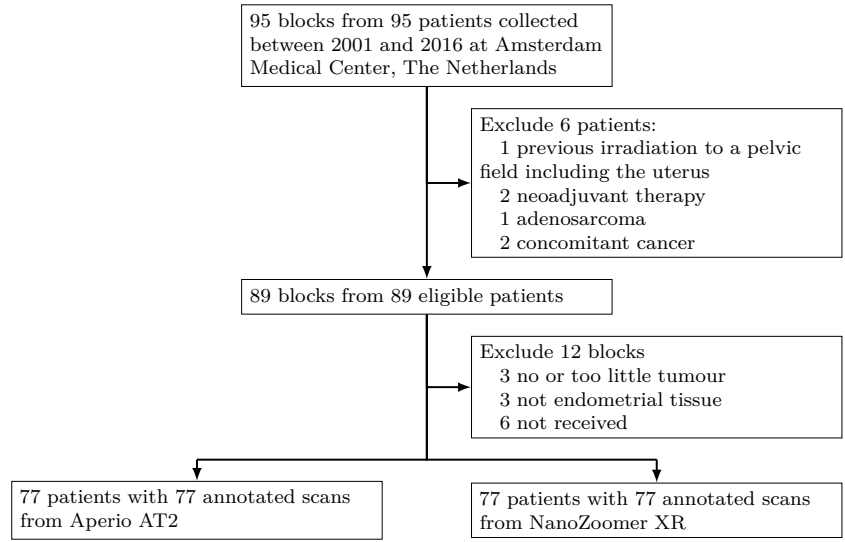

**Protocol Figure 9:** Flow from requested blocks to annotated scans for the VEn1 cohort

### 1.2.3 VEn2 — Endometrial carcinoma

177 blocks from 152 patients collected between 1999 and 2018 at Innsbruck Medical University, Austria. 19 blocks were excluded for no or too little tumour in the received block. 6 blocks were too thin to section. 14 patients had all blocks excluded for no or too little tumour, 4 patients had all blocks excluded for too thin block, and one additional patient had one block excluded for both reasons leaving no blocks left. This leaves 133 patients with 152 annotated scans from both the Aperio AT2 and NanoZoomer XR scanner (see protocol figure 10).

Scans from the NanoZoomer XR scanner were manually annotated by a pathologist (MP) and transferred to the corresponding Aperio AT2 scans.

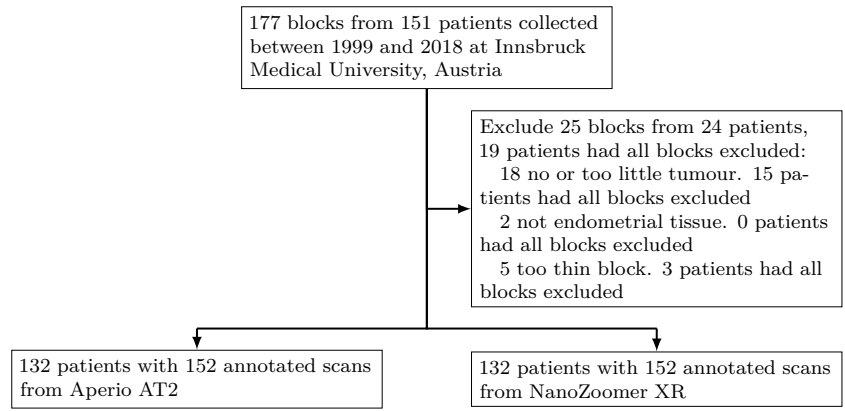

**Protocol Figure 10:** Flow from requested blocks to annotated scans for the VEn2 cohort

### 1.2.4 VLu1 — Lung carcinoma

A consecutive series of 633 patients with stage I to III non-small cell lung carcinoma (NSCLC) operated between 1990 and 2010 at the University Hospital of Northern Norway and Nordland Hospital, Norway.[16, 17] H&E stained tissue slides were received at ICGI for scanning, and after exclusions (see protocol figure 11) we had 522 eligible patients with 522 annotated scans from both the Aperio AT2 scanner and the NanoZoomer XR scanner.

NanoZoomer XR scans were manually annotated by a pathologist (MP), and they were automatically transferred to the corresponding Aperio AT2 scans. Large areas of fibrous tissue and necrosis were not annotated.

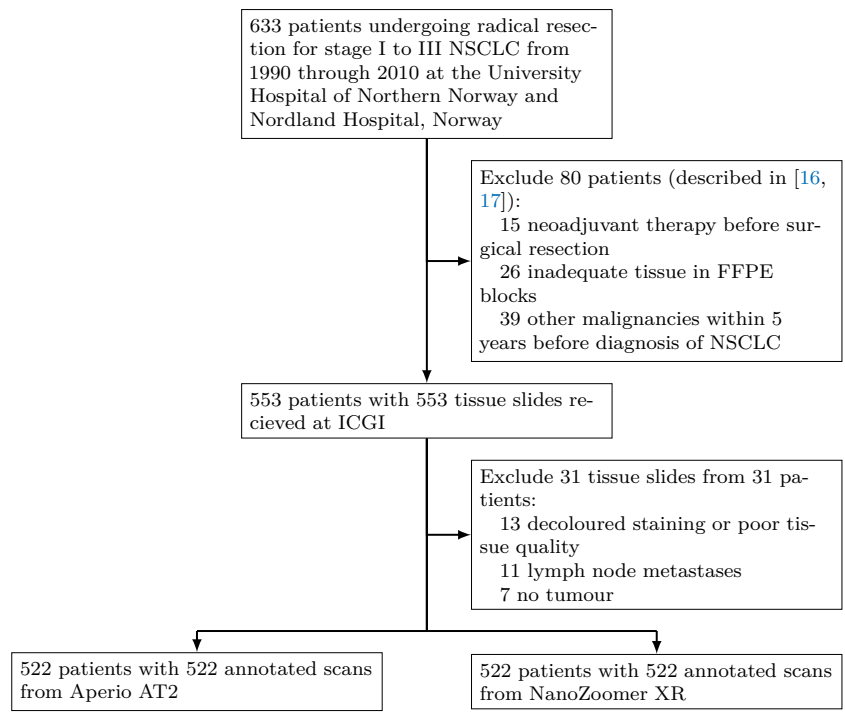

**Protocol Figure 11:** Flow from original study recruitment to annotated scans for the VLu1 cohort

### 1.2.5 VPr1 — Prostate carcinoma

The cohort comprised 287 patients who underwent RP between 2001 and 2006 at the Norwegian Radium Hospital (now a part of Oslo University Hospital), Norway. All patients were operated by one surgeon (Bjørn Brennhovd).[18]

After exclusions (see protocol figure 12), 259 eligible patients remained from which three blocks

were selected. The first and the second block represented the highest Gleason score and the largest tumour area, respectively. The third block was selected randomly from the remaining blocks with a tumour area $> 5\,mm^2$ on a diagnostic H&E section.

We obtained three sections from all 259 eligible patients, resulting in 777 sections scanned on both the Aperio AT2 scanner and the NanoZoomer XR scanner.

The origin of the digital annotations are manual annotations made directly on the cover slip with a marker. These marked slides were scanned with a NanoZoomer HT (Hamamatsu Photonics, Japan) scanner. Digital annotations were generated on these scans by drawing inside the area delineated by the analog marker, this was done either manually or automatically. The resulting digital annotations were transferred to scans from the Aperio AT2 scanner before they were manually verified and potentially corrected by a pathologist (MP). Finally, the verified annotations were transferred to scans from the NanoZoomer XR scanner. Large areas of benign epithelium and stroma were not annotated.

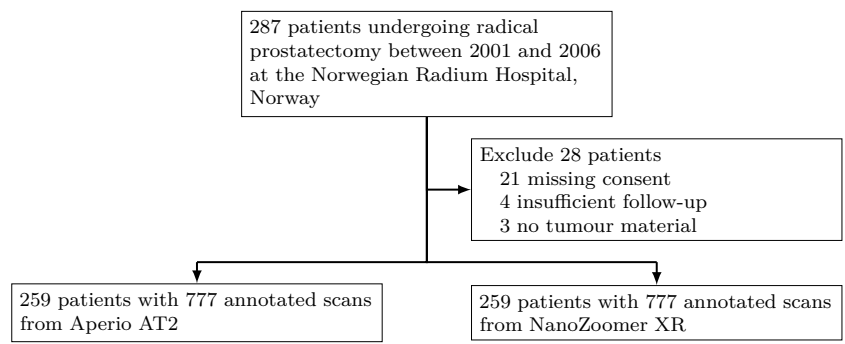

**Protocol Figure 12:** Flow from operated patients to annotated scans for the VPr1 cohort

### 1.2.6 VBr1 — Breast carcinoma

This cohort comprises 348 patients registered with breast cancer between 1990 and 1998 at Stavanger University Hospital, Norway.[19, 20, 21, 22] 320 H&E stained tissue sections prepared as slides from 320 patients were received at ICGI and scanned. After exclusions (see protocol figure 13) we were left with 310 slides from 310 patients scanned on both the Aperio AT2 scanner and the NanoZoomer XR scanner.

Scans from the Aperio AT2 scanner were manually annotated by a pathologist (MP) and transferred to the corresponding NanoZoomer XR scans. Both infiltrating tumour areas and intraductal carcinoma were annotated.

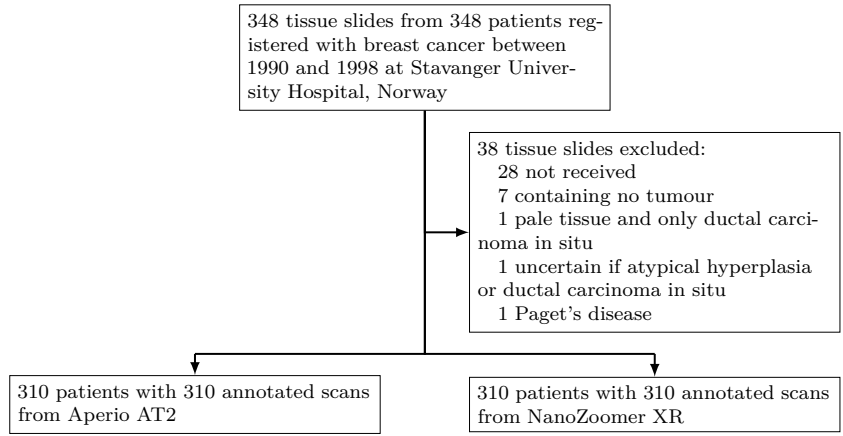

**Protocol Figure 13:** Flow from requested tissue slides to annotated scans for the VBr1 cohort

### 1.2.7   VBr2 — Breast carcinoma

This cohort comprises 339 patients registered with breast cancer between 2000 and 2004 at Stavanger University Hospital, Norway.[23] 315 H&E stained tissue sections prepared as slides from 315 patients were received at ICGI and scanned. After exclusions (see protocol figure 14) we were left with 304 slides from 304 patients scanned on both the Aperio AT2 scanner and the NanoZoomer XR scanner.

Scans from the Aperio AT2 scanner were manually annotated by a pathologist (MP), and annotations were transferred to the corresponding NanoZoomer XR scans. Both infiltrating tumour areas and intraductal carcinoma were annotated.

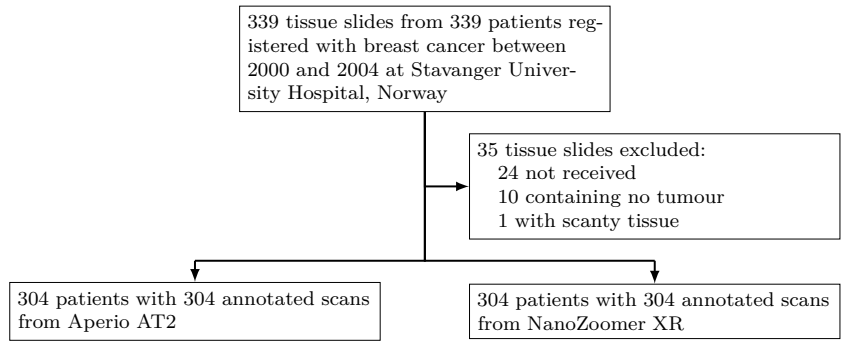

**Protocol Figure 14:** Flow from requested tissue slides to annotated scans for the VBr2 cohort

### 1.2.8 VUr1 — Urothelial carcinoma

357 patients diagnosed with urothelial carcinoma of the bladder between 1992 and 2010 at Stavanger University Hospital, Norway.[24, 25] From this, 333 slides from 333 patients were received at ICGI and scanned on the Aperio AT2 and the NanoZoomer XR scanner. 10 slides contained two parallel tissue sections, in which case only the best tissue section of the two parallels was scanned. One scan was excluded since it contained squamous epithelium and soft tissue but no tumour tissue, leaving us with 332 scans from 332 patients (see protocol figure 15).

All Aperio AT2 scans were manually annotated by a pathologist (MP), and the annotations were transferred to the NanoZoomer XR scanner. Urothelial tumours with or without infiltration were annotated.

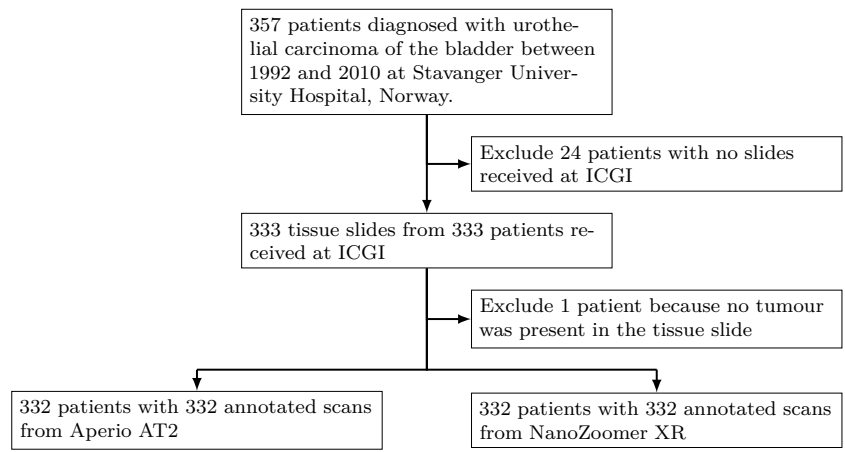

**Protocol Figure 15:** Flow from eligible patients to annotated scans for the VUr1 cohort

## 1.3  Material analyses

### 1.3.1  Baseline characteristics

Baseline characteristics for all cohorts grouped by cancer type are given in protocol tables 3 to 8. Data are given as *median (interquartile range)* or *count (percentage)*. Time to *event* statistics are based only on patients with the respective event. "Age" is age at a given starting point, which is the same starting point as time to event durations and follow-up time durations starts from. This starting point can vary between cohorts and will be specified.

**Protocol Table 3:** Baseline characteristics in colorectal carcinoma cohorts. Starting point is at randomisation for VCo1 and at surgery for the other cohorts.

|  | DCo1 | DCo2 | DCo3 | VCo1 |
|---|---|---|---|---|
| Patient count | 206 | 578 | 765 | 1132 |
| Age |  |  |  |  |
|   Years | 71 (61 − 78) | 73 (63 − 79) | 64 (58 − 71) | 65 (59 − 71) |
|   Missing | 2 (1%) | 0 | 0 | 0 |
| Sex |  |  |  |  |
|   Female | 106 (51%) | 290 (50%) | 270 (35%) | 480 (42%) |
|   Male | 98 (48%) | 288 (50%) | 495 (65%) | 652 (58%) |
|   Missing | 2 (1%) | 0 | 0 | 0 |
| Cancer-specific death |  |  |  |  |
|   False | 144 (70%) | 377 (65%) | 645 (84%) | 961 (85%) |
|   True | 60 (29%) | 201 (35%) | 120 (16%) | 157 (14%) |
|   Missing | 2 (1%) | 0 | 0 | 14 (1%) |
| Time to cancer-specific death |  |  |  |  |
|   Years | 0.7 (0.3 − 2.3) | 2.8 (1.7 − 5.1) | 3.0 (2.2 − 4.1) | 2.7 (1.7 − 3.6) |
| Follow-up time |  |  |  |  |
|   Years | 3.5 (1.0 − 5.0) | 6.6 (2.9 − 9.4) | 5.1 (4.1 − 5.8) | 4.6 (3.3 − 5.1) |
|   Missing | 2 (1%) | 0 | 0 | 0 |
| pT stage |  |  |  |  |
|   pT1 | 2 (1%) | 27 (5%) | 12 (2%) | 18 (2%) |
|   pT2 | 24 (12%) | 103 (18%) | 52 (7%) | 71 (6%) |
|   pT3 | 164 (80%) | 414 (72%) | 527 (69%) | 588 (52%) |
|   pT4 | 13 (6%) | 34 (6%) | 150 (20%) | 404 (36%) |
|   Missing | 3 (1%) | 0 | 24 (3%) | 51 (5%) |
| pN stage |  |  |  |  |
|   pN0 | 128 (62%) | 388 (67%) | 367 (48%) | 406 (36%) |
|   pN1 | 60 (29%) | 152 (26%) | 260 (34%) | 515 (45%) |
|   pN2 | 14 (7%) | 37 (6%) | 113 (15%) | 185 (16%) |
|   Missing | 4 (2%) | 1 (<1%) | 25 (3%) | 26 (2%) |
| Stage |  |  |  |  |
|   I | 8 (4%) | 112 (19%) | 0 | 0 |
|   II | 99 (48%) | 277 (48%) | 379 (50%) | 406 (36%) |
|   III | 52 (25%) | 189 (33%) | 386 (50%) | 726 (64%) |
|   IV | 45 (22%) | 0 | 0 | 0 |
|   Missing | 2 (1%) | 0 | 0 | 0 |
| Histological grade |  |  |  |  |
|   1 | 9 (4%) | 58 (10%) | 64 (8%) | 46 (4%) |
|   2 | 173 (84%) | 452 (78%) | 608 (79%) | 855 (76%) |
|   3 | 21 (10%) | 63 (11%) | 76 (10%) | 173 (15%) |
|   Missing | 3 (1%) | 5 (1%) | 17 (2%) | 58 (5%) |

**Protocol Table 4:** Baseline characteristics in endometrial carcinoma cohorts. Starting point is at surgery.

|  | DEn1 | VEn1 | VEn2 |
|---|---|---|---|
| Patient count | 1241 | 77 | 132 |
| Age |  |  |  |
|     Years | 68 (61 − 76) | 67 (60 − 73) | 68 (61 − 75) |
|     Missing | 5 (<1%) | 0 | 0 |
| Histological type |  |  |  |
|     Adenosquamous carcinoma | 0 | 2 (3%) | 0 |
|     Carcinosarcoma | 83 (7%) | 17 (22%) | 4 (3%) |
|     Clear cell carcinoma | 35 (3%) | 14 (18%) | 4 (3%) |
|     Endometrioid carcinoma | 825 (66%) | 11 (14%) | 98 (74%) |
|     Mucinous carcinoma | 7 (1%) | 0 | 0 |
|     Neuroendocrine carcinoma | 11 (1%) | 0 | 0 |
|     Papillary serous carcinoma | 133 (11%) | 31 (40%) | 24 (18%) |
|     Squamous cell carcinoma | 0 | 2 (3%) | 0 |
|     Mixed with clear cell or papillary serous carcinoma | 67 (5%) | 0 | 0 |
|     Mixed without clear cell or papillary serous carcinoma | 53 (4%) | 0 | 0 |
|     Undifferentiated carcinoma | 14 (1%) | 0 | 2 (2%) |
|     Unclassifiable | 9 (1%) | 0 | 0 |
|     Missing | 4 (<1%) | 0 | 0 |
| Cancer-specific death |  |  |  |
|     False | 974 (78%) | 56 (73%) | 115 (87%) |
|     True | 263 (21%) | 21 (27%) | 17 (13%) |
|     Missing | 4 (<1%) | 0 | 0 |
| Time to cancer-specific death |  |  |  |
|     Years | 2.5 (1.4 − 4.2) | 1.9 (1.3 − 3.4) | 3.7 (2.7 − 5.8) |
| Overall death |  |  |  |
|     False | 756 (61%) | 41 (53%) | 94 (71%) |
|     True | 481 (39%) | 36 (47%) | 38 (29%) |
|     Missing | 4 (<1%) | 0 | 0 |
| Time to overall death |  |  |  |
|     Years | 3.7 (1.8 − 6.7) | 2.0 (1.3 − 3.4) | 5.0 (2.7 − 8.3) |
| Recurrence |  |  |  |
|     False | 896 (72%) | 54 (70%) | 105 (80%) |
|     True | 341 (27%) | 23 (30%) | 27 (20%) |
|     Missing | 4 (<1%) | 0 | 0 |
| Time to recurrence |  |  |  |
|     Years | 1.3 (0.7 − 2.4) | 1.2 (1.0 − 1.5) | 1.2 (0.8 − 2.6) |
| Follow-up time |  |  |  |
|     Years | 7.6 (5.0 − 11.8) | 3.6 (2.5 − 4.7) | 4.5 (2.0 − 8.4) |
|     Missing | 4 (<1%) | 0 | 0 |
| FIGO stage |  |  |  |
|     1a | 548 (44%) | 20 (26%) | 48 (36%) |
|     1b | 296 (24%) | 12 (16%) | 32 (24%) |
|     2 | 79 (6%) | 14 (18%) | 19 (14%) |
|     3 | 0 | 0 | 8 (6%) |
|     3a | 28 (2%) | 1 (1%) | 0 |
|     3b | 12 (1%) | 0 | 0 |
|     3c1 | 106 (9%) | 18 (23%) | 15 (11%) |
|     3c2 | 72 (6%) | 8 (10%) | 9 (7%) |
|     4 | 0 | 0 | 1 (1%) |
|     4a | 0 | 1 (1%) | 0 |
|     4b | 96 (8%) | 3 (4%) | 0 |
|     Missing | 4 (<1%) | 0 | 0 |
| Histological grade |  |  |  |
|     1 | 467 (38%) | 0 | 26 (20%) |
|     2 | 286 (23%) | 1 (1%) | 64 (48%) |
|     3 | 179 (14%) | 75 (97%) | 42 (32%) |
|     Missing | 309 (25%) | 1 (1%) | 0 |

**Protocol Table 5:** Baseline characteristics in lung carcinoma cohorts. Starting point is at surgery for DLu1 and at diagnosis for VLu1.

| | DLu1 | VLu1 |
|---|---|---|
| Patient count | 933 | 522 |
| Age | | |
|    Years | 68 (62 – 73) | 68 (60 – 73) |
|    Missing | 11 (1%) | 0 |
| Sex | | |
|    Female | 460 (49%) | 168 (32%) |
|    Male | 462 (50%) | 354 (68%) |
|    Missing | 11 (1%) | 0 |
| Histological type | | |
|    Adenocarcinoma | 521 (56%) | 226 (43%) |
|    Adenosquamous carcinoma | 16 (2%) | 3 (1%) |
|    Bronchioloalveolar carcinoma | 8 (1%) | 0 |
|    Carcinoid | 42 (5%) | 0 |
|    Large cell carcinoma | 29 (3%) | 0 |
|    Large cell neuroendocrine carcinoma | 6 (1%) | 0 |
|    Salivary gland type lung carcinoma | 5 (1%) | 0 |
|    Squamous cell carcinoma | 287 (31%) | 289 (55%) |
|    Undifferentiated carcinoma | 4 (<1%) | 3 (1%) |
|    Mixed | 2 (<1%) | 0 |
|    Other | 2 (<1%) | 1 (<1%) |
|    Missing | 11 (1%) | 0 |
| Cancer-specific death | | |
|    False | 635 (68%) | 316 (61%) |
|    True | 287 (31%) | 206 (39%) |
|    Missing | 11 (1%) | 0 |
| Time to cancer-specific death | | |
|    Years | 2.2 (1.2 – 3.7) | 1.7 (0.9 – 3.2) |
| Follow-up time | | |
|    Years | 4.6 (2.5 – 6.8) | 3.6 (1.4 – 7.4) |
|    Missing | 11 (1%) | 0 |
| pT stage | | |
|    pT1 | 323 (35%) | 171 (33%) |
|    pT2 | 433 (46%) | 196 (38%) |
|    pT3 | 137 (15%) | 98 (19%) |
|    pT4 | 29 (3%) | 57 (11%) |
|    Missing | 11 (1%) | 0 |
| pN stage | | |
|    pN0 | 671 (72%) | 366 (70%) |
|    pN1 | 184 (20%) | 102 (20%) |
|    pN2 | 67 (7%) | 54 (10%) |
|    Missing | 11 (1%) | 0 |
| Stage | | |
|    I | 507 (54%) | 224 (43%) |
|    II | 273 (29%) | 170 (33%) |
|    III | 132 (14%) | 128 (25%) |
|    IV | 10 (1%) | 0 |
|    Missing | 11 (1%) | 0 |

**Protocol Table 6:** Baseline characteristics in prostate carcinoma cohorts. Starting point is at surgery.

| | DPr1 | DPr2 | VPr1 |
|---|---|---|---|
| Patient count | 328 | 254 | 259 |
| Age | | | |
|     Years | 64 (61 − 68) | 62 (58 − 67) | 62 (59 − 66) |
|     Missing | 0 | 1 (<1%) | 0 |
| Overall death | | | |
|     False | 257 (78%) | 176 (69%) | 200 (77%) |
|     True | 71 (22%) | 77 (30%) | 59 (23%) |
|     Missing | 0 | 1 (<1%) | 0 |
| Time to overall death | | | |
|     Years | 8.6 (6.3 − 12.3) | 9.6 (5.7 − 12.8) | 9.8 (7.2 − 13.5) |
| Biochemical recurrence | | | |
|     False | 215 (66%) | 98 (39%) | 188 (73%) |
|     True | 113 (34%) | 155 (61%) | 71 (27%) |
|     Missing | 0 | 1 (<1%) | 0 |
| Time to biochemical recurrence | | | |
|     Years | 0.8 (0.0 − 3.9) | 2.9 (1.5 − 5.3) | 3.4 (1.2 − 5.6) |
| Follow-up time | | | |
|     Years | 11.0 (8.8 − 13.9) | 11.2 (7.7 − 14.4) | 9.6 (8.5 − 12.3) |
|     Missing | 0 | 1 (<1%) | 0 |
| pT stage | | | |
|     pT2 | 178 (54%) | 54 (21%) | 159 (61%) |
|     pT3 | 136 (41%) | 168 (66%) | 96 (37%) |
|     pT4 | 0 | 26 (10%) | 2 (1%) |
|     pTx | 14 (4%) | 4 (2%) | 0 |
|     Missing | 0 | 2 (1%) | 2 (1%) |
| Gleason grade | | | |
|     2+3 | 0 | 0 | 3 (1%) |
|     3+3 | 63 (19%) | 11 (4%) | 126 (49%) |
|     3+4 | 147 (45%) | 92 (36%) | 83 (32%) |
|     3+5 | 1 (<1%) | 3 (1%) | 0 |
|     4+3 | 88 (27%) | 77 (30%) | 25 (10%) |
|     4+4 | 13 (4%) | 41 (16%) | 15 (6%) |
|     4+5 | 4 (1%) | 26 (10%) | 1 (<1%) |
|     5+4 | 1 (<1%) | 3 (1%) | 2 (1%) |
|     5+5 | 0 | 0 | 1 (<1%) |
|     Missing | 11 (3%) | 1 (<1%) | 3 (1%) |

**Protocol Table 7:** Baseline characteristics in breast carcinoma cohorts. Starting point is at diagnosis.

|  | VBr1 | VBr2 |
|---|---|---|
| Patient count | 310 | 304 |
| Age |  |  |
|   Years | 56 (50 – 64) | 57 (47 – 64) |
| Histological Type |  |  |
|   Invasive ductal carcinoma | 235 (76%) | 284 (93%) |
|   Invasive lobular carcinoma | 20 (6%) | 6 (2%) |
|   Lobular carcinoma in citu | 24 (8%) | 0 |
|   Medullary carcinoma | 5 (2%) | 8 (3%) |
|   Tubular carcinoma | 12 (4%) | 0 |
|   Other | 13 (4%) | 6 (2%) |
|   Missing | 1 (<1%) | 0 |
| Distant metastases |  |  |
|   False | 255 (82%) | 220 (72%) |
|   True | 55 (18%) | 74 (24%) |
|   Missing | 0 | 10 (3%) |
| Time to distant metastases |  |  |
|   Years | 5.3 (2.6 – 11.7) | 1.5 (0.7 – 3.3) |
| Local recurrence |  |  |
|   False | 279 (90%) | 274 (90%) |
|   True | 31 (10%) | 20 (7%) |
|   Missing | 0 | 10 (3%) |
| Time to local recurrence |  |  |
|   Years | 9.3 (4.3 – 16.5) | 8.3 (2.8 – 13.3) |
| Follow-up time distant metastases |  |  |
|   Years | 14.9 (8.4 – 21.2) | 12.4 (4.0 – 15.2) |
|   Missing | 0 | 10 (3%) |
| Follow-up time local recurrence |  |  |
|   Years | 13.3 (6.2 – 21.0) | 11.9 (3.8 – 15.1) |
|   Missing | 0 | 11 (4%) |
| Oestrogen recetor (ER) |  |  |
|   Negative | 40 (13%) | 153 (50%) |
|   Borderline | 11 (4%) | 0 |
|   Positive | 258 (83%) | 146 (48%) |
|   Missing | 1 (<1%) | 5 (2%) |
| Progesterone receptor (PR) |  |  |
|   Negative | 60 (19%) | 166 (55%) |
|   Borderline | 55 (18%) | 2 (1%) |
|   Positive | 194 (63%) | 98 (32%) |
|   Missing | 1 (<1%) | 38 (12%) |
| Lymph node status |  |  |
|   Negative | 310 (100%) | 216 (71%) |
|   Positive | 0 | 53 (17%) |
|   Missing | 0 | 35 (12%) |
| Nottingham prognostic index |  |  |
|   3–5 | 109 (35%) | 39 (13%) |
|   6–7 | 134 (43%) | 115 (38%) |
|   8–9 | 64 (21%) | 134 (44%) |
|   Missing | 3 (1%) | 16 (5%) |

**Protocol Table 8:** Baseline characteristics in urothelial carcinoma cohort. Starting point is at diagnosis.

| | VUr1 |
|---|---|
| Patient count | 332 |
| Age | |
|     Years | 72 (62 – 80) |
|     Missing | 2 (1%) |
| Sex | |
|     Female | 84 (25%) |
|     Male | 248 (75%) |
| Recurrence | |
|     False | 173 (52%) |
|     True | 159 (48%) |
| Time to recurrence | |
|     Years | 1.2 (0.6 – 2.0) |
| Stage progression | |
|     False | 307 (92%) |
|     True | 25 (8%) |
| Time to stage progression | |
|     Years | 2.0 (0.7 – 3.0) |
| Follow-up recurrence | |
|     Years | 5.8 (2.6 – 8.2) |
| Follow-up stage progression | |
|     Years | 7.2 (4.4 – 10.1) |
| pT stage | |
|     pTa | 255 (77%) |
|     pTis | 1 (<1%) |
|     pT1 | 76 (23%) |
| Histological grade | |
|     1 | 65 (20%) |
|     2 | 155 (47%) |
|     3 | 111 (33%) |
|     Missing | 1 (<1%) |
| Metastasis | |
|     False | 321 (97%) |
|     True | 11 (3%) |
| Multifocal | |
|     False | 197 (59%) |
|     True | 107 (32%) |
|     Missing | 28 (8%) |

**1.3.2 Survival characteristics**

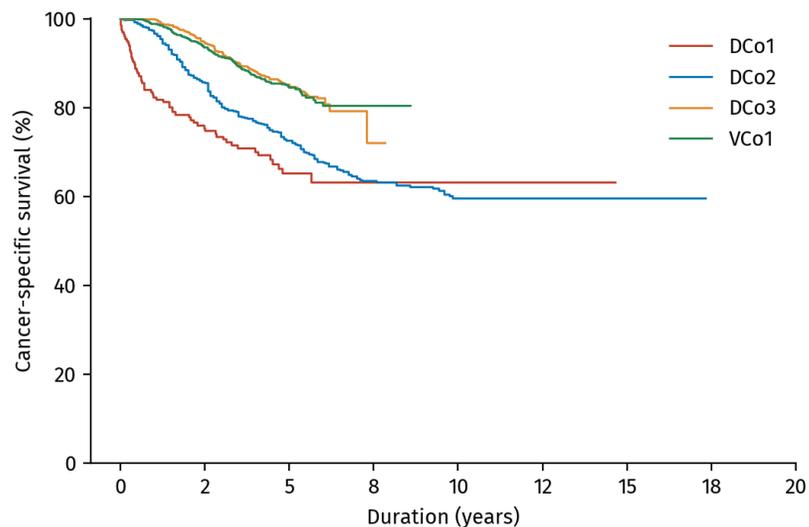

| | | | | | | | | | |
|---|---|---|---|---|---|---|---|---|---|
| Number at risk | | | | | | | | | |
| DCo1 (missing *n*=2) | 204 | 122 | 50 | 10 | 5 | 5 | 0 | 0 | 0 |
| DCo2 | 578 | 467 | 360 | 231 | 123 | 55 | 10 | 0 | 0 |
| DCo3 | 765 | 689 | 424 | 4 | 0 | 0 | 0 | 0 | 0 |
| VCo1 (missing *n*=14) | 1118 | 989 | 425 | 8 | 0 | 0 | 0 | 0 | 0 |

**Protocol Figure 16:** Kaplan-Meier analysis of survival in colorectal carcinoma materials. Duration is years since randomisation for VCo1 and years since surgery for the other cohorts.

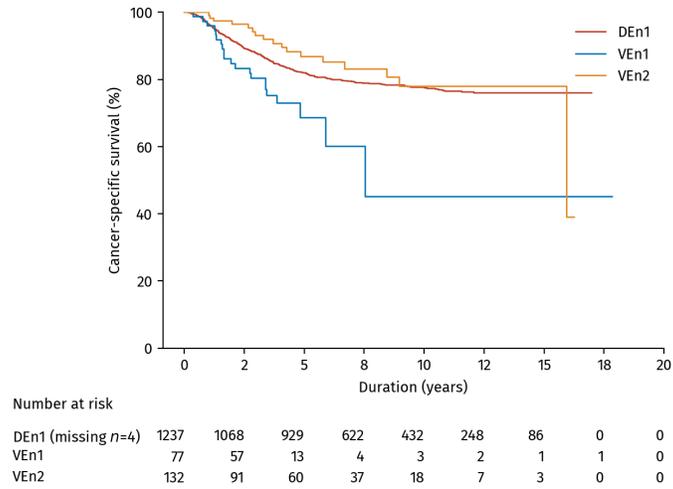

**(a)** Cancer-specific survival

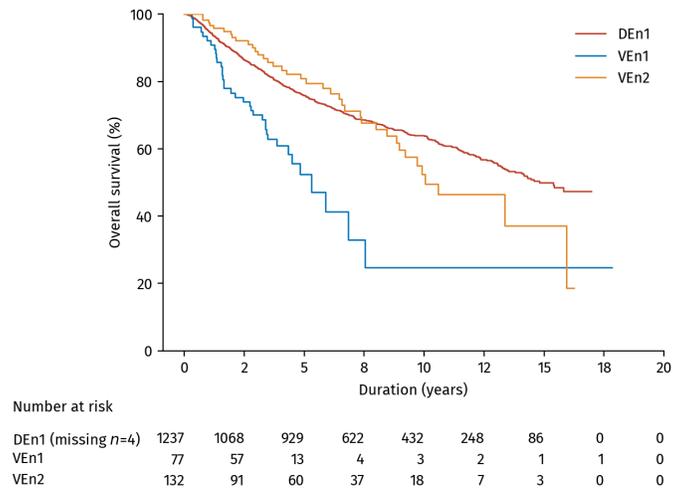

**(b)** Overall survival

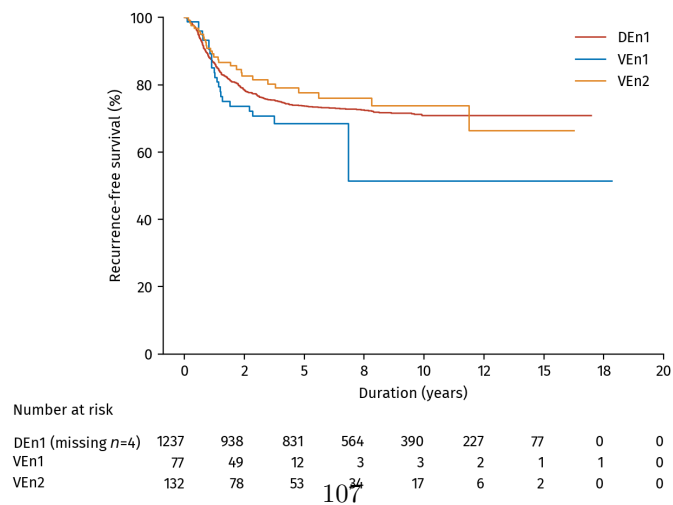

107

**(c)** Recurrence-free survival

**Protocol Figure 17:** Kaplan-Meier analysis of survival in endometrial carcinoma materials. Duration is years since surgery.

24                                                                              74

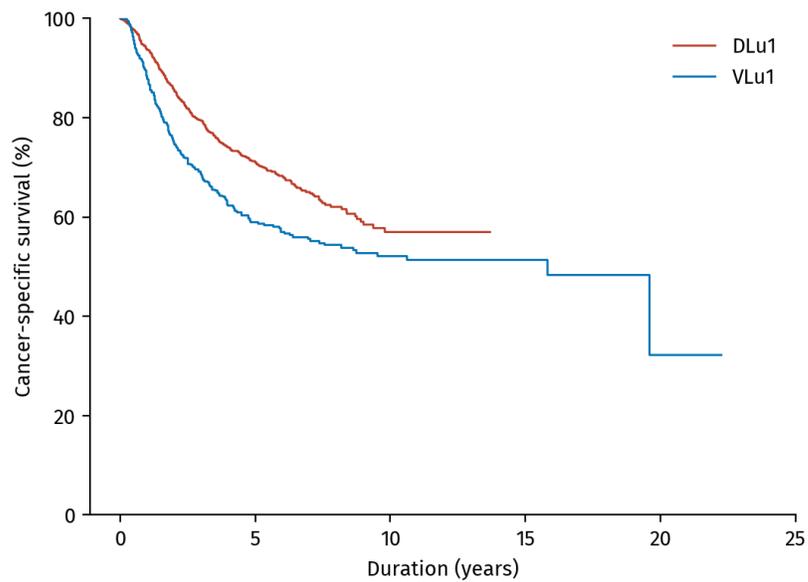

**Protocol Figure 18:** Kaplan-Meier analysis of survival in lung carcinoma materials. Duration is years since surgery for DLu1 and years since diagnosis for VLu1.

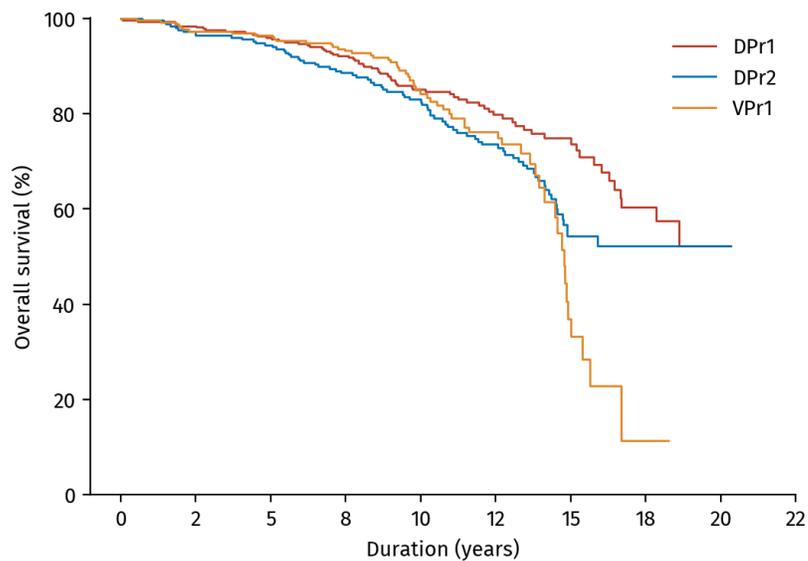

**(a)** Overall survival

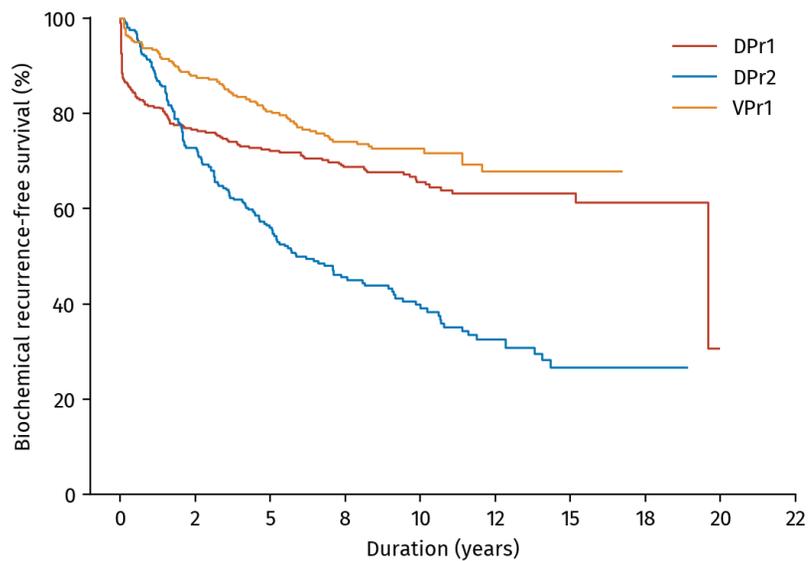

**(b)** Biochemical recurrence-free survival

**Protocol Figure 19:** Kaplan-Meier analysis of survival in prostate carcinoma materials. Duration is years since surgery.

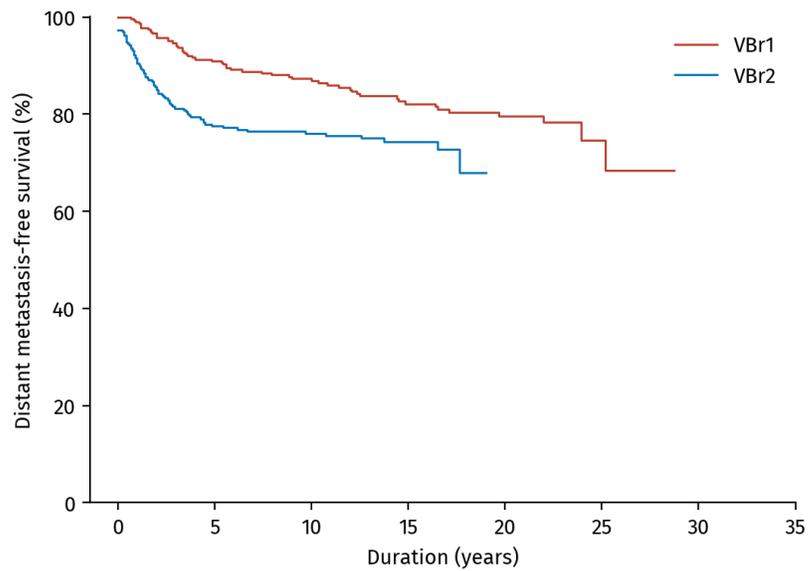

**(a)** Distant metastases-free survival

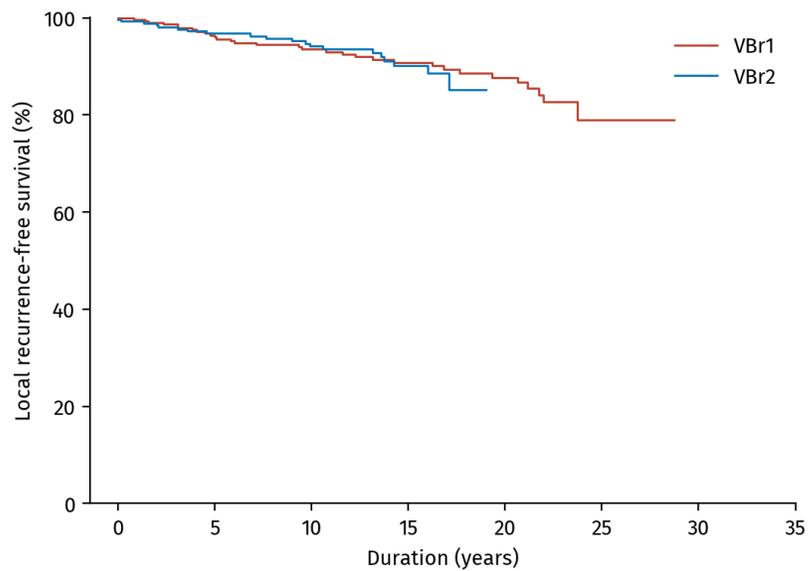

**(b)** Local recurrence-free survival

**Protocol Figure 20:** Kaplan-Meier analysis of survival in breast carcinoma materials. Duration is years since diagnosis. Note different follow-up times between the different event types. 110

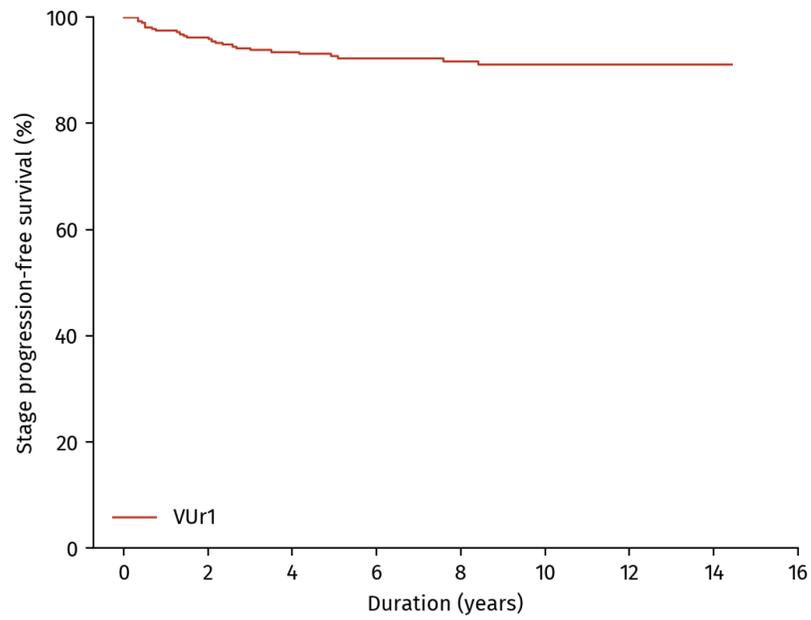

**(a)** Stage progression-free survival

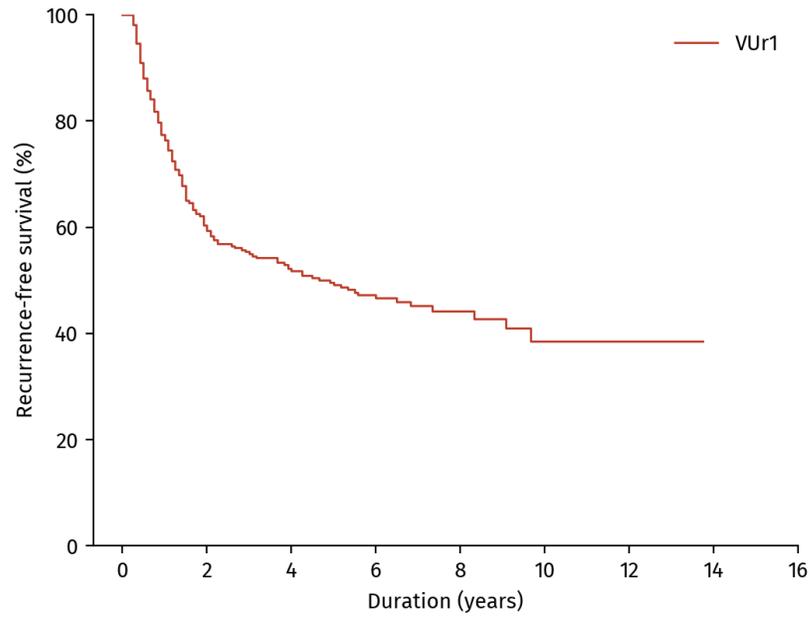

**(b)** Recurrence-free survival

**Protocol Figure 21:** Kaplan-Meier analysis of survival in urothelial carcinoma material. Duration is years since diagnosis. Note different follow-up times between the event types.

### 1.3.3 Colour statistics

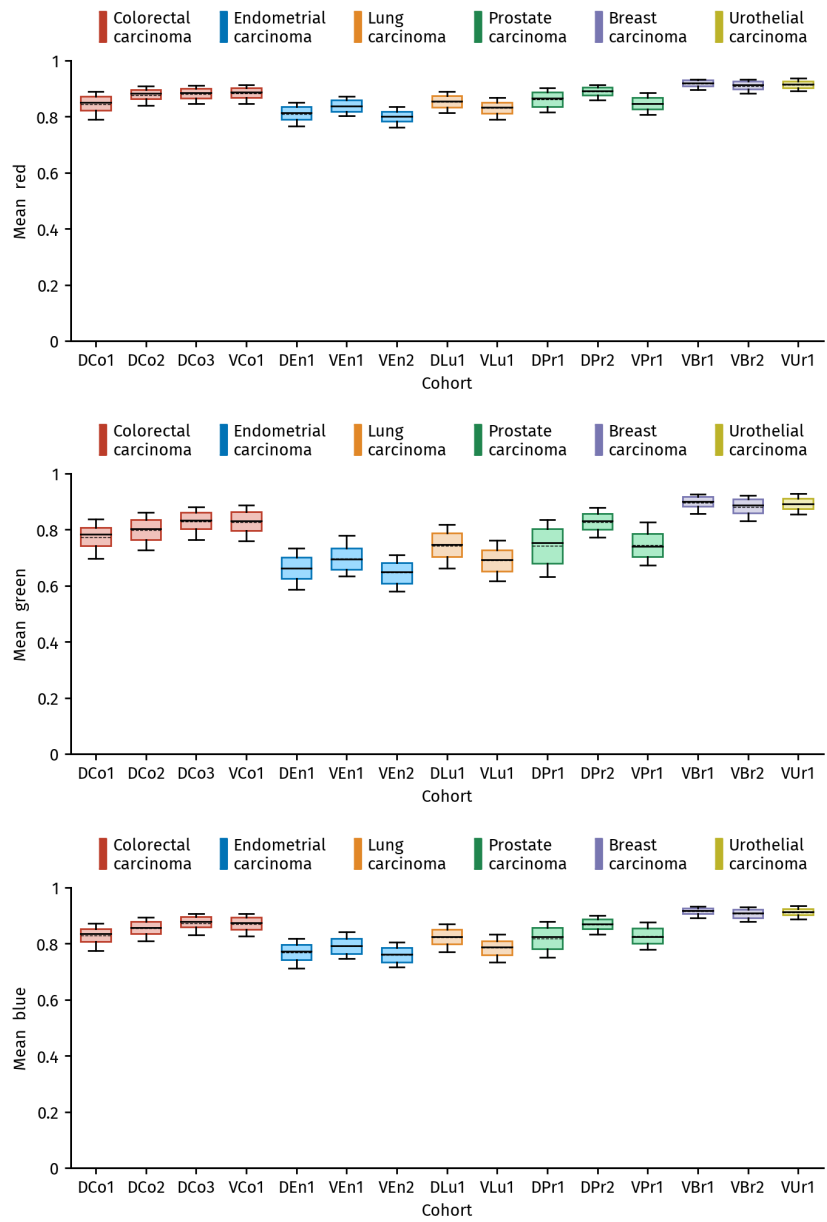

**Protocol Figure 22:** Colour channel mean value in full Aperio AT2 scans downscaled to 5 μm per pixel

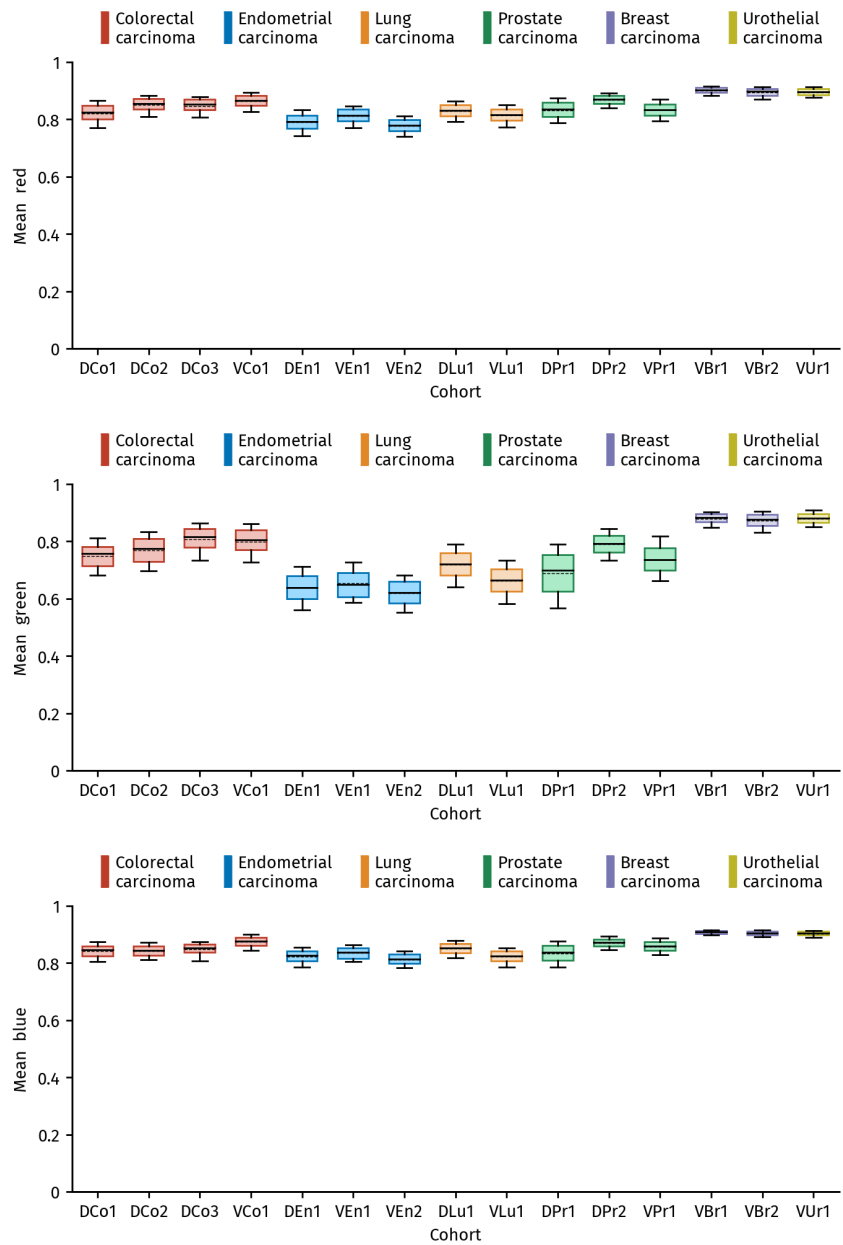

**Protocol Figure 23:** Colour channel mean value in full NanoZoomer XR scans downscaled to 5 μm per pixel

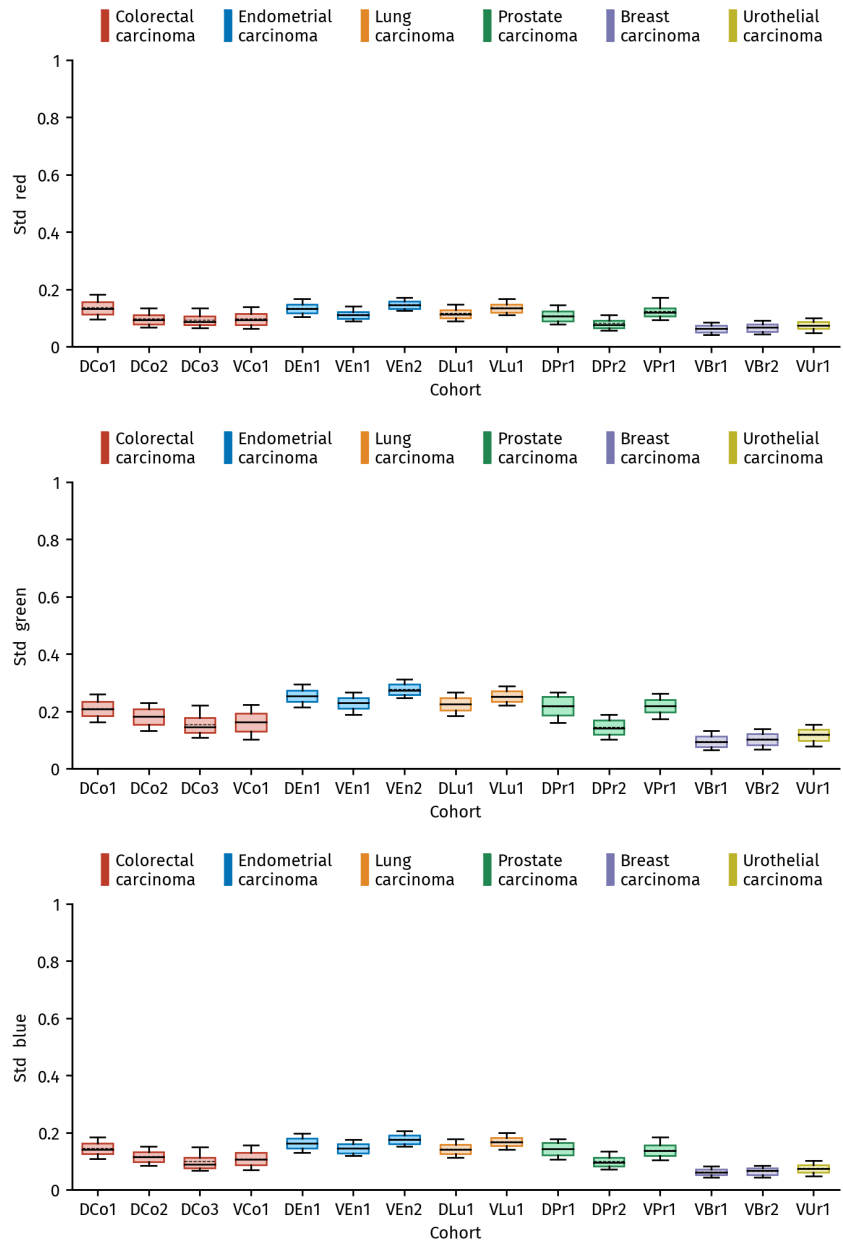

**Protocol Figure 24:** Colour channel standard deviation in full Aperio AT2 scans downscaled to 5 µm per pixel

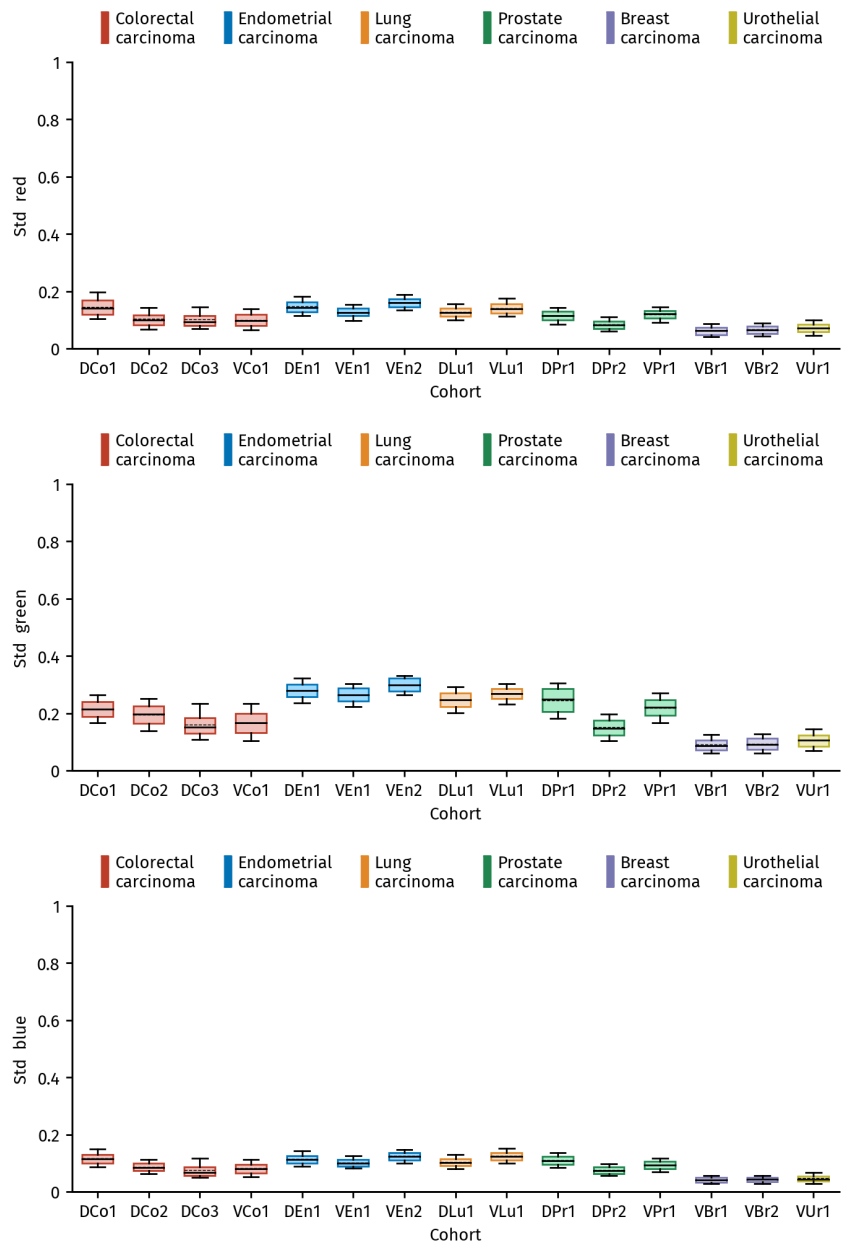

**Protocol Figure 25:** Colour channel standard deviation in full NanoZoomer XR scans downscaled to 5 µm per pixel

## 2 Methods

This section give a detailed explanation of how the segmentation method was developed and how it is applied. The software and hardware development environment is described in section 2.1. Steps necessary for preparing the method are described in section 2.2 while the segmentation method itself is described in section 2.3. How we evaluate the performance of the segmentation result is explained in section 2.4.

### 2.1 Programming environment

Most programs used in this project are implemented in the *Python* programming language. For method validation, programs were run in a *Docker container* based on the `pytorch/pytorch:1.11.0-cuda11.3-cudnn8-runtime` image. The network optimisation was run in a Docker container based on the image `nvcr.io/nvidia/pytorch:22.02-py3`. Additional python packages used are listed in protocol table 9.

**Protocol Table 9:** Python packages and their versions

| Package | Version |
|---|---|
| albumentations | 1.1.0 |
| matplotlib | 3.5.1 |
| opencv_python_headless | 4.5.5.64 |
| openslide_python | 1.1.2 |
| pandas | 1.4.1 |
| segmentation_models_pytorch | 0.2.1 |
| scikit-image | 0.19.2 |
| timm | 0.4.12 |
| toml | 0.10.2 |
| torchinfo | 1.6.3 |

Segmentation network processing was done on graphical processing units (GPUs). We used an Nvidia DGX machine with 8 A100 40 GB SXM GPUs, driver version 470.57.02 and CUDA version 11.4 for the network optimisation. For the validation, we used computers with Nvidia Titan RTX 24 GB GPU cards with driver version 465.19.01 and CUDA version 11.3.

### 2.2 Method development

The segmentation method use a convolutional neural network which needs to be optimised to this particular task of tumour segmentation. All necessary steps needed for preparation are described in this section and are summarised next:

1. Read input scans and downsample them to resolution 1 μm per pixel (section 2.2.1)
2. Partition each scan into tiles with $2048 \times 2048$ pixels (section 2.2.2)
3. Balance the development dataset (section 2.2.3)
4. Exclude background tiles (section 2.2.5)
5. Augment the development dataset (section 2.2.6)
6. Standardise input images (section 2.2.7)
7. Optimise the segmentation network (section 2.2.8)

### 2.2.1 Downsampling

Each scan used in this study is downsampled to a target resolution of 1 µm per pixel (MPP). For reference, the highest resolution of many scans is labelled 40× magnification which corresponds to about 0.25 MPP depending on the scanner vendor and model. As an example, we have scans from Aperio AT2 with a 40× magnification with resolution 0.2530 MPP and scans from NanoZoomer XR with a 40× magnification with resolution 0.2267 MPP (rounded to four decimal places).

The target downsampling factor is found by dividing the target MPP by the MPP at the highest resolution level of the scan (*level 0*). The level 0 MPP is accessed from the scan by `OpenSlide` using the `PROPERTY_NAME_MPP_X` and `PROPERTY_NAME_MPP_Y`. In case the directional level 0 MPP are different*, the target downsampling factor will also be different in the two directions. If these two properties are not available in the scan, it is not included in the study.

For neural network optimisation and application, we read tile regions from the scan file one by one rather than the entire scan. Each tile is read from the scan at the pyramid level with a corresponding downsampling factor smaller than or equal to the target downsampling factor (or the smallest of the two directional target downsampling factors if they are different). Unless the target downsampling factor is equal to the reading downsampling factor, the size of the read tile will be larger than the target size. The enlarged tile is therefore downsampled to the target size so that the resulting resolution is equal to the target resolution. Downsampling to a target size (instead of to a target factor) also ensures that the resulting tile has the exact height and width we desire (and not e.g. off-by-one due to rounding). This final resizing is performed using `OpenCVs` `resize` function with the `INTER_AREA` interpolation option. This ensures that no upsampling is performed, but may result in tiles being read from the scan at different resolutions depending on the scanner model and settings.

For background exclusion, performance evaluation and display purposes, we use the downsampled scan as a single image, and in these cases the scan is downsampled to a resolution of 5 MPP (about 2× magnification). Extracting the image from the scan file is done as for the tiles explained in the previous paragraph, with the exception that the target resolution is different and that the entire scan is read all at once in stead of in smaller regions.

### 2.2.2 Tiling

Since the downscaled scans are too large to process at a resolution of 1 MPP, they are partitioned into a set of tiles. The horizontal and vertical spatial dimensions are split in the same way, and the procedure for computing tile start and end coordinates is listed as python code in protocol listing 1.

The scan is partitioned into overlapping tiles if the scan dimension is not an integer multiple of the tile dimension and the minimum overlap is not specified to be 0. The amount of overlap is equal between all tile columns in the horizontal direction, except for between the rightmost tile columns which may overlap more, so that the rightmost tile column aligns with the right scan boundary. The same is true in the vertical direction where tile rows overlap with the same amount except perhaps for between the bottommost tile rows. With the procedure shown in protocol listing 1, we can also specify the minimum number of overlapping pixels along a dimension.

---

*In this study, none of the included scans had different directional level 0 MPP

```python
from typing import List, Tuple
import numpy as np


def find_overlap(full_size: int, part_size: int, min_overlap: int) -> Tuple[float, int]:
    assert full_size > part_size, "The part is larger than the whole"
    num_parts = int(np.ceil(full_size / part_size))
    overlap = (part_size * num_parts - full_size) / (num_parts - 1)
    if overlap < min_overlap:
        assert part_size > min_overlap, "Part size must be greater than minimal overlap"
        num_parts = int(np.ceil((full_size - min_overlap) / (part_size - min_overlap)))
        overlap = (part_size * num_parts - full_size) / (num_parts - 1)
    return overlap, num_parts


def partition(full_size: int, part_size: int, min_overlap: int) -> List[range]:
    """
    Divide a full line into parts where the line have size full_size and the parts have
    size part_size (except when full_size < part_size).

    Return a list of part start (inclusive) and stop (exclusive) points on the full line
    """
    ranges: List[range] = []
    if full_size > part_size:
        overlap, num_parts = find_overlap(full_size, part_size, min_overlap)
        fractional_part = overlap - np.floor(overlap)
        num_ceils = int(np.floor(num_parts * fractional_part))
        for k in range(num_parts):
            if k <= num_ceils:
                int_overlap = int(np.ceil(overlap))
            else:
                int_overlap = int(np.floor(overlap))
            if k == 0:
                start = 0
            else:
                start = ranges[k - 1].stop - int_overlap
            ranges.append(range(start, start + part_size))
    else:
        ranges.append(range(0, full_size))
    return ranges
```

**Protocol Listing 1:** Divide with overlap

Tiles used for network optimisation have a target spatial dimension of $2\,048 \times 2\,048$ pixels and are sampled from the scan with a minimum overlap of 0 pixels. Tiles used for network inference have a target spatial dimension of $7\,680 \times 7\,680$ pixels with a minimum overlap of $1\,024$ pixels.

Scan tiles are written as `jpg` files with 95% quality while annotation mask tiles are written as `png` files. Full scans at 5 MPP are written as `png` files.

### 2.2.3 Dataset balancing

The development set was balanced w.r.t. cancer type by oversampling the minority groups on a tissue slide level. Tissue slides were selected multiple times at random without replacement so that no slides were selected $n+1$ times before all slides had been selected $n$ times. This resulted in $3\,519$ sections sampled from each cancer type (the same number of sections included in lung carcinoma, which was the majority group). Counting scans from both scanners, the result was $7\,030$ scans from colorectal carcinoma and $7\,038$ scans from each of the other cancer types. See protocol table 10 for an overview of the number of scans for each cohort. Note that since the selection was done on a slide level, and cohorts DCo2 and DCo3 had fewer NanoZoomer XR scans than Aperio AT2, there are slightly fewer scans from NanoZoomer XR than from Aperio AT2 in protocol table 10 for these two cohorts.

**Protocol Table 10:** Number of annotated scans in the training cohorts after balancing with oversampling

| Cancer type | Cohort | Scans Aperio AT2 | NanoZoomer XR |
|---|---|---|---|
| Colorectal carcinoma | DCo1 | 471 | 471 |
|  | DCo2 | 1 309 | 1 303 |
|  | DCo3 | 1 739 | 1 737 |
|  | Sum | 3 519 | 3 511 |
| Endometrial carcinoma | DEn1 | 3 519 | 3 519 |
| Lung carcinoma | DLu1 | 3 519 | 3 519 |
| Prostate carcinoma | DPr1 | 1 981 | 1 981 |
|  | DPr2 | 1 538 | 1 538 |
|  | Sum | 3 519 | 3 519 |
| Sum |  | 14 076 | 14 068 |

### 2.2.4 Background segmentation

A simple method is employed to segment the white background in an image from the rest. This background mask is used to alter both predicted and reference segmentation masks. This is useful when large background regions are inside the annotated region (one example being holes from *tissue microarray* acquisition) without being manually annotated as background. These regions are clearly not cancerous tissue, and should not be annotated as such neither by the reference nor by the prediction.

Note that this segmentation is quite sensitive in that it will mark most tissue as foreground, also adipose tissue that is often left out when applying threshold methods based on image brightness or saturation or similar. But it may also include artefacts such as pen markings, air bubbles, dust, glass cracks, etc. But since the mask is used to exclude white background tiles used in training, it can be an advantage that foreground elements other than tissue is included. The method with the stated parameter values assumes images of H&E-stained tissue with 5 MPP resolution.

Canny edge detection is performed on the input colour image, using the `OpenCV Canny` implementation.[26] We use a $3 \times 3$ *Sobel* filter for the gradient computation, and thresholds of 10 and 50 for the lower and upper thresholds in the hysteresis. This produce a mask with lots of foreground pixels in regions with structure and lots of background pixels in homogeneous regions. This foreground mask is refined by first removing small background regions. The mask first

undergoes morphological closing (`openCV morphologyEx`) with a square $9 \times 9$ structure element before background regions with an area smaller than $10\,000$ pixels are filled in with the function `remove_small_holes` from the `scikit-image` python library. An area of $10\,000$ pixels at 5 MPP resolution corresponds to a square region of $0.5\,\text{mm} \times 0.5\,\text{mm}$.

Finally, small foreground regions are removed from the mask. Morphological opening is applied on the mask using the `openCV` function `morphologyEx` with the same $9 \times 9$ structure element before foreground regions with an area smaller than $1\,600$ pixels are erased using the function `remove_small_objects` from the `scikit-image` python library. An area of $1\,600$ pixels corresponds to a square region of $0.2\,\text{mm} \times 0.2\,\text{mm}$ at 5 MPP resolution.

This method is simple to implement, very robust, and quite fast, spending around one second per image on a single CPU core on consumer-grade hardware. An example of a downscaled scan from colorectal carcinoma scanned with Aperio AT2 and manually annotated is shown in protocol figure 26.

With this we can classify every pixel as either white background, foreground without annotation and foreground with annotation. This content classification is summarised for all scans in all cohorts used in this study in protocol figures 27 to 30.

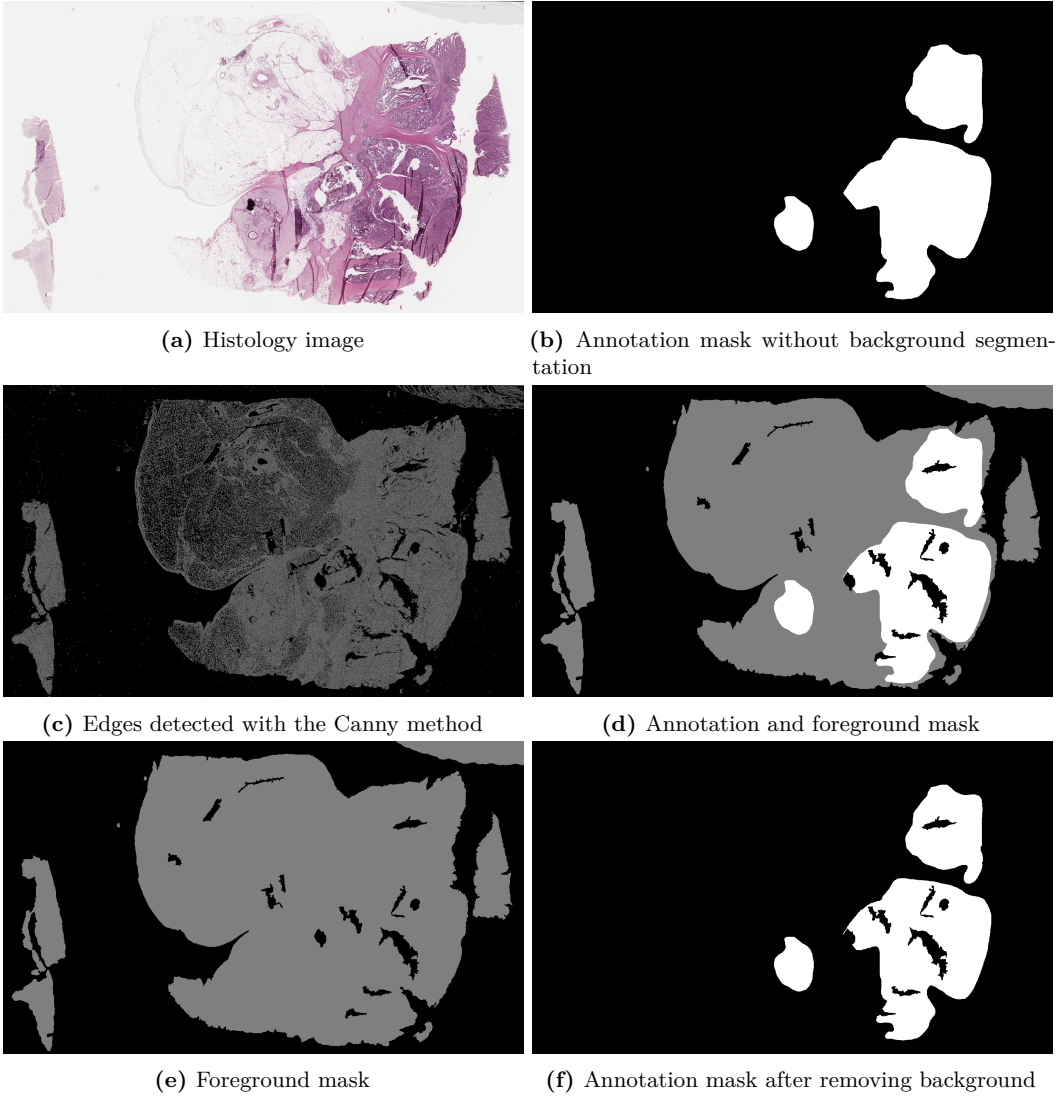

**(a)** Histology image

**(b)** Annotation mask without background segmentation

**(c)** Edges detected with the Canny method

**(d)** Annotation and foreground mask

**(e)** Foreground mask

**(f)** Annotation mask after removing background

**Protocol Figure 26:** Scan and segmentation mask with background (black), foreground (gray) and tumour annotation (white)

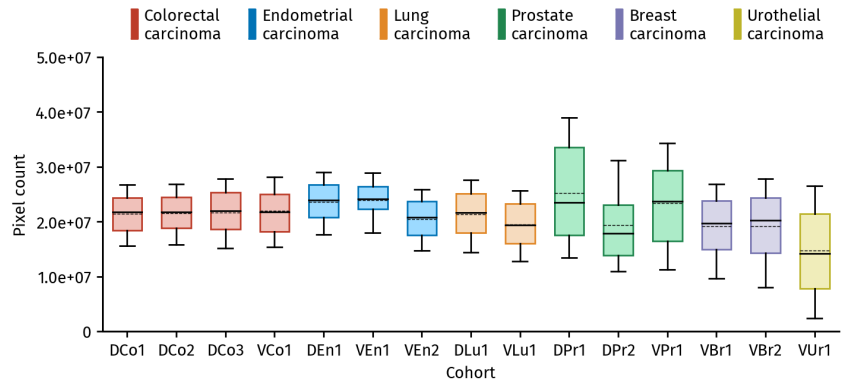

**(a)** Image area

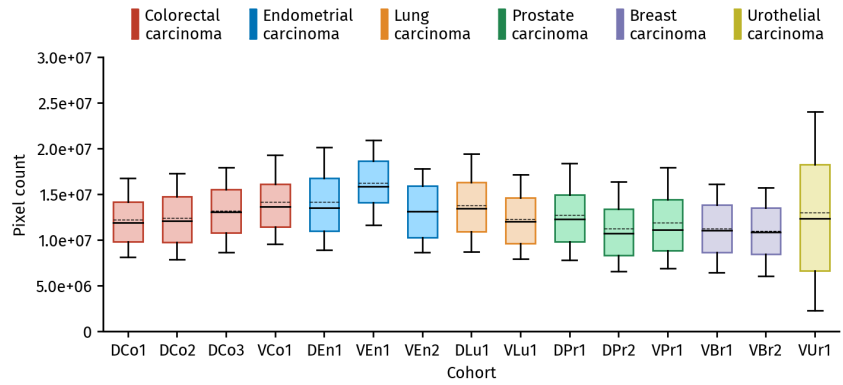

**(b)** Foreground area

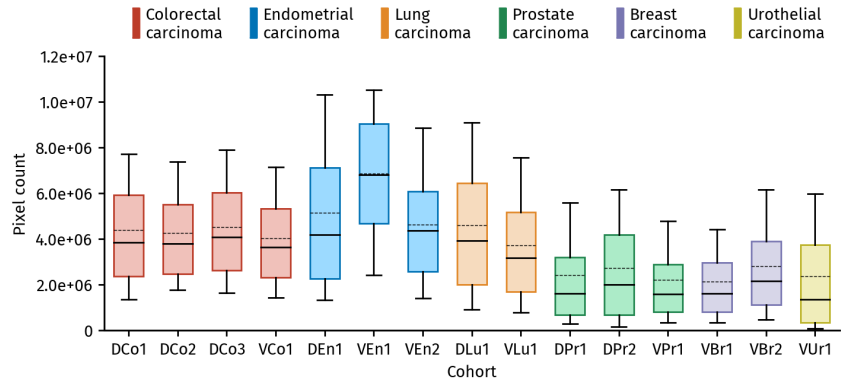

**(c)** Annotation area

**Protocol Figure 27:** Area in number of pixels at resolution 5 µm per pixel in Aperio AP2 scans. "Foreground" is foreground without annotation and "Annotation" is foreground with annotation. Background exclusion masks are applied on all images. Note the difference in vertical axis range between subplots.

122

39                                                    89

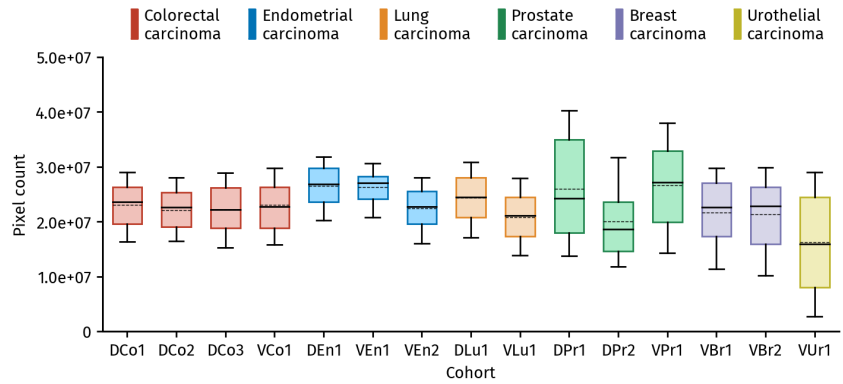

**(a)** Image area

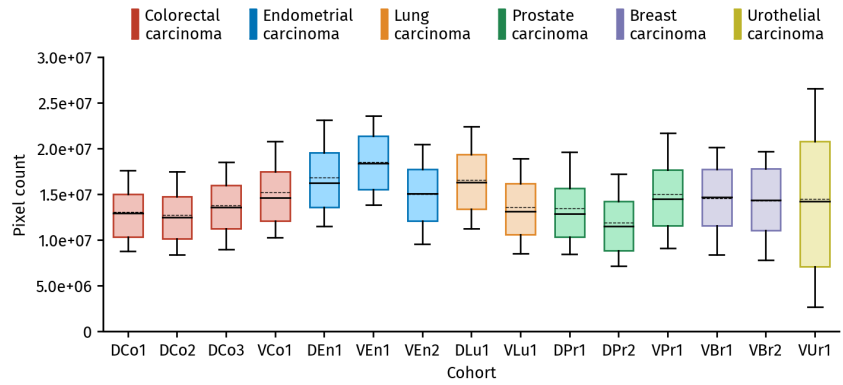

**(b)** Foreground area

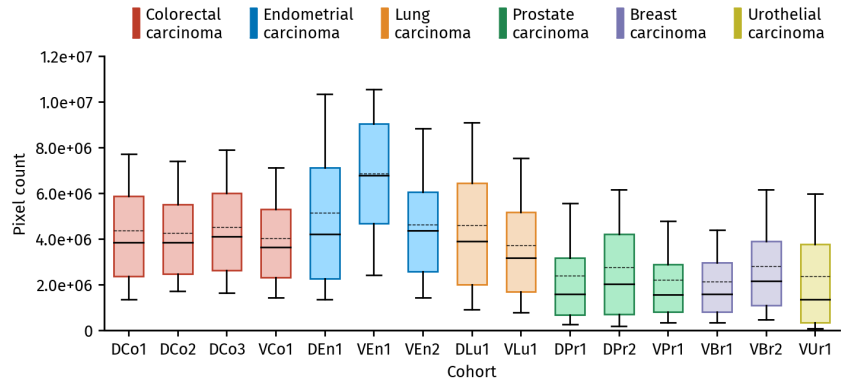

**(c)** Annotation area

**Protocol Figure 28:** Area in number of pixels at resolution $5\,\mu m$ per pixel in NanoZoomer XR scans. "Foreground" is foreground without annotation and "Annotation" is foreground with annotation. Background exclusion masks are applied on all images. Note the difference in vertical axis range between subplots.

40                    90

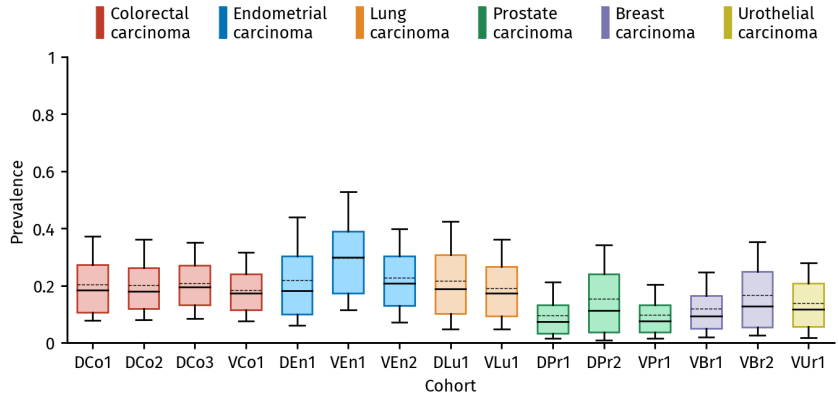

**(a)** Annotation prevalence in image

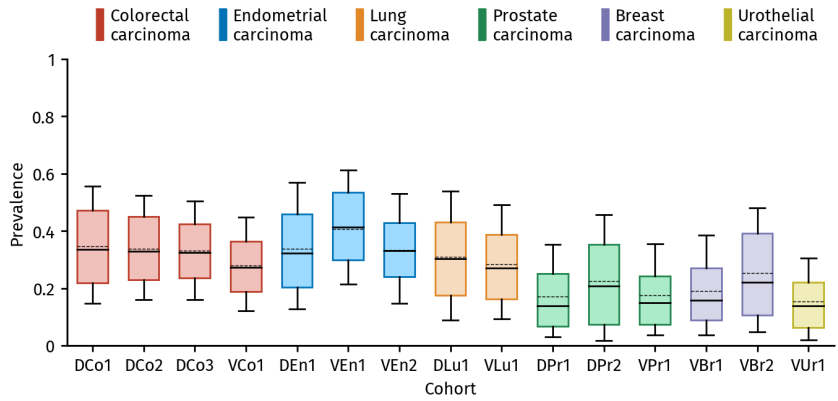

**(b)** Annotation prevalence in foreground

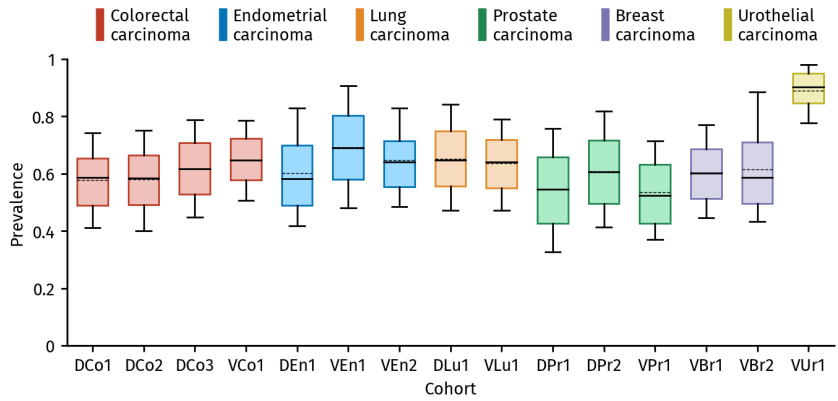

**(c)** Foreground prevalence in image

**Protocol Figure 29:** Prevalence at resolution 5 µm per pixel in Aperio AT2 scans. "Foreground" is foreground without annotation and "Annotation" is foreground with annotation. Background exclusion masks are applied on all images.

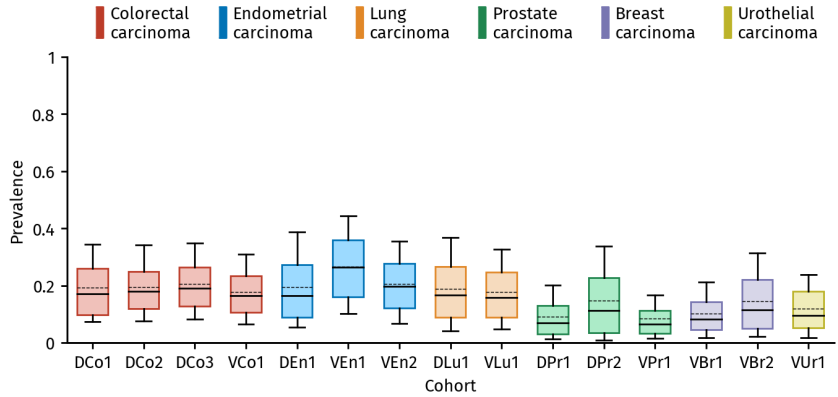

**(a)** Annotation prevalence in image

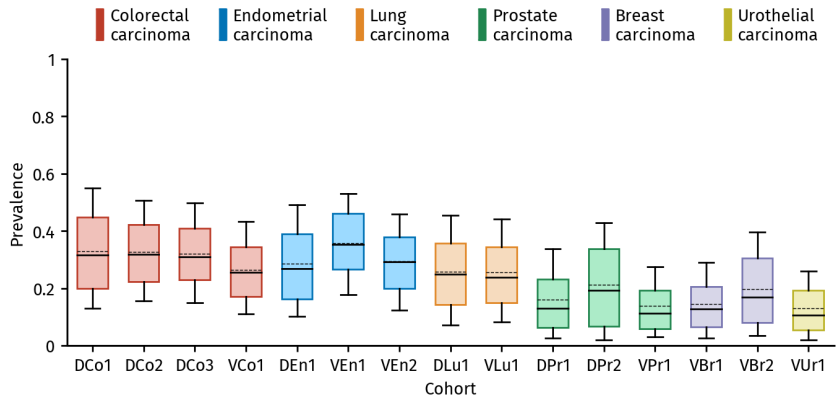

**(b)** Annotation prevalence in foreground

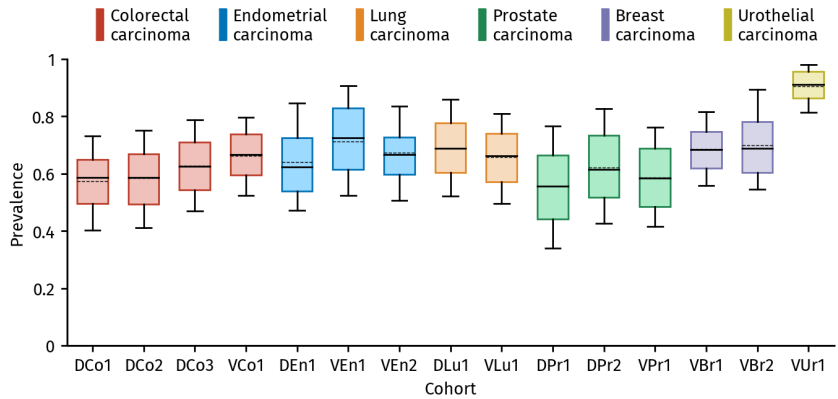

**(c)** Foreground prevalence in image

**Protocol Figure 30:** Prevalence at resolution 5 μm per pixel in NanoZoomer XR scans. "Foreground" is foreground without annotation and "Annotation" is foreground with annotation. Background exclusion masks are applied on all images.

### 2.2.5 Background tile exclusion

Tiles containing too much white background are removed from the development set. The background segmentation is performed on 5 MPP full images as described in section 2.2.4 and transferred to the 1 MPP tiles. Specifically, we include all tiles that contain some tumour annotated regions, and for those that don't, we keep those with a background fraction smaller than 50%.

In total this reduces the number of unique tiles from 3 080 330 to 2 144 651 or from 4 233 081 to 2 902 032 non-unique tiles in the balanced dataset (see protocol table 11 and protocol figure 31).

**Protocol Table 11:** Number of unique tiles in development cohorts before and after background exclusion

| Cohort | Aperio AT2 Tiles before | Tiles after | % | NanoZoomer XR Tiles before | Tiles after | % | Both scanners Tiles before | Tiles after | % |
|--------|----------|-----------|-------|-------------|------------|-------|-------------|-----------|-------|
| DCo1 | 28 911 | 19 429 | 67.20 | 30 552 | 20 245 | 66.26 | 59 463 | 39 674 | 66.72 |
| DCo2 | 81 651 | 54 356 | 66.57 | 82 294 | 55 104 | 66.96 | 163 945 | 109 460 | 66.77 |
| DCo3 | 108 429 | 76 511 | 70.56 | 109 720 | 78 609 | 71.65 | 218 149 | 155 120 | 71.10 |
| DEn1 | 514 341 | 340 922 | 66.28 | 566 105 | 394 211 | 69.64 | 1 080 446 | 735 133 | 68.04 |
| DLu1 | 492 150 | 364 029 | 73.97 | 551 767 | 422 192 | 76.52 | 1 043 917 | 786 221 | 75.31 |
| DPr1 | 158 001 | 91 707 | 58.04 | 163 527 | 96 500 | 59.01 | 321 528 | 188 207 | 58.54 |
| DPr2 | 95 015 | 63 855 | 67.21 | 97 867 | 66 981 | 68.44 | 192 882 | 130 836 | 67.83 |
| Sum | 1 478 498 | 1 010 809 | 68.37 | 1 601 832 | 1 133 842 | 70.78 | 3 080 330 | 2 144 651 | 69.62 |

126

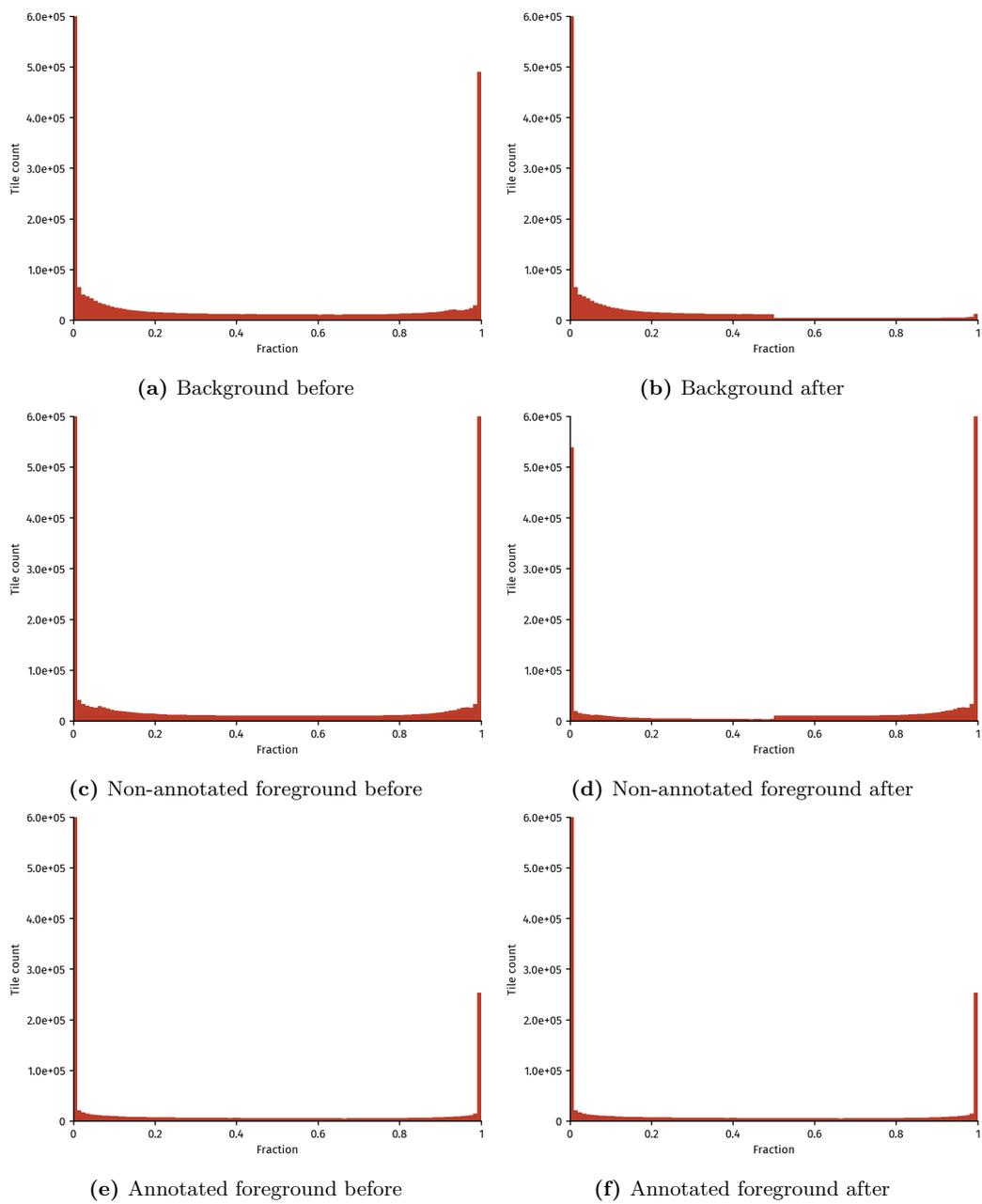

**Protocol Figure 31:** Distribution of number of tiles with a certain fraction of either background (top row) non-annotated foreground (middle row) and annotated foreground (bottom row) before (left column) and after (right column) background exclusion.

### 2.2.6 Dataset augmentations

Image tiles are read as *RGB* with 8 bits values per channel, cast to 32 bits floating point values, and then preprocessed before they enter the segmentation network. We artificially augment the training dataset by distorting images using the `albumentations` library.[27] The operations are listed in protocol listing 2 in the order they are applied. Note that while the tiles are sampled at a size of $2\,048 \times 2\,048$ pixels, they are cropped to a size of $1\,536 \times 1\,536$ pixels before they enter the network.

Image distortions are only applied during network optimisation, and not when the fixed network is applied.

```
1  HorizontalFlip(p=0.5)
2  RandomRotate90(p=0.5)
3  PadIfNeeded(min_height=1536, min_width=1536, border_mode=cv2.BORDER_CONSTANT, value=0)
4  RandomCrop(height=1536, width=1536)
5  RandomBrightnessContrast(
6      brightness_limit=[-0.2, 0.2], contrast_limit=[-0.2, 0.2], brightness_by_max=True, p=1
7  )
8  HueSaturationValue(
9      hue_shift_limit=[-26, 26], sat_shift_limit=[-26, 26], val_shift_limit=[-26, 26], p=1
10 )
```

**Protocol Listing 2:** Image distortions

### 2.2.7 Image value standardisation

Before the image enters the network, the image values are divided by 255 before the image is centred around the development dataset mean value and scaled with the development dataset standard deviation. This standardisation is applied both during network optimisation and inference.

The dataset mean value for an image channel is computed as

$$\mu = \frac{1}{m} \sum_{i=1}^{m} \mu_i$$

$$= \frac{1}{m} \sum_{i=1}^{m} \frac{1}{n_i} \sum_{j=1}^{n_i} x_{ij}$$

where $x_{ij}$ is the value at pixel $j$ in image $i$ for the image channel and $\mu_i$ is the mean value in image $i$. $n_i$ is the number of pixels in image $i$, and $m$ is the number of images in the dataset. Similarly, the dataset variance for a single channel is estimated as

$$\sigma^2 = \frac{1}{m} \sum_{i=1}^{m} \sigma_i^2$$

$$= \frac{1}{m} \sum_{i=1}^{m} \frac{1}{n_i - 1} \sum_{j=1}^{n_i} (x_{ij} - \mu_i)^2.$$

We use $\sigma = \sqrt{\sigma^2}$ as the estimate for the dataset standard deviation. For both estimates $\mu$ and $\sigma$, the final result is divided by 255 before it is applied.

When applied on all unique $2\,048 \times 2\,048$-sized tiles in the development dataset at resolution 1 MPP without distortions, we get the result shown in protocol table 12 and protocol figures 32

and [33]. Colour mean and standard deviation distributions for the all scans at resolution 5 MPP are shown in protocol figures [22] to [25].

**Protocol Table 12:** Colour statistics for all unique tiles in the development set. Here, all 8-bit integer colour channel values are cast to 32 bit floating point values before the per-image statistics are computed. These values are then averaged over all tiles and the result is divided by 255.

| Colour channel | Mean value | Standard deviation |
|---|---|---|
| Red | 0.8297992 | 0.1051075 |
| Green | 0.7106879 | 0.1543867 |
| Blue | 0.8241846 | 0.0991757 |

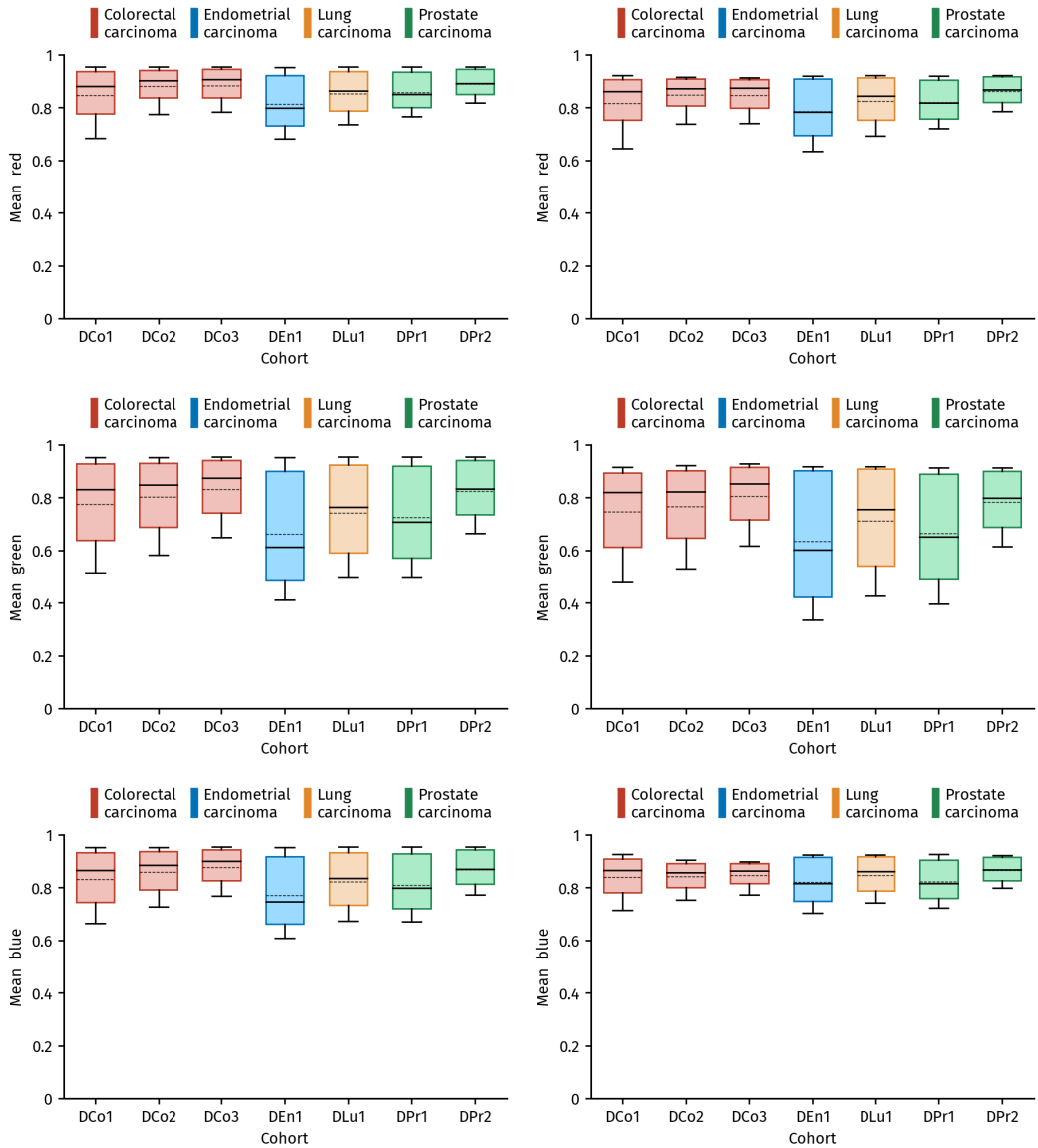

**Protocol Figure 32:** Colour channel mean value for all unique tiles in the development set. Tiles have resolution 1 MPP and a size of $2048 \times 2048$ pixels. Aperio AT2 in the left column and NanoZoomer XR in the right column.

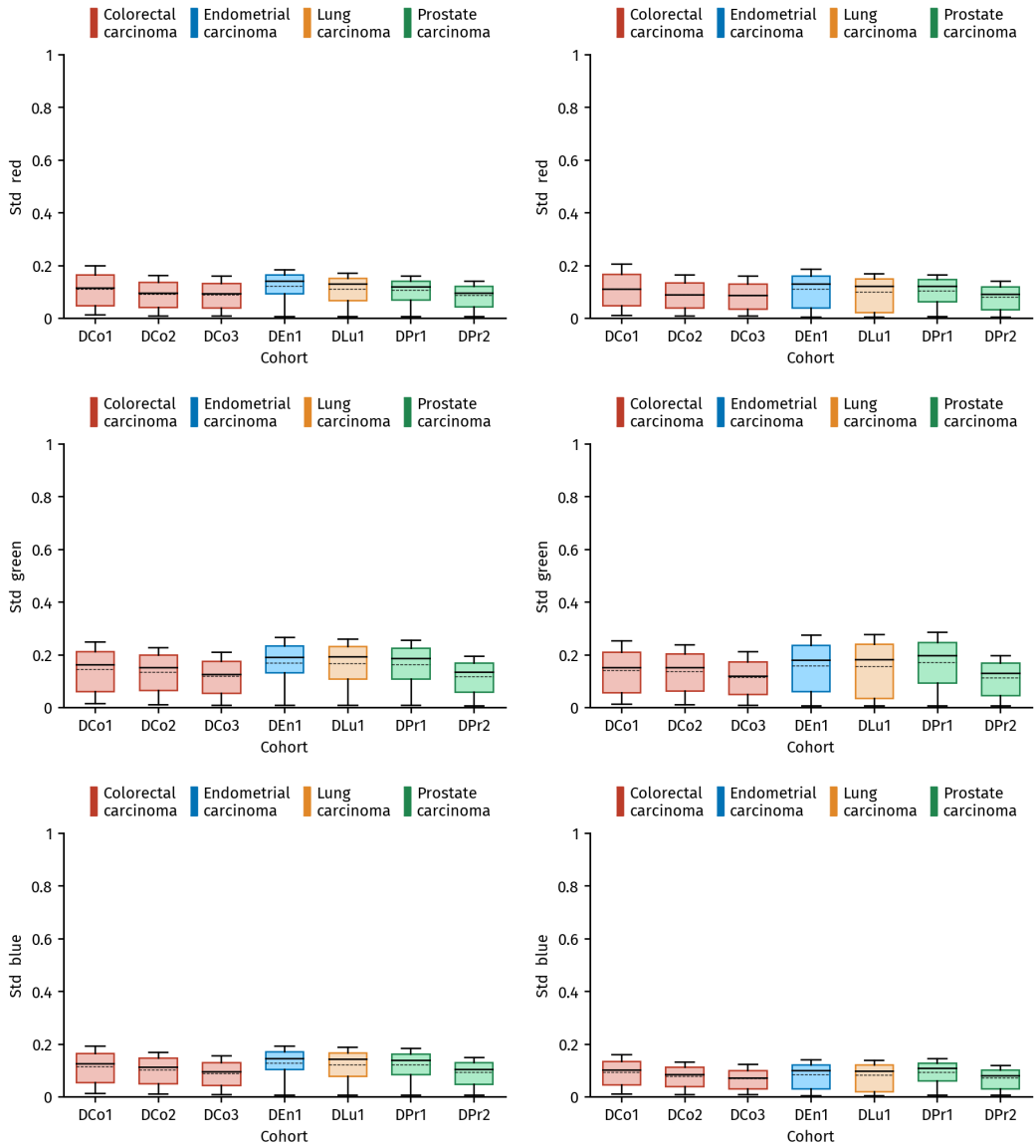

**Protocol Figure 33:** Colour channel standard deviation for all unique tiles in the development set. Tiles have resolution 1 MPP and a size of $2048 \times 2048$ pixels. Aperio AT2 in the left column and NanoZoomer XR in the right column.

### 2.2.8 Segmentation network

The segmentation network is an encoder-decoder network developed using the `PyTorch v1.11` machine learning framework.[28]

For the encoder we used a *Normalising-free Network* (NFNet), a modern classification network designed to achieve state-of-the-art performance without using batch normalisation.[29, 30] More specifically, we use the `eca_nfnet_l3` implementation provided by the `timm version 0.4.12` package.[31]

This implementation differ from the one described by Brock and colleagues in that it has 4, 8, 24, and 12 blocks for the four stages, respectively. The *Squeeze and Excitation* module is replaced by the *Efficient Channel Attention* module.[32, 33] It also use *SiLu* activation functions instead of *GeLu*.[34]

The decoder is the decoder from the *DeepLabV3+* segmentation network, and the implementation is from the `segmentation_models_pytorch` python package.[35, 36] We modified the DeeplLabV3+ decoder to be free of batch normalization, following the NFNet encoder. To achieve this, we simply replaced every batch norm layer with a group norm layer with groups of size 8.[37]

The network consist in total of 73 472 403 adjustable parameters to be optimised (computed by `torchinfo`). An overview of the architecture can be seen in protocol figure 34.

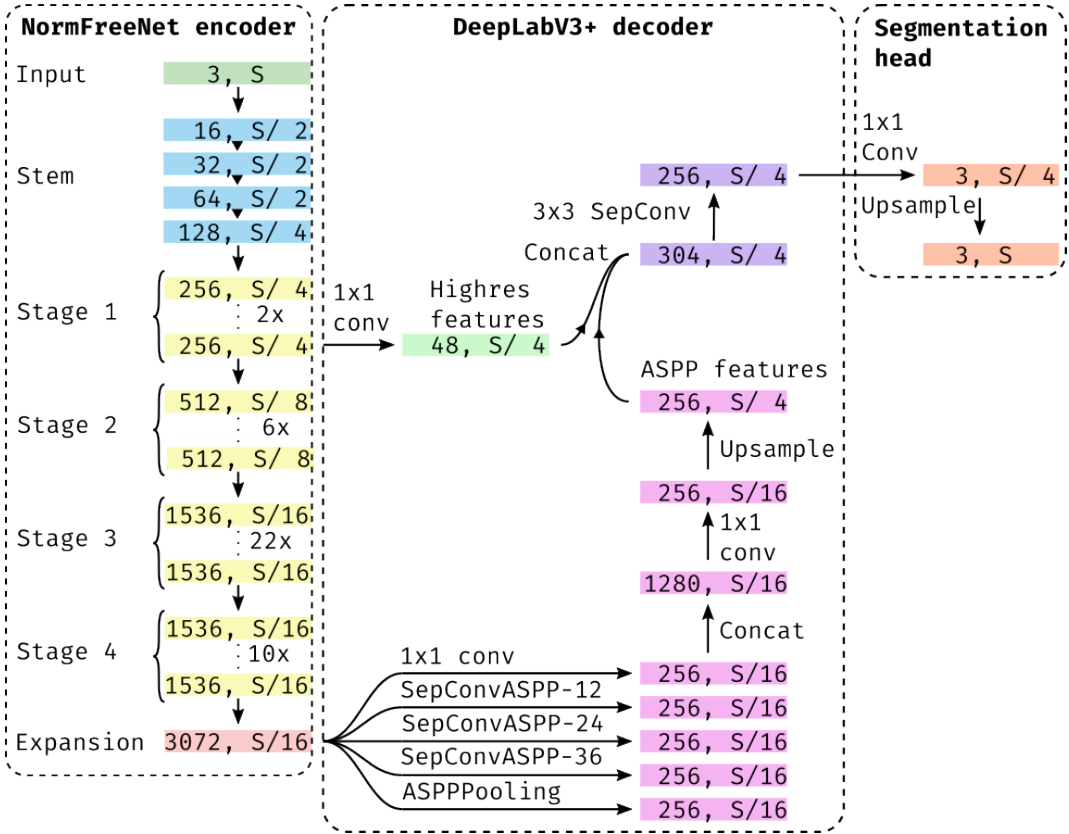

**Protocol Figure 34:** Segmentation network architecture. Each coloured block represent an element of the network and is described with two symbols separate by a comma. The first is the number of output channels and the second is the spatial size relative to the input size $S$. The first block has three channels since we use RGB-images, while the last block has three channels since we predict three classes.

132

### 2.2.9 Network optimisation

Below we describe how the network was optimised, but in general we follow the procedure described in by Brock and colleagues in their NFNet paper, with some exceptions.[29] We do *not* use *Adaptive Gradient Clipping* as we did not see any benefit for it in our case, perhaps because of our small batch size (24 images). We also do not use moving averages of the model parameters.

The objective is to minimise the difference between the output of the segmentation network and the reference segmentation by iteratively modifying the adjustable parameters of the segmentation network. The difference to be minimised is captured by the loss function $l = l_1 + l_2$, where $l_1$ is the so-called *Dice-loss* (`DiceLoss` from `segmentation_models_pytorch` with `mode="multiclass"`), and $l_2$ is a so-called *top-90 Cross Entropy* loss function. The top-90 Cross Entropy at a particular step is computed by first computing the per-pixel cross entropy for all pixels in the mini batch of this step and then averaging the cross entropy value over pixels in the top 90 percentile. That is, when computing the mean cross entropy, we are ignoring 10% of pixels with the lowest cross entropy value.

We predict three classes, and the reference is segmented into background, non-annotated foreground, and tumour-annotated foreground. We also experimented with using just two classes, tumour-annotated foreground and everything else, but we did not notice any important difference in performance.

The convolution weights in the encoder are initialised with *normal* initialisation while the biases are initialised to zero.

$$X \sim \mathcal{N}(0, \sigma^2), \text{ where } \sigma = \sqrt{\frac{1}{c_i h w}}.$$

The convolution weights and biases used in the decoder and segmentation head are initialised with *uniform* initialisation

$$X \sim \mathcal{U}(-a, a), \text{ where } a = \sqrt{\frac{1}{c_i h w}}.$$

In the above equations, $c_i h w$ is the volume of the input feature maps in the convolutional layer (number of input channels times the height times the width), often called *fan in*.[38]

At each iteration (or step), the adjustable network parameters are updated according to the *Stochastic Gradient Descent* optimisation method with *Nesterov momentum* 0.9.[39] The optimisation is regularised with a *weight decay* value of $2 \times 10^{-5}$ with the exceptions described by Brock and colleagues.[29]

A batch of 24 images is randomly selected without replacement from the development dataset and processed at each step. When the dataset is exhausted we say that an *epoch* is complete, and the selection is reset. The whole batch is processed by the segmentation network before the output is compared with the corresponding reference segmentation batch with the objective function. The batch of 24 is distributed on 8 GPUs with 3 tiles per GPU using `pytorch`s `DistributedDataParallel`

The step length is initialised to $1.0 \times 10^{-4}$ and incremented by $1.0 \times 10^{-4}$ every 10th step until step 1 000 when the step length has reached $1.0 \times 10^{-2}$. After this warm up period, the step length follows a cosine annealing schedule until termination (see protocol figure 35).[40]

The optimisation is carried out for 500 000 steps (or 4.14 epochs) before termination. Since we have 2 902 032 tiles in the dataset and 24 tiles per batch, we have 120 918 steps per epoch. The model at step 500 000 is selected as the model used in the segmentation method.

We employ *Automatic mixed precision* both during optimisation of the network and when applying it. This is provided by the `torch.cuda.amp` module in the `pytorch` python package.

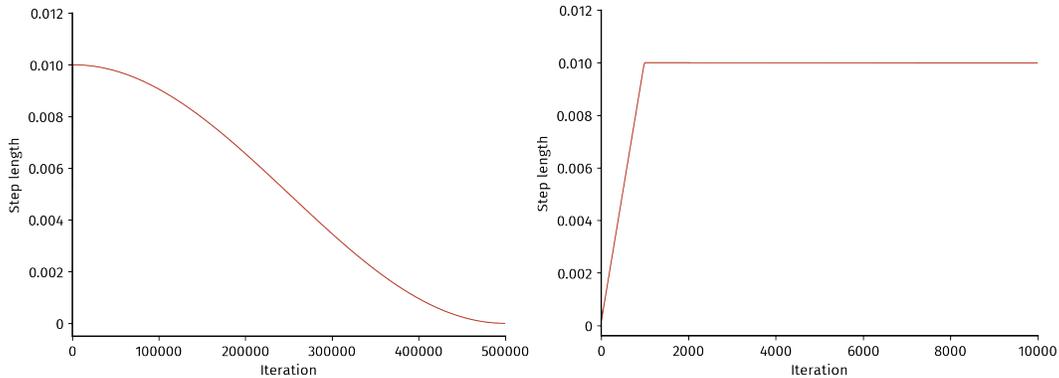

**Protocol Figure 35:** Step length with respect to optimisation iteration. Left panel shows all 500 000 steps while the right panel only shows the first 10 000 steps.

## 2.3  Method application

Application of the method on a single input scan can be summarised as

1. Read the input scan at 1 MPP resolution (section 2.3.1)
2. Partition the downsampled scan into overlapping tiles (section 2.3.1)
3. Apply the optimised segmentation network on each tile (section 2.3.2)
4. Construct a probability image from the segmentation network tiles (section 2.3.3)
5. Post-process to yield a final segmentation mask (section 2.3.4)

### 2.3.1  Downsampling and tiling

Scan reading and downsampling is done as described in section 2.2.1. Tiling is done as described in section 2.2.2, with tile size of $7\,680 \times 7\,680$ pixels with a minimum overlap of $1\,024$ pixels in each direction (see example in protocol figure 36).

### 2.3.2  Neural network

Input images are processed with the optimised segmentation network after the following operations are applied on the input image

1. Read image as RGB with 8-bit values in each channel
2. Zero-pad image so that both the image height and width are divisible by 16. This step is not necessary for this particular setup since we have tiles with size $7\,680 \times 7\,680$, but is included for making the method applicable in the general case with varying input sizes.
3. Scale image values to $(0, 1)$ by dividing by 255
4. Subtract image by development dataset mean (protocol table 12)
5. Divide image by development dataset standard deviation (protocol table 12)

The resulting prediction from the segmentation network is an image with one channel per output class, where only the channel corresponding to the tumour class is used further. Its values are floats where pixel value 0 indicates negative prediction and 1 indicate positive prediction. The image values are multiplied by 255 before the image is quantised to 8 bits. The padding (if any) is removed before the prediction is written as a `png` image.

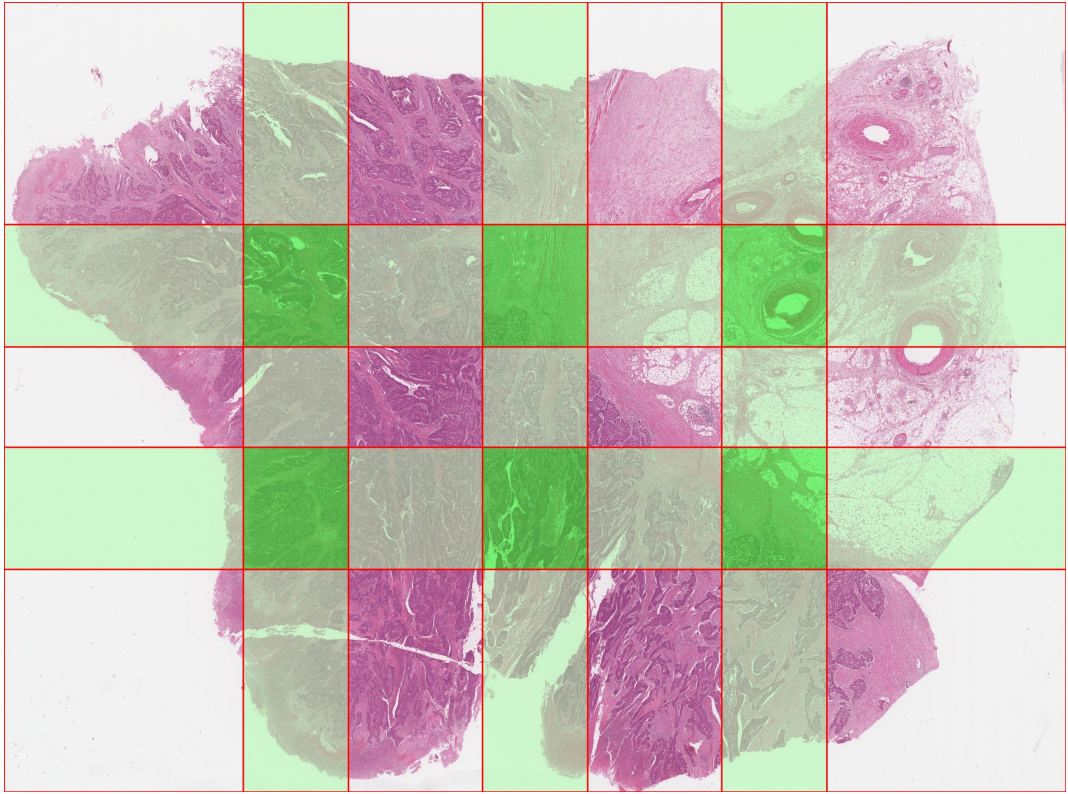

**Protocol Figure 36:** Tiling with size $7\,680 \times 7\,680$ pixels and minimum overlap of $1\,024$ pixels at resolution 1 MPP. Red shows the tile contour. Tile interiors are shown in green with opacity increasing with the number of overlapping tiles: transparent for no overlapping, light green for two overlapping, and darker green for four overlapping tiles.

### 2.3.3 Reconstruction from tiles

The final reconstructed image $f$ is computed as $f = \sum_i w_i g_i$ where $f, w_i, g_i$ are $m \times n$ matrices and $i$ iterate over all tiles. $g_i$ represent a single tile output from the segmentation network, and has the output tile value in the tile location and value zero everywhere else. $w_i$ represent a single weight tile which has values in the corresponding tile location and value zero everywhere else. $w_i$ have values in $[0, 1]$ and $g_i$ have integer values in $[0, 255]$ since they have been written as 8-bit $\texttt{png}$ files by the segmentation network. The values of $f$ are quantised to integer values by rounding with the tie-breaking rule of rounding half to even before $f$ is written as $\texttt{png}$.

The weight tiles are constructed so that the sum weight image $s = \sum_i w_i$ with shape $m \times n$ will have value 1 in all pixels. In the rest of this explanation a *weight tile* and *image tile* will refer only to the part of $w_i$ and $g_i$ that correspond to the location of each tile, respectively.

The tile weights are constructed in three phases, and an example result is shown in protocol figure 37. First, initial weight tiles are computed for each image tile. These weight tiles are weighted by distance in overlapping regions. A sum image the same size of $f$ is constructed by adding all initial weight tiles $w$ at their locations within this sum image. Each initial weight tile is normalised by dividing it by the tile cropped out from its location within the sum image. The next two paragraphs explain the construction of the initial weight tiles.

An initial weight tile $w$ is computed as the element-wise product of four side-specific weight tiles: $w_t$ weighting overlaps at the top of $w$, $w_b$ weighting overlaps at the bottom of $w$, $w_l$ weighting

overlaps at the left of $w$ and $w_r$ weighting overlaps at the right of $w$.

In order to compute a side-specific weight tile, e.g. $w_r$, the smallest leftmost coordinate of all overlapping tiles with a leftmost coordinate greater than the leftmost coordinate in $w$ is recorded. The region between this recorded coordinate and the rightmost coordinate of $w$ defines the overlapping area to the right in $w$. All pixels in $w_r$ to the left of this overlapping area are given value 1, and all other pixels are giving a value decreasing linearly with the distance from the left overlapping border: $v = 1 - \frac{d}{1+l}$ where $v$ is the result value, $d$ is the distance from the left overlapping border, and $l$ is the length of the overlapping region. Both $d$ and $l$ are measured in pixels. The procedure and weighting is similar for the other side-specific weight tiles.

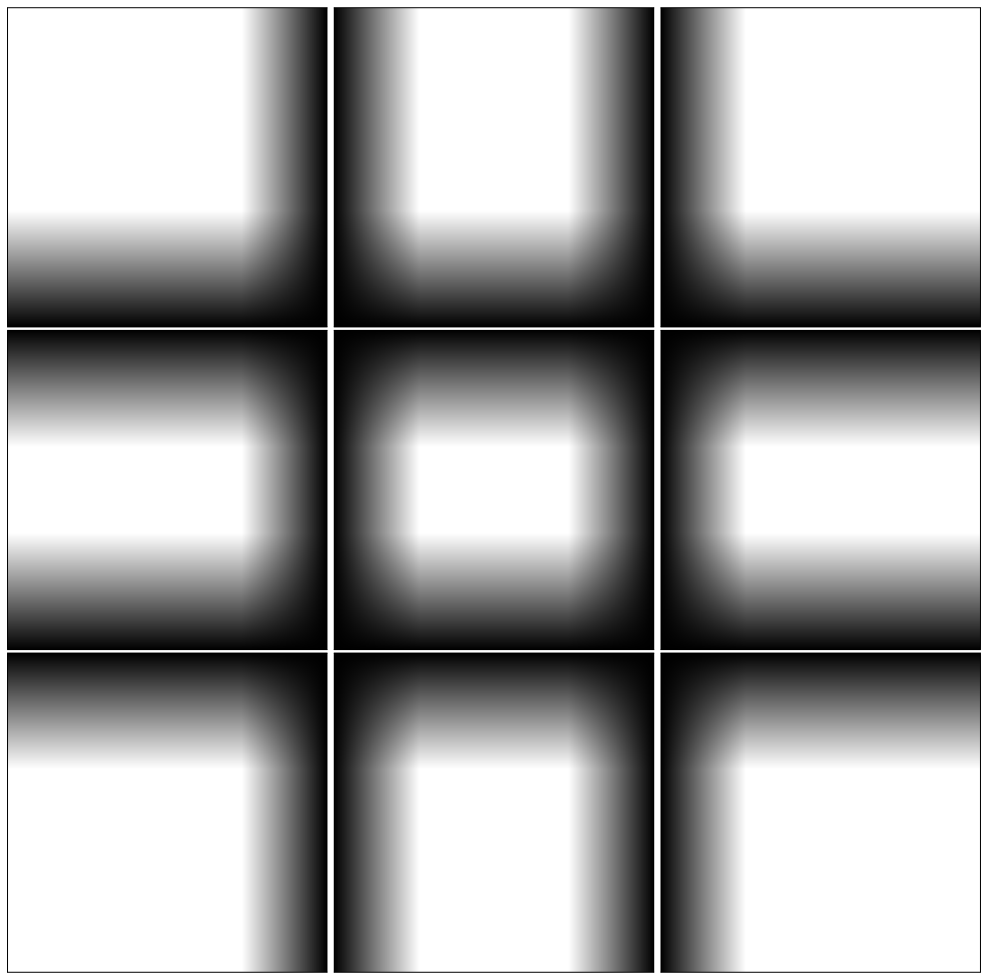

**Protocol Figure 37:** Weight tiles for merging overlapping tiles. The top left tile is overlapping with the tile to its right, the tile below, and to the tile in the middle. The centre tile is overlapping with all other tiles. The weight tile sub-images are arranged as their corresponding tiles. Note that frames are added in the above figure for clarity, but they are not present in the weight tiles.

### 2.3.4 Result post-processing

Post-processing is used to transform the segmentation network output probability maps to binary foreground and background masks. The process comprise three steps

1. Smooth the probability map
2. Binarise the smoothed probability map
3. Clean the binarised mask

The merged probability map from section 2.3.3 has the same size as the 1 MPP scan image they originate from. Before further post-processing, this probability map is downsampled by five times in both horizontal and vertical directions (corresponding to the scan image at 5 MPP).

We apply smoothing of the probability map both to get a smooth segmentation boundary in the final segmentation, and to reduce the impact of noise in the post-processing. For the sake of efficiency, the probability map is further downsampled before smoothing and upsampled again after smoothing is done. The downsampling factor is set to 0.2 for both the vertical and the horisontal direction unless the resulting image has an area less than $10^6$, in which case then the image is resized to have an area of $10^6$. This threshold is arbitrarily chosen as a safeguard against very small scans. Specifically, the new height and width is found by multiplication with a factor $\max\{0.2, \sqrt{10^6/(hw)}\}$ where $hw$ is the area of the input. Then the resulting float value is floored to get an integer value. The image then undergoes median blurring with an aperture size of 9 using `OpenCVs medianBlur` function. Next, the result is further smoothed using `OpenCVs GaussianBlur` function with a kernel size of $5 \times 5$. Finally, the smoothed probability map is upsampled back to the original size corresponding to the 5 MPP scan image.

The smooth probability map is then dichotomised into foreground and background using a hysteresis threshold method. The lower threshold value is set to 85 (1/3 of 255) and the higher threshold value is set to 229 ($\approx 90\%$ of 255).

Finally, foreground regions in the mask are pruned with the following procedure. For each connected foreground region in the foreground mask, collect the values the region cover in the smooth probability map. If the 95th percentile value of this collection is greater than 229, the corresponding region is kept as foreground, else it is labelled background.

All pixels not foreground in both the foreground mask from the probability map and the foreground mask from the scan image (section 2.2.4) are labelled as background. The resulting mask is further processed by removing small background regions and then small foreground regions as explained in section 2.2.4 for the foreground mask.

## 2.4 Performance evaluation

To measure the similarity between the reference and predicted segmentation, we use different metrics to highlight different similarities.

Since we employ the same background exclusion on both reference and prediction masks, it is of little interest to count true negative pixels in the white background area of a scan. We therefore excluded background in the performance evaluation. True negatives are therefore pixels that are marked as background in the prediction and neither as tumour nor background in the reference mask.

### 2.4.1 Overlap counting

For simple overlap comparison, we partition the pixels based on how they overlap in the reference and predicted segmentation:

137

**Protocol Table 13:** Confusion matrix

|  |  | Prediction | |  |
|---|---|---|---|---|
|  |  | Background | Foreground |  |
| Reference | Background | $TN$ | $FP$ | $RN$ |
|  | Foreground | $FN$ | $TP$ | $RP$ |
|  |  | $PN$ | $PP$ | $N$ |

$N$ = Pixel count in the image after excluding white background

$RP = |\{x : x \text{ is foreground in reference}\}|$

$RN = |\{x : x \text{ is background in reference}\}|$

$PP = |\{x : x \text{ is foreground in prediction}\}|$

$PN = |\{x : x \text{ is background in prediction}\}|$

$TP = |\{x : x \text{ is foreground in reference and prediction}\}|$

$FN = |\{x : x \text{ is foreground in reference and background in prediction}\}|$

$FP = |\{x : x \text{ is background in reference and foreground in prediction}\}|$

$TN = |\{x : x \text{ is background in reference and prediction}\}|$

These counts comprise a contingency table termed a *confusion matrix* (protocol table 13).

We can derive different metrics from the confusion matrix to measure different features of the segmentation result. Some common metrics that are used in this work are presented in the following.

True positive rate or sensitivity or recall measures the fraction of reference foreground pixels that are correctly marked as foreground

$$TPR = \frac{TP}{TP + FN} \tag{1}$$

False negative rate measures the fraction of reference foreground pixels that are wrongly marked as background

$$FNR = \frac{FN}{TP + FN} \tag{2}$$

True negative rate or specificity measures the fraction of reference background pixels that are correctly marked as background

$$TNR = \frac{TN}{TN + FP} \tag{3}$$

False positive rate measures the fraction of reference background pixels that are wrongly marked as foreground

$$FPR = \frac{FP}{TN + FP} \tag{4}$$

Positive predictive value or precision measures the fraction of predicted foreground pixels that are correctly marked as foreground

$$PPV = \frac{TP}{TP + FP} \qquad (5)$$

Negative predictive value measures the fraction of predicted background pixels that are correctly marked as background

$$NPV = \frac{TN}{TN + FN} \qquad (6)$$

Informedness

$$BIN = TPR + TNR - 1 \qquad (7)$$

Markedness

$$BMA = PPV + NPV - 1 \qquad (8)$$

Matthew's correlation coefficient is the geometric mean of informedness and markedness

$$MCC = \frac{TP \times TN - FN \times FP}{\sqrt{(TP + FN)(TP + FP)(TN + FN)(TN + FP)}} \qquad (9)$$

Sørensen-Dice similarity coefficient or $F_1$ score is the harmonic mean of the true positive rate and the positive predictive value

$$DSC = \frac{2TP}{2TP + FN + FP} \qquad (10)$$

# 3 Analyses

## 3.1 Primary analysis

The primary analysis of this study is the performance assessment of an automatic method tasked to segment tumour regions from non-tumour regions in WSIs of H&E-stained tissue sections known to contain cancerous regions. The performance is evaluated against manual segmentations in the validation cohorts described in section 1.2 using only scans from the Aperio AT2 scanner.

The single segmentation method is as described in section 2 and developed using images from cohorts described in section 1.1.

The primary analysis of each validation cohort described in section 1.2 is the segmentation method's performance measured using the Dice similarity coefficient (eq. (10)) averaged over the images in the cohort with associated 95% confidence interval. The results will also be presented as a box plot showing mean value (which is the primary metric), median value, interquartile range, whiskers and outliers for each cohort.

## 3.2 Secondary analyses

### 3.2.1 Different performance evaluation metrics of primary result

In the corresponding manner as done in the primary analysis for the Dice similarity coefficient, report the following segmentation performance evaluation metrics:

- Prevalence (amount of reference positive / total)
- Bias (amount of predicted positive / total)
- True positive rate (eq. (1))
- False negative rate (eq. (2))
- True negative rate (eq. (3))
- False positive rate (eq. (4))
- Positive predictive value (eq. (5))
- Negative predictive value (eq. (6))
- Informedness (eq. (7))
- Markedness (eq. (8))
- Matthews correlation coefficient (eq. (9))

### 3.2.2 Primary result on scans from the NanoZoomer XR

Repeat the primary analysis and the analysis in section 3.2.1 but on scans from NanoZoomer XR instead of Aperio AT2.

### 3.2.3 Single cancer type training

Repeat the primary analysis and the analysis in section 3.2.1 on methods that are developed exactly as the method analysed in the primary analysis except that the methods only have been trained on a subset of the original training set. The original training set is partitioned into one subset per cancer type, comprising scans only from that cancer type. Therefore, four segmentation methods are analysed, one for each of the cancer types

- Colorectal carcinoma
- Endometrial carcinoma

- Lung carcinoma
- Prostate carcinoma

All four additional segmentation methods will be analysed on all validation cohorts.

### 3.2.4  Primary analysis replication

Repeat the primary analysis and the analysis in section 3.2.1 on methods trained exactly as the one in the primary analysis, except for different random seed which will affect the neural network parameter initialisation and the image input pipeline. Both two additional methods will be analysed on all validation cohorts.

143



\end{document}